\documentclass[a4paper,11pt]{article}
\pdfoutput=1

\usepackage{jheppub}
\usepackage[T1]{fontenc}
\usepackage{tensor}
\usepackage{subcaption}
\hypersetup{bookmarksnumbered}
\usepackage{bbold}
\usepackage{float}

\newcommand{\be}{\begin{equation}}
\newcommand{\ee}{\end{equation}}

\newcommand{\bea}{\begin{eqnarray}}
\newcommand{\eea}{\end{eqnarray}}

\newcommand{\ba}{\begin{equation}\begin{aligned}}
\newcommand{\ea}{\end{aligned}\end{equation}}

\newcommand{\II}{{\mathcal{I}}}
\newcommand{\JJ}{{\mathcal{J}}}

\newcommand{\RR}{{\mathcal{R}}}
\newcommand{\XX}{{\mathcal{X}}}
\newcommand{\YY}{{\mathcal{Y}}}

\newcommand{\grho}{g_{\pi\pi\rho}}

\def\PP{\mathbb{P}}

\def\MM{\mathcal{M}}

\def\legP{\mathcal{P}}
\def\avg#1{\Big\langle#1\Big\rangle}
\def\tr#1{\text{Tr}\left(#1\right)}

\title{\boldmath Bootstrapping Pions at Large $N$}

\author[a,b]{Jan Albert}
\author[a]{and Leonardo Rastelli}

\affiliation[a]{C. N. Yang Institute for Theoretical Physics, Stony Brook University,\newline Stony Brook, NY 11794-3840, U.S.A.}

\affiliation[b]{Simons Center for Geometry and Physics, Stony Brook University,\newline Stony Brook, NY 11794-3636, U.S.A.}

\preprint{YITP-SB-2022-07}

\abstract{We revisit from a modern bootstrap perspective the longstanding problem of solving QCD in the large $N$  limit. We
derive universal bounds on the effective field theory of massless pions by imposing the full set of positivity constraints that follow from $2 \to 2$ scattering. Some features of our exclusion plots have intriguing connections with hadronic phenomenology. 
The exclusion boundary exhibits a sharp kink, raising the tantalizing scenario that  large $N$  QCD may sit at this kink. We critically examine this possibility, developing in the process 
 a partial analytic understanding of the geometry of the bounds.
}

\begin{document} 
\maketitle

\section{Introduction}

Solving large $N$ QCD is a longstanding open problem. 
 It has been apparent since the seminal work of 't Hooft~\cite{tHooft:1973alw} that the generalization of QCD to $N$ colors and fixed number of quarks should admit a string theory description, which becomes perturbative as $N~\to~\infty$. 't~Hooft's prophecy has been fully realized (within the standard framework of critical superstring theory, no less) for the maximally supersymmetric cousin of QCD~\cite{Maldacena:1997re, Gubser:1998bc, Witten:1998qj} and several related models. But despite  interesting attempts (see e.g.~\cite{Polyakov:1997tj, Polyakov:1998ju}), we seem still far from a concrete worldsheet description of ordinary  large~$N$ QCD.

From a  {\it spacetime} perspective, the formulation of the problem has also been clear for decades. At $N~=~\infty$,
the spectrum of QCD consists of infinite towers of stable color-singlet hadrons~\cite{tHooft:1973alw, Witten:1979kh}; 
to leading large $N$ order, hadronic  scattering amplitudes are meromorphic functions that satisfy standard crossing and unitarity constraints and have
 well-understood high-energy limits.

\subsection*{Carving out the space of large $N$ gauge theories}

It seems very natural to revisit this classic problem in the spirit of the modern bootstrap program. 
We have an infinite-dimensional set of observables (hadronic masses and spins, and their cubic on-shell couplings) obeying an infinite-dimensional set of constraints (crossing and unitarity for all possible four-point scattering amplitudes). We expect  these bootstrap equations  to admit many solutions, one for each consistent large $N$ confining gauge theory. While it is   {a priori}
unclear how to directly zoom in on the solution that corresponds to large $N$ QCD,  
we may  simply proceed to {\it carve out} the space of consistent possibilities, a strategy that has proved enormously successful
in the conformal bootstrap~\cite{Rattazzi:2008pe, Poland:2018epd, Poland:2022qrs}.
Does large $N$ QCD  sit at a special point of the exclusion boundary (as is serendipitously the case for the 3D Ising CFT~\cite{El-Showk:2012cjh})?  Could further physical input (e.g.~suitable spectral assumptions) narrow down the set of possibilities to a small numerical island?

The idea of bootstrapping the hadronic S-matrix is of course an ancient one. It predated QCD and led, via  the discovery of the Veneziano amplitude~\cite{Veneziano:1968yb}, to the development of string theory itself.
 The S-matrix bootstrap program (for general  QFTs) has undergone a recent renaissance (see~e.g.~\cite{Paulos:2016fap, Paulos:2016but, Paulos:2017fhb, He:2018uxa, Cordova:2018uop, Homrich:2019cbt, Bercini:2019vme, Cordova:2019lot, Hebbar:2020ukp, Guerrieri:2020kcs, He:2021eqn}, and~\cite{Kruczenski:2022lot} for an overview). Emulating the modern conformal bootstrap, these new developments emphasize the role of theory space and rely on powerful numerical optimization methods. These ideas have been applied to real-world QCD in~\cite{Guerrieri:2018uew, Guerrieri:2020bto}, with very intriguing results. Large $N$ leads to a major conceptual simplification.
  While the analytic
 structure of the  finite $N$ hadronic S-matrix is still far from understood, the analyticity properties at large $N$ are  uncontroversial  (see~e.g.~\cite{Veneziano:2017cks} for a recent discussion). Any $2\to 2$ connected amplitude is expected to be a meromorphic function of the Mandelstam invariants, with obvious crossing and unitarity properties; the high energy Regge behavior is controlled by the pomeron trajectory for   glueball scattering and by the rho trajectory for meson scattering.

 In this paper, we further specialize to the mesons. At leading large $N$ order, they form a consistent subsector, as only other mesons appear as intermediate states in meson-meson scattering amplitudes. (In the language of string theory, we would be  studying tree-level open string amplitudes.)
 We consider the chiral limit of vanishing quark masses. 
 Assuming that the  standard pattern of chiral symmetry breaking persists\footnote{A very safe assumption, confirmed by lattice studies for increasing values of $N$, see e.g.~\cite{Lucini:2012gg,DeGrand:2016pur,Hernandez:2019qed,Perez:2020vbn,Baeza-Ballesteros:2022azb}. General arguments in its favor were given in \cite{Coleman:1980mx, Veneziano:1980xs}.} at large $N$, the lowest lying mesons are the massless ``pions'', in the adjoint representation of the $SU(N_f)$ flavor group.  It would be straightforward to generalize our analysis to include a non-zero quark mass, but we are in fact hoping that if any analytic clues are to be found, they will show up in the zeroth order approximation to QCD, which is large $N$ in the chiral limit.

 \subsection*{Effective field theory and positivity}
There is  a neat way to organize our bootstrap problem, using the language of effective field theory (EFT). 
 We introduce a cut-off scale $M$, and divide the mesons into light states with masses smaller than $M$, and heavy states with masses larger than $M$. In principle, if we knew the full large $N$ theory,
 the EFT of the light states would arise by integrating out  the heavy states at tree level. (Recall that meson three-point vertices scale as $1/\sqrt{N}$, so at leading large $N$ order we are always justified in using the tree-level approximation.)  Instead, we decide to be agnostic about the heavy data (either than they satisfy the usual axioms) and to constrain the low-energy EFT
 by imposing its compatibility with ``healthy'' 
 scattering of the light states. 
 The computational complexity grows with the number of light states, so in the simplest setup, 
 the only light states are the massless pions; 
the  cut-off scale $M$ can then be identified with the mass of the first massive state that appears in pion-pion scattering -- in  QCD, this would be the rho vector meson. The next step is to include the rho among the light states, and so on.

It has long  been appreciated that {\it not anything goes} in EFT. For an EFT to arise as the low-energy approximation of a unitary and causal quantum field theory, its Wilson coefficients must obey certain inequalities.
These ``positivity'' bounds  have a long history originating precisely in pion physics (see~\cite{Martin1969,Pham:1985cr,Ananthanarayan:1994hf,Pennington:1994kc,Comellas:1995hq,Dita:1998mh} for some early references) and have been the subject of intense study since their significance was emphasized in~\cite{Adams:2006sv}, see~e.g.~\cite{Manohar:2008tc, Mateu:2008gv, Nicolis:2009qm, Baumann:2015nta, Bellazzini:2015cra, Bellazzini:2016xrt, Cheung:2016yqr, Bonifacio:2016wcb,  Cheung:2016wjt, deRham:2017avq, Bellazzini:2017fep, deRham:2017zjm, deRham:2017imi, Hinterbichler:2017qyt, Bonifacio:2017nnt, Bellazzini:2017bkb, Bonifacio:2018vzv, deRham:2018qqo, Zhang:2018shp, Bellazzini:2018paj, Bellazzini:2019xts, Melville:2019wyy, deRham:2019ctd, Alberte:2019xfh, Alberte:2019zhd, Bi:2019phv, Remmen:2019cyz, Ye:2019oxx, Herrero-Valea:2019hde, Bellazzini:2020cot, Tolley:2020gtv, Caron-Huot:2020cmc, Arkani-Hamed:2020blm, Sinha:2020win, Trott:2020ebl,Wang:2020jxr,Zhang:2020jyn,Zhang:2021eeo,Du:2021byy,Davighi:2021osh, Chowdhury:2021ynh,Henriksson:2021ymi,Bern:2021ppb,Caron-Huot:2021rmr,Li:2021lpe,deRham:2022hpx,Caron-Huot:2022ugt,Henriksson:2022oeu}. To get started, one must make canonical assumptions about the S-matrix (such as analyticity, crossing,
boundedness and a positive partial wave decomposition), 
which are believed to encode the fundamental principles of unitarity and causality. (As we have already remarked, all requisite axioms are very clear in our large $N$ setup.)
The basic strategy is then
to write a
(suitably subtracted) dispersion relation for the $2 \to 2$ amplitude,
which expresses low-energy
parameters in terms of an unknown but positive UV spectral density. Remarkably, one finds two-sided bounds that enforce the standard Wilsonian power counting, with higher dimensional operators suppressed by the appropriate powers of the UV cut-off.

Given the venerable history in the application of positivity bounds to pion physics~\cite{Martin1969,Pham:1985cr,Ananthanarayan:1994hf,Pennington:1994kc,Comellas:1995hq,Dita:1998mh,Manohar:2008tc,Mateu:2008gv,Alvarez:2021kpq,Guerrieri:2018uew,Guerrieri:2020bto,Bose:2020cod,Bose:2020shm,Zahed:2021fkp}, it may come as a surprise that we have something new to say. In fact, the full set of inequalities for Wilson coefficients that follow from $2\to 2$ scattering have been derived only very recently~\cite{Tolley:2020gtv, Caron-Huot:2020cmc, Arkani-Hamed:2020blm}. A key insight of these papers is that low-energy crossing symmetry implies the existence of an infinite set of ``null constraints'' for the heavy data, which in turn
can be used to optimize the numerical bounds for the low-energy parameters. These methods are ideally suited for our problem, because at large $N$ we can treat the EFT at tree level and hence derive rigorous bounds on the Wilson coefficients.

\subsection*{Results}
After generalizing the approach of~\cite{Tolley:2020gtv, Caron-Huot:2020cmc} to the kinematic setup 
of large $N$ pion-pion scattering, we obtain novel numerical bounds for the leading higher-dimensional operators of the chiral Lagrangian. Our bounds  are new because they take into account the full set of null constraints, up to numerical convergence. 
Let us highlight in this introduction a couple of our key results, to be discussed at length in the main text.
\begin{figure}[ht]
\centering
\includegraphics[scale=1.1]{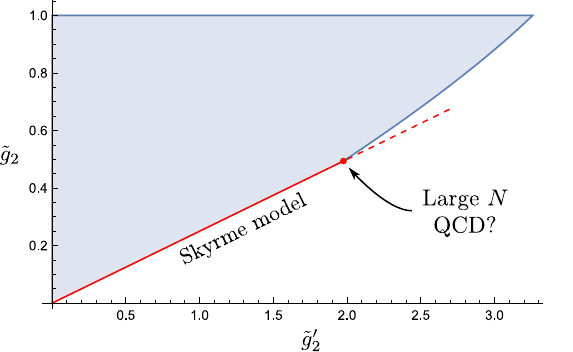}
\caption{Exclusion plot in the space of four-derivative couplings. Healthy theories must lie inside the colored region.}
\label{fig:introPlot}
\end{figure}

We display in figure \ref{fig:introPlot} 
the exclusion plot in  the space of the two four-derivative couplings (normalized by the pion decay constant and in units of the UV cut-off $M$, see (\ref{eq:gtilde})).  As expected, the 
allowed region is compact. Strikingly, the lower exclusion boundary displays a  prominent kink. Unfortunately, we  lack reliable data to place 
large $N$ QCD on this plot. There are a few lattice studies of mesons in large $N$ QCD (see~e.g.~\cite{Lucini:2012gg,DeGrand:2016pur,Hernandez:2019qed,Perez:2020vbn,Baeza-Ballesteros:2022azb}), but to the best of our knowledge the four-derivative Wilson coefficients have not been determined. The best we can currently do is to compare with real world, see figure~\ref{fig:exclPlot_experiment}. Apart from systematic errors due to finite $N$ (three of course) and the non-zero pion mass, experimental uncertainties are also quite large. Within these large errors  real-world  QCD is compatible with our bounds and perhaps prefers to sit near the lower boundary, but it seems challenging to draw any sharper conclusion. Curiously, the
bottom part of the lower bound is a straight segment  (shown in red in figure~\ref{fig:introPlot}) whose slope agrees precisely with the  choice made in the Skyrme model, i.e.~with the combination of $O(p^4)$ terms that has at most two time derivatives.

\begin{figure}[ht]
\centering
\includegraphics[scale=1.1]{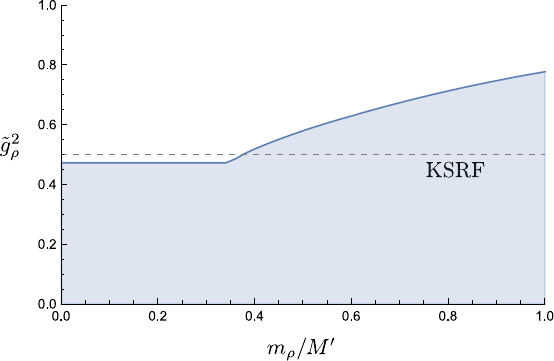}
\caption{Upper bound on $\grho$ (normalized by $f_\pi$ and $m_\rho$, see \eqref{eq:normcouplingsRho}) as a function of the gap after the rho.}
\label{fig:introPlot2}
\end{figure}

It is a  well-known experimental fact (also confirmed by large $N$ lattice studies) that the lightest resonance  in pion-pion scattering is the 
 rho vector meson.\footnote{This is strictly true only at large $N$, as we review in appendix \ref{app:internalstates}.} If we make this spectral assumption, including the rho among the light states, the new cut-off $M'$ is the mass of the next meson (in QCD, this is the $f_2(1270)$ state). This allows us to put an 
 upper bound for the $g_{\pi \pi \rho}$ coupling, as shown in figure~\ref{fig:introPlot2}. Again, we see an interesting kink. We also find another curious connection with another  bit of ``voodoo QCD''.\footnote{Expression attributed by R.L. Jaffe to Bjorken, who coined it to denote a few mysteriously successful phenomenological models of  the strong interactions.} The value of  $g_{\pi \pi \rho}$ at the plateau is rather close
 (but not equal) to the one that corresponds to the  phenomenologically successful KSRF relation~\cite{Kawarabayashi:1966kd,Riazuddin:1966sw}. 

What is the physical significance of the kinks that we find in our plots?
Do they  correspond to large $N$ QCD (or, possibly, to some other interesting large $N$ gauge  theory)? Following \cite{Caron-Huot:2020cmc}, we are able to easily find simple extremal (unphysical) amplitudes that ``explain'' a large portion of the allowed 
region in figure~\ref{fig:introPlot}, but leave out the most interesting part, a sliver where the kink sits.
Several numerical experiments (playing with gaps and spectral assumptions)
lead us to the hypothesis that two simple UV completions
of the tree-level
rho exchange amplitude might in fact explain the entire geometry of the bounds;
the kink would arise because the two UV completions exchange dominance there.
However we only succeeded in finding  analytic expressions
that rule in a part but not the entirety of the sliver, notably leaving out the kink, see figure~\ref{fig:rulingin2}. This may indicate a lack of imagination on our part, or that the physics of the kink is richer. We  hasten to add that even if our hypothesis is correct (that there exist simple unphysical amplitudes  sitting at the exclusion boundary in a two-dimensional projection of the full parameter space), this would not rule out that an actual physical theory such as QCD may saturate the same bounds. 

The question of whether large $N$ QCD sits at the kink remains open. This paper is just a first step in a systematic program; we are optimistic to be able to address this and many other questions in future work.

\bigskip
\noindent 
The detailed organization of the paper is best  apprehended from the table of contents. In section \ref{sec:setup}, we  set up the problem, spelling out our analyticity, unitarity and Regge boundedness assumptions for pion scattering at large $N$. In section \ref{sec:disprels}, we use dispersion relations to derive  ``positive'' sum rules for the low-energy parameters in the pion EFT. In section \ref{sec:bounds} we use semidefinite programming methods to carve out the space of consistent low-energy parameters. We also compare our results with the previous literature and with the real world. In section \ref{sec:rho},
we include an explicit rho vector meson in the pion-pion amplitude, and derive bounds for the $g_{\pi \pi \rho}$ coupling. In section \ref{sec:rulingin} we look for an analytic (or at least conceptual) understanding of the geometry of our bounds.
We conclude in section~\ref{sec:outlook} with a brief discussion and  directions for future work. In appendix~\ref{app:internalstates} we review the standard nomenclature for mesons and the selection rules that apply to pion-pion scattering at large~$N$. In appendix~\ref{app:Nf} we give the generalization of our kinematic setup to a general number $N_f$ of fundamental quarks.

\section{Setup and assumptions}\label{sec:setup}
Consider four-dimensional $SU(N)$ Yang-Mills theory with $N_f$ fundamental massless Dirac fermions,
in the standard large $N$ 't Hooft limit~\cite{tHooft:1973alw}.
We will make the usual (uncontroversial)  assumption that the theory remains confining at large $N$, so that that the asymptotic states are color-singlet glueballs, mesons and (heavy) baryons \cite{Witten:1979kh}. Here we will be concerned with the mesons, and in particular with the scattering of the lowest-lying ones. We further assume the standard pattern of chiral symmetry breaking, so that the lightest mesons are the  massless Goldstone bosons $\pi^a$ ($a = 1, 2, \ldots, N_f^2-1$),  in the adjoint representation of $SU(N_f)$. For $N_f=2$, this is the isospin triplet of pions and for $N_f=3$, the octet of pions, kaons and the eta. By a slight abuse of terminology, we will refer to them 
as pions regardless of $N_f$.

\subsection{Parametrizations and crossing symmetry}  \label{sec:parametriz}
\begin{figure}[htb]
\centering
\includegraphics[scale=0.4]{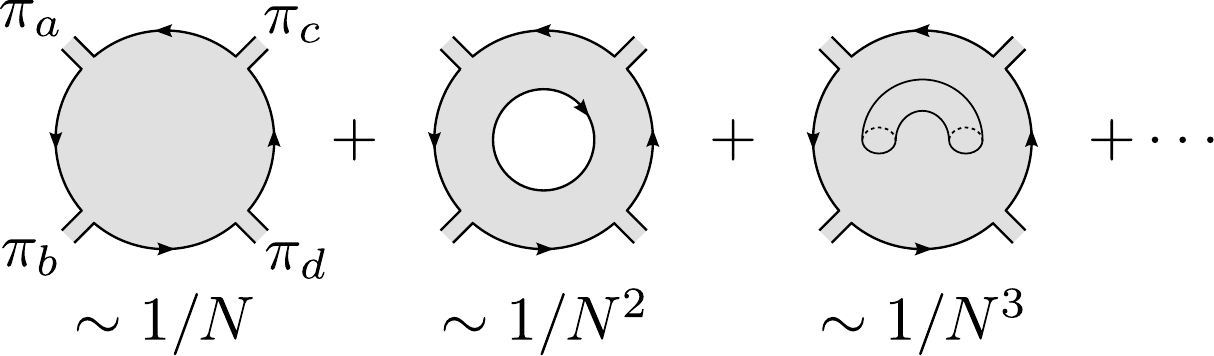}
\caption{As $N\to \infty$ the diagrams contributing to $\pi\pi\to\pi\pi$ scattering arrange in a topological expansion in powers of $1/N$. At leading order, only planar diagrams with the topology of a disk survive. Quark loops are suppressed by $1/N$ and handles (which make the diagram non-planar) by $1/N^2$. Solid oriented lines denote quark propagators, the shaded area represents all possible (planar) dressings with gluons.}
\label{fig:LargeNexp}
\end{figure}

In the $1/N$ expansion, quark loops and non-planar diagrams are down by powers of $1/N$, so at leading order only diagrams with the topology of a disk delimited by an external quark loop contribute to meson scattering~\cite{tHooft:1973alw,Witten:1979kh} (see figure \ref{fig:LargeNexp}). The only dependence on the flavor indices of these diagrams comes as a single trace of the $SU(N_f)$ generators. Therefore, at leading large $N$ order the scattering amplitude for the process $\pi^a\pi^b\to \pi^c\pi^d$ can be parametrized  as\footnote{Our conventions for the $SU(N_f)$ generators in the defining representation $T_a$ are
$$\tr{T_aT_b}=\frac{1}{2}\delta_{ab}\,,\qquad \left[T_a,T_b\right]=if\indices{_{ab}^c}\, T_c\,, \qquad \tr{T_a\left\{T_b,T_c\right\}}=\frac{1}{2}d_{abc}\,.$$
Adjoint indices $a,b,c,\ldots$ can be raised and lowered with the Kronecker delta $\delta^a_b$  and can thus be treated as equivalent.}
\begin{align} \label{eq:Tcp}
{\cal T}_{ab}^{cd}  = \,& 4\left[\tr{T_aT_bT^cT^d}+\tr{T_aT^dT^cT_b}\right] M(s, t)\nonumber \\ 
+\,& 4\left[\tr{T_aT^cT^dT_b}+\tr{T_aT_bT^dT^c}\right] M(s, u)\nonumber \\
+\,& 4\left[\tr{T_aT^dT_bT^c}+\tr{T_aT^cT_bT^d}\right] M(t, u)\,.
\end{align}
Since we are scattering identical particles, the basic amplitude $M(s,u)$ enjoys $s\leftrightarrow u$ crossing symmetry, i.e.\
\begin{equation}
    M(s, u) = M(u, s)\,. 
\end{equation}
It does not, however, enjoy full $s\leftrightarrow t \leftrightarrow u$ crossing symmetry because the outer quark loop fixes the ordering of the external states. This parametrization is very natural in the language of tree-level string theory, where $M(s,u)$ is the basic disk amplitude (such as the Beta function in the case of the Veneziano amplitude) and the flavor structure is introduced via Chan-Paton factors. But at this stage, \eqref{eq:Tcp} is purely a kinematic statement.

Although we will mostly use \eqref{eq:Tcp}, another parametrization that will prove useful is
\begin{align} \label{eq:T}
{\cal T}_{ab}^{cd} \,&= A(s| t, u) \left(\frac{2}{N_f}\delta_{ab} \delta^{cd}+d_{abe}d^{cde}\right)\nonumber\\
&\;+ A(t| s, u) \left(\frac{2}{N_f}\delta_{a}^d \delta^{c}_b + d\indices{_a^d_e}d\indices{_b^{ce}}\right)\nonumber\\
&\;+ A(u| s, t) \left(\frac{2}{N_f}\delta_{a}^c \delta^{d}_b + d\indices{_a^c_e}d\indices{^d_b^e}\right) \,.
\end{align}
Crossing symmetry is now the statement that
\begin{equation}
    A(s| t, u) = A(s| u, t)  \,.
\end{equation}
Using
\begin{align}
    \tr{T_aT_bT^cT^d}+\tr{T_aT^dT^cT_b}=&\;\frac{1}{2N_f}\left(\delta_{ab} \delta^{cd}+\delta_{a}^d \delta^{c}_b-\delta_{a}^c \delta^{d}_b\right)\nonumber\\
    &+\;\frac{1}{4}\left(d_{abe}d^{cde}+d\indices{_a^d_e}d\indices{_b^{ce}}-d\indices{_a^c_e}d\indices{^d_b^e}\right)\,,
\end{align}
we see that the two parametrizations are related by
\begin{equation} \label{eq:AtoM}
A(s | t, u) =  M(s, t) + M(s, u)  - M(t, u) \,, \quad  2 M(s, u) = A(s |t, u) + A(u | s, t) \,.
\end{equation}

\subsection*{The $\boldsymbol{\eta'}$ meson}
A subtlety of the large $N$ expansion is that the axial anomaly is suppressed in this limit (see e.g.\ \cite{Witten:1979vv,Veneziano:1979ec, Leutwyler:1997yr,Kaiser:2000gs}). The axial $U(1)$ is then non-anomalous at $N=\infty$ and its spontaneous breaking brings in a new Goldstone boson; the ---now massless--- $\eta'$ meson. More precisely, the pattern of chiral symmetry breaking gets upgraded from $SU(N_f)_\text{L}\times SU(N_f)_\text{R}\times U(1)_\text{V}\to U(N_f)_\text{V}$ to $U(N_f)_\text{L}\times U(N_f)_\text{R}\to U(N_f)_\text{V}$, so the $\eta'$ meson becomes degenerate with the pions forming a multiplet of $U(N_f)$. For this reason, to incorporate the $\eta'$ in pion scattering we just have to consider the additional generator $T_0=\frac{1}{\sqrt{2N_f}}\mathbb{1}$,\footnote{The normalization of this generator proportional to the identity is chosen such that $\tr{T_0T_0}=\frac{1}{2}$ to match the normalization of the other generators.} associated to the determinant of the matrices in $U(N_f)$, for any leg involving an $\eta'$. Thus, scattering processes between pions and the $\eta'$ can be described with the same basic amplitudes as above. Indeed, defining the fully $s\leftrightarrow t\leftrightarrow u$ symmetric amplitude
\begin{equation}
    A_{stu}\equiv A(s|t,u) + A(t|s,u) + A(u|s,t) = M(s, t) + M(s, u) + M(t, u)\,,
\end{equation}
we have respectively for the processes $2\eta'\to2\eta'$, $2\eta'\to\eta'\pi$, $2\eta'\to2\pi$, $\eta'\pi\to 2\pi$, the amplitudes
\begin{equation}
    {\cal T}_{00}^{00}=\frac{2}{N_f}A_{stu}\,,\quad
    {\cal T}_{00}^{0d}=0\,,\quad
    {\cal T}_{00}^{cd}=\frac{2}{N_f}A_{stu}\delta^{cd}\,,\quad
    {\cal T}_{0b}^{cd}  = \sqrt{\frac{2}{N_f}}A_{stu}d\indices{_b^{cd}}\,.
\end{equation}
Since including the $\eta'$ does not bring any additional information, we will only consider the scattering of $SU(N_f)$ pions from this point forward.

\subsection{Analyticity and Zweig's rule}\label{sec:zweig}
We can further rewrite the amplitude ${\cal T}_{ab}^{cd}$ as a sum over the $SU(N_f)$ irreducible representations (irreps) that can appear as intermediate states. For the sake of clarity, we restrict to $N_f=2$ for the following discussion, but our conclusions will actually apply to any $N_f$. The derivation for general $N_f$ is given in appendix \ref{app:Nf}. The $2\to 2$ scattering of $SU(2)$ pions splits into three isospin channels, $\II=1\otimes 1=0\oplus 1\oplus 2$. So we can rewrite the amplitude as
\begin{equation}\label{eq:Idecomp}
    {\cal T}   =   \sum_{\II=0}^2   \MM^{\II} (s | t, u) \PP^{\II}_s\,,
\end{equation}
where $\PP^{\II}_s$ are the $s$-channel isospin projectors
\begin{equation}
    \PP^{0}_s=\frac{1}{3}\delta_{ab}\delta^{cd}\,,\quad
    \PP^{1}_s=\frac{1}{2}\left(\delta_a^c\delta_b^d-\delta_a^d\delta_b^c\right)\,,\quad
    \PP^{2}_s=\frac{1}{2}\left(\delta_a^c\delta_b^d+\delta_a^d\delta_b^c-\frac{2}{3}\delta_{ab}\delta^{cd}\right)\,,
\end{equation}
and the amplitudes in the different channels are related to the previous parametrizations by
\begin{subequations}\label{eq:MIs}
\begin{align}
    \MM^{0} (s | t, u)=\;& 3 A(s|t,u) + A(t|s,u) + A(u|s,t) = 3M(s,t) + 3M(s,u) - M(t,u)\,,\\
    \MM^{1} (s | t, u)=\;& A(u|s,t) - A(t|s,u) = 2\Big(M(s,u) - M(s,t)\Big)\,,\\
    \label{eq:MI2}
    \MM^{2} (s | t, u)=\;& A(t|s,u) + A(u|s,t) = 2M(t,u)\,.    
\end{align}
\end{subequations}
Note that under $t\leftrightarrow u$ crossing $\MM^{\II} (s | t, u)$ is symmetric for $\II=0,2$ and antisymmetric for $\II=1$. Under $s\leftrightarrow t$ crossing they mix into one another. 

At leading order in $1/N$, the pion scattering amplitude reduces to an infinite sum of tree diagrams corresponding to the exchange of physical mesons \cite{Witten:1979kh}. Thus, for fixed $u<0$, $M(s,u)$ is a meromorphic function with poles on the real $s$ axis. Also at $N=\infty$, mesons are exactly $q\bar q$ states, i.e.\ there are no exotic mesons. Since massless quarks have isospin $\II=\frac{1}{2}$, this implies that the physical intermediate states in large $N$ pion scattering can only carry two possible representations; $\II= \frac{1}{2} \otimes \frac{1}{2}=0\oplus 1$.

Diagrams contributing to the $\II=2$ channel either come from the exchange of exotic mesons or are such that the initial and final states can be separated by cutting only internal gluon lines. The former cannot happen at large $N$, and the latter are processes suppressed by Zweig's rule, which becomes exact in the limit $N\to\infty$. In the language of string theory, these two types of process correspond respectively to exchanging multiple open strings or exchanging a closed string and are therefore down by powers of the string coupling constant. Of course, this should only be taken as a picture, we are emphatically not making any dynamical assumption that the theory must be a theory of strings -- we are just imposing large $N$ selection rules.

In summary, the isospin-two amplitude cannot have physical poles at large $N$. In other words, \ $\MM^{2} (s | t, u)$ for fixed $u<0$ must be analytic on the real $s>0$ axis. From \eqref{eq:MI2} we conclude that \textit{$M(s,u)$ (for fixed $u<0$) only has poles on the positive real $s$ axis}, i.e.~it does not have $t$-channel poles. Indeed, a pole at negative $s$ would correspond to a pole for positive $t$ which would contribute a physical state in the partial wave expansion of $\MM^{2}(s|t,u)$.

\subsection{Unitarity}\label{sec:unitarity}
Each $\MM^{\II} (s| t, u)$ admits an s-channel partial wave expansion,\footnote{In our conventions for the Mandelstam invariants, 
\be
s = - (p_1 + p_2)^2 \, , \quad t = - (p_1 - p_4)^2\, , \quad u = - (p_1 - p_3)^2 \,,
\ee
the scattering angle $\theta$ for $1 \to 3$ is given by $\cos (\theta) = 1 + \frac{2u}{s}$.
In some of the literature the definitions of $t$ and $u$ are interchanged.}
\be  \label{partialwaves}
  \MM^{\II} (s | t, u)  =
s^{\frac{4-D}{2}} \sum_{J} n^{(D)}_J c^{(\II)}_{J}  (s) \legP_J\left(1+\frac{2u}{s}\right)\,,
\ee
where the normalization constant is chosen as\footnote{We  follow the normalization conventions of \cite{Correia:2020xtr, Caron-Huot:2020cmc}. In this paper we are ultimately interested in $D=4$, but we keep $D$ general for as long as possible, anticipating  future applications of our program to other spacetime dimensions.}
\begin{equation}
    n_J^{(D)}\equiv \frac{2^{D} \pi^{\frac{D-2}{2}}}{\Gamma(\tfrac{D-2}{2})} (J+1)_{D-4} (2J+D-3)\,,
\end{equation}
so that the Gegenbauer polynomials are defined by
\be \legP_J(x) \equiv {}_2F_1\left(-J,J+D-3,\frac{D-2}{2},\frac{1-x}{2}\right)\, .\ee
With these normalizations, unitarity for the amplitude $\MM^{\II} (s | t, u)$ implies $|1+ic^{(\II)}_J (s) |^2\leq 1$ in the physical regime $s>0$. 
Defining the spectral densities $\rho^{(\II)}_J (s) = {\rm Im}\, c^{(\II)}_J(s)$, this condition translates into
\begin{equation}\label{eq:untryrhoI}
    2 \geq \rho^{(\II)}_J(s) \geq 0 \, ,  \qquad s > 0 \,.
\end{equation}

For the full amplitude ${\cal T}_{ab}^{cd}$ to define a unitary $S$-matrix, we need each of the isospin amplitudes \eqref{eq:MIs} to be separately unitary. As discussed above, the $\II=2$ amplitude is analytic in the physical region, hence $\rho^{(2)}_J (s)=0$ for $s>0$ (at leading order in large $N$). Using the identity $\legP_J(-x)=(-1)^J\legP_J(x)$ and the symmetry properties of \eqref{eq:MIs} under $t\leftrightarrow u$ crossing, we see that $\rho^{(0)}_J (s)$ is only nonvanishing for $J$ even and $\rho^{(1)}_J (s)$ for $J$ odd. Thus, if we expand the basic amplitude $M(s,u)$ as
\be \label{eq:ImMbasic}
{\rm Im} \,  M(s,u)  = s^{\frac{4-D}{2}} \sum_{J} n^{(D)}_J \rho_J(s)\, \legP_J\left( {1+\frac{2u}{s}} \right) \, ,
\ee
with
\begin{subequations}
\bea
 \rho_J(s) &  =  & \frac{1}{6}\rho^{(0)}_J(s)  \qquad s > 0 \,, \;J \;{\rm even}\, ,\\
\rho_J(s) &  =  & \frac{1}{4}\rho^{(1)}_J(s)  \qquad s > 0 \,, \;J \;{\rm odd}\, ,
\eea
\end{subequations}
unitarity of the full scattering amplitude implies positivity of $M(s,u)$. That is,
\begin{equation}\label{eq:unitarity}
    \rho_J(s) \geq 0 \, ,  \qquad s > 0 \,, \; \text{all }J\,.
\end{equation}

For a meromorphic function, the only contribution to the spectral density comes from the isolated poles, which give delta functions,
\begin{equation}\label{eq:meromorphic}
    M(s,u)\sim g_{\pi\pi\alpha}^2\frac{m_\alpha^2\legP_J\left(1+\frac{2u}{s}\right)}{m_\alpha^2-s}\,,\quad \longrightarrow \quad 
    n_J^{(D)}\rho_J(s)\sim\pi \,g_{\pi\pi\alpha}^2\delta(s-m_\alpha^2)\,m_\alpha^{D-2}\,.
\end{equation}
Then, \eqref{eq:unitarity} implies the positivity of the coupling constants squared, $g_{\pi\pi\alpha}^2\geq 0$. We will not use the upper bounds on $g_{\pi\pi\alpha}^2$ coming from \eqref{eq:untryrhoI} because meson interactions scale as $g_{\pi\pi\alpha}\sim 1/\sqrt{N}$ and thus die at large $N$. (This is precisely what allows us to treat the meson theory at tree level in the first place.) The positivity condition will be enough. However, we will mostly use it in the form of \eqref{eq:unitarity} because we will parametrize our ignorance of the meson spectrum by a whole cut on the real $s$ axis rather than isolated poles.

\subsection{Regge behavior}
At finite $N$,  pion scattering  is believed to  have Regge behavior controlled by the pomeron trajectory, which needs to have intercept above one to account for the rise with energy of the total cross section, see e.g.~the discussion in \cite{Brower:2006ea}.
The pomeron trajectory corresponds to glueball states and is  suppressed at large $N$.  In string theory language, the pomeron is a closed string trajectory, and contributes to an open string amplitude via a subleading (non-planar) topology. The dominant trajectories after the pomeron are the rho trajectory
and the $P'$ trajectory (associated to the $f_2$(1270) resonance). In the analysis of appendix B of \cite{Pelaez:2004vs} these two trajectories are taken to have the same intercept, $\alpha_\rho (0) = \alpha_{P'} (0) \cong 0.52$. So
 to leading order in $N$  it seems safe to assume that the Regge behavior of ${\cal T}_{ab}^{cd}$ is strictly better than spin one. 
 This translates into two conditions on $M(s, u)$,
\be \label{eq:Regge}
\lim_{|s| \to \infty} \frac{M(s, u)}{s} = 0 \, ,  \quad  \lim_{|s| \to \infty} \frac{M(s, -s -u)}{s} = 0 \, ,  \quad u \lesssim 0 \, .
\ee
They are respectively the limits at fixed $u$ and at fixed $t$ of $M(s,u)$, which are independent when the amplitude is not fully crossing symmetric.

We will study these amplitudes around $u\sim 0$. For the first amplitude in \eqref{eq:Regge} this corresponds to the usual \textit{forward limit} $\cos \theta\sim 1$, for the second one, this is the ``backward limit'' $\cos \theta \sim -1$. It would be very interesting to study scattering at fixed angle $\theta$ (away from $0$ and $\pi$) since this could discern between standard string amplitudes, which decay exponentially at high energies, and amplitudes in QCD-like theories, which are expected to decay as $1/s^2$ by the Brodsky-Farrar counting rules \cite{Brodsky:1973kr,Lepage:1980fj}. Also at fixed angle, albeit in the unphysical regime $s,u\gg 1$, scattering amplitudes of large-$N$ confining gauge theories are known to follow a universal behavior \cite{Caron-Huot:2016icg}. Unfortunately, fixed-angle scattering amplitudes fall outside the scope of our methods.

\subsection{An example: The Lovelace-Shapiro amplitude}
A remarkable function that obeys all of our assumptions for $M(s, u)$ at large $N$ is the Lovelace-Shapiro amplitude \cite{Lovelace:1968kjy,Shapiro:1969km} (see  \cite{Bianchi:2020cfc} for a recent discussion),
\be \label{eq:LS}
M_{\rm LS}(s, u)   =- \frac{\Gamma(1/2 - \alpha' s)\Gamma(1/2 - \alpha' u)}{\Gamma(- \alpha' (s + u))}\,.
\ee
For fixed $u$, this amplitude has  poles only for positive $s$, with a positive-definite partial wave expansion.\footnote{Positivity is by no means obvious and, at present, it can only be understood from its relation to the NSR string, as explained in \cite{Bianchi:2020cfc}. See \cite{Arkani-Hamed:2022gsa} for recent developments on  a more direct understanding of unitarity of string amplitudes.} As discussed, this ensures that it does not have physical poles in the $\II=2$ channel. Note also the Adler zero $M_{\rm LS}(s, u) \sim \pi  \alpha'(s+u)$ for $s, u\to 0$, which agrees with the fact that Goldstone bosons have derivative couplings.

\section{Sum rules from dispersion relations} \label{sec:disprels}
The Regge behavior \eqref{eq:Regge} implies that for fixed $u<0$, all the isospin amplitudes $\MM^{\II}(s|t,u)$ grow more slowly than $\sim s$ at infinity. Thus, we can obtain three sets of $k$-subtracted dispersion relations by taking contour integrals in $s$ around infinity for each of these amplitudes,
\begin{equation}
    \frac{1}{2 \pi i } \oint_\infty ds'\,\frac{\MM^{\II}(s'|-s'-u, u)}{s'^{k+1}} = 0\, ,\quad k=1,2, \dots\,, \; \II=0,1,2\,.
\end{equation}
By \eqref{eq:MIs}, these relations can be traded for the following dispersion relations of the basic amplitude $M(s,u)$,
\begin{equation}\label{eq:disprels}
    \frac{1}{2 \pi i } \oint_\infty ds'\,\frac{M(s', u)}{s'^{k+1}}=
    \frac{1}{2 \pi i } \oint_\infty ds'\,\frac{M(-s'-u, u)}{s'^{k+1}}=
    \frac{1}{2 \pi i } \oint_\infty ds'\,\frac{M(s', -s'-u)}{s'^{k+1}}=0\,.
\end{equation}
For $N_f>2$, instead of the three isospin channels we should consider all the channels discussed in appendix \ref{app:Nf}, but that does not bring in additional constraints since all the amplitudes can be written in terms of the three crossed versions of $M(s,u)$ (see \eqref{eq:channels}). The three sets from \eqref{eq:disprels} are all we need independently from the number of flavors. In fact, we only need the first and last one of them. Under the change of variables $s'\to-s'-u$, the second set of dispersion relations in \eqref{eq:disprels} reduces to the first one (albeit with the subtraction shifted from 0 to $-u$) and it is thus redundant.

\begin{figure}
\centering
\begin{subfigure}[b]{0.75\textwidth}
   \includegraphics[width=\linewidth]{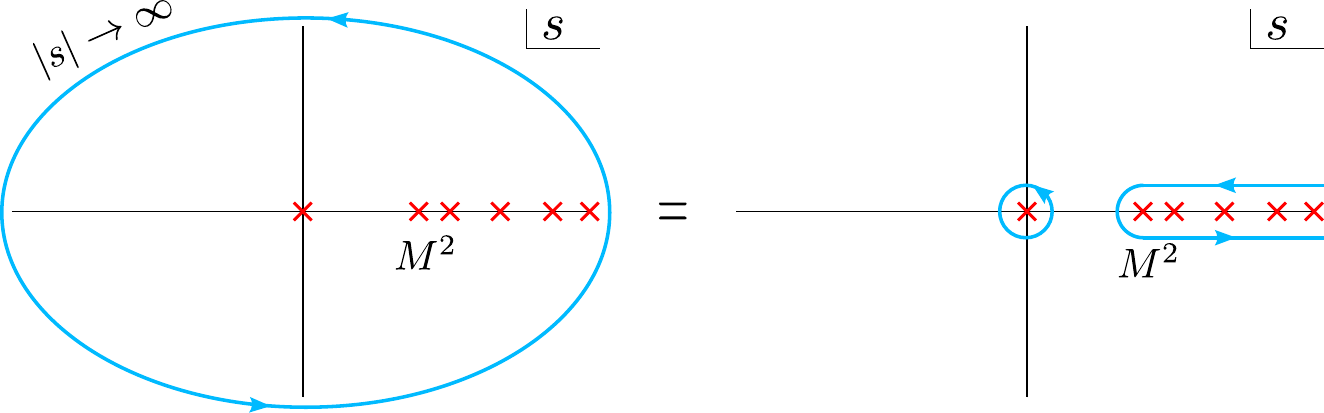}
   \caption{SU dispersion relations.}
\end{subfigure}
\begin{subfigure}[b]{0.75\textwidth}
   \includegraphics[width=1\linewidth]{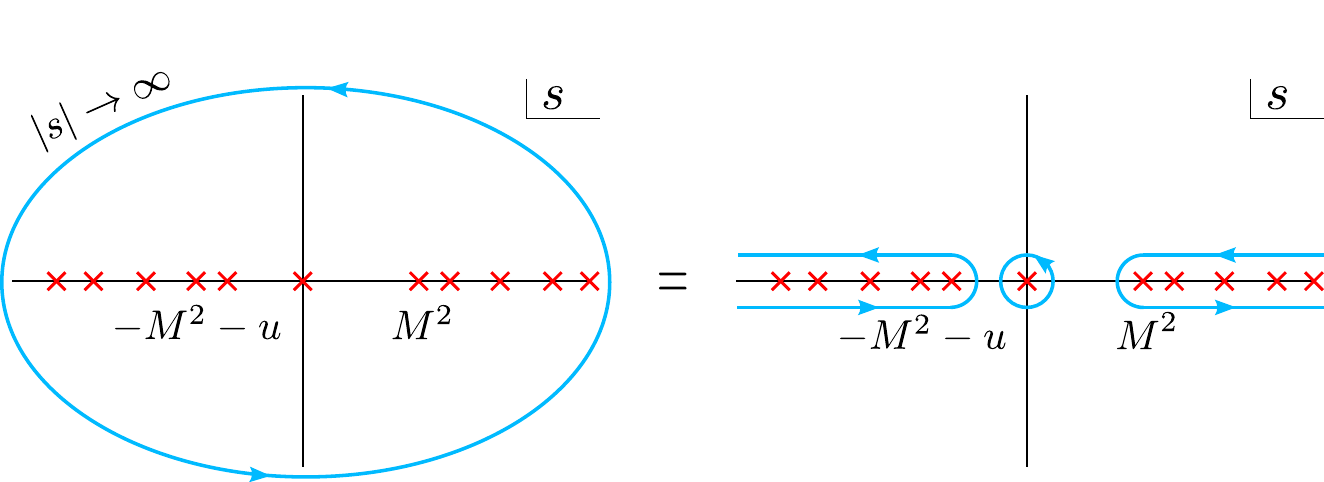}
   \caption{ST dispersion relations.}
\end{subfigure}
\caption{\label{fig:contours}Contour deformations for the two independent sets of dispersion relations. The high-energy poles correspond to the  meson spectrum.}
\end{figure}

As discussed above, $M(s,u)$ has physical poles in $s$ and $u$ (related by crossing symmetry) but not in $t$. Thus, for fixed $u<0$, $M(s,u)$ only has poles on the positive side of the real $s$ axis while $M(s,-s-u)$ has poles on both sides. Using that the amplitude is analytic away from the real $s$ axis, we can safely deform the contours towards the real axis as shown in figure \ref{fig:contours}. We will separate low and high energies by a cutoff scale $M^2$ such that all the poles lie above the cutoff and $M(s,u)$ is analytic below it. One should think of $M^2$ as the mass of the first exchanged meson in the spectrum. Then, with the contour deformations of figure \ref{fig:contours}, the two sets of independent dispersion relations relate low and high energies by
\begin{subequations}\label{eq:disprels2}
\begin{equation}\label{eq:disprelSU}
    \text{SU:}\quad\frac{1}{2 \pi i } \oint_0 ds'\,\frac{M_{\text{low}}(s', u)}{s'^{k+1}} =
    \frac{1}{\pi} \int_{M^2}^\infty ds' \frac{\text{Im}\,M_{\rm high} (s', u) }{ s'^{k+1} } \,,  \qquad k=1,2, \dots
\end{equation}
\begin{align}\label{eq:disprelST}
    \text{ST:}\quad\frac{1}{2 \pi i } \oint_0 ds'\,\frac{M_{\text{low}}(s', -s'-u)}{s'^{k+1}} &= 
    \frac{1}{\pi} \int_{M^2}^\infty ds' \frac{\text{Im}\,M_{\rm high} (s', -s'-u) }{ s'^{k+1} } \quad\qquad k=1,2, \dots\,\nonumber\\
    &+\frac{1}{\pi} \int_{M^2}^\infty ds' (-1)^k\frac{\text{Im}\,M_{\rm high} (s', -s'-u) }{ (s'+u)^{k+1} }\,,
\end{align}
\end{subequations}
where we have used that the discontinuity of $M(s,u)$ is equal to its imaginary part, $\text{Im}\,M(s,u)=\frac{1}{2i}\left[M(s+i\epsilon,u)-M(s-i\epsilon,u)\right]$. These two independent sets of dispersion relations are in essence ``fixed $u$'' and ``fixed $t$'' dispersion relations. There are two of them because the amplitude is $s\leftrightarrow u$ crossing symmetric. For a fully $s\leftrightarrow t\leftrightarrow u$ symmetric theory, both sets would be equivalent.

\subsection{Low energy: Effective field theory}
As usual, we describe the physics at low energy with an effective field theory (EFT). The standard EFT for pion scattering is the chiral Lagrangian \cite{Gasser:1983yg,Gasser:1984gg}. The spontaneously broken chiral symmetry determines the $n$-point vertices,
$n>4$, in terms of the 4-point  interactions (with arbitrarily many derivatives),
but  since we are only studying pion $2 \to 2$ scattering, we will be blind to this fact. The  only input about the Goldstone boson nature of the pions that we can try to impose  (apart that they are massless) is that they are derivatively coupled, but in fact the need to perform at least one substraction means that the quartic non-derivative coupling drops out from our sum rules.

Like any other EFT, the chiral Lagrangian comprises all the terms that are consistent with the symmetries of the theory, but with unfixed coefficients. If we knew the full underlying theory, we could compute these coefficients by integrating out the higher mesons in the spectrum, but since that is not the case, the coefficients must be fixed differently. The usual way of doing so is by comparison to experiment or from lattice computations. Our approach will be to \textit{bootstrap} them, i.e.\ we will ask what values of these coefficients are compatible with an underlying theory satisfying the assumptions of section \ref{sec:setup}.

\begin{figure}[htb]
\centering
\includegraphics[scale=0.4]{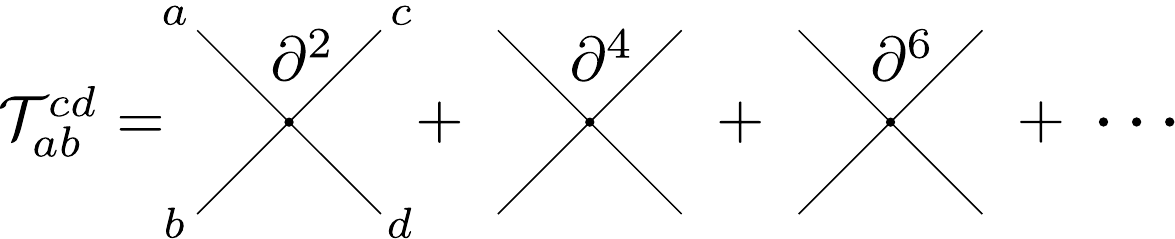}
\caption{The EFT amplitude for $2\to 2$ pion scattering is given by an infinite sum of four-point vertices with unknown coefficients. The sum can be arranged into a series with increasing number of derivatives in the interaction.}
\label{fig:EFTdiagrams}
\end{figure}

Instead of using the EFT Lagrangian, which is defined up to field redefinitions and integration by parts, we define the low-energy coefficients directly at the level of the amplitude, which is unambiguous. We defer the relation to the chiral Lagrangian to section \ref{sec:whereartthou}. At the level of the amplitude, integrating out the heavy mesons leaves a series of four-point interaction vertices as in figure \ref{fig:EFTdiagrams}. Thus, for energies below the cutoff we can describe $M(s,u)$ by
\bea\label{eq:Mlow}
{M}_{\rm low} (s, u)  & =  & \sum_{n=1}^\infty \sum_{\ell = 0}^{[n/2]} g_{n, \ell} \, (s^{n-\ell} u^\ell + u^{n-\ell} s^\ell) \\
& = & g_{1, 0} (s+u) + g_{2, 0} (s^2 + u^2) + 2 g_{2, 1} \, su +  g_{3, 0} (s^3 + u^3) + g_{3, 1} (s^2 u + u^2 s) + \dots\,, \nonumber
\eea
where $g_{n, \ell}$ are the low energy coefficients that we will bound. They represent the (unknown) coupling constants of the four-point vertices of figure \ref{fig:EFTdiagrams}. Equation \eqref{eq:Mlow} is essentially the crossing-symmetric Taylor expansion of the full amplitude $M(s,u)$ around $s\sim u\sim 0$, which converges as long we do not step on any singularity, i.e.\ at energies below $M^2$. Note also that \eqref{eq:Mlow} incorporates the Adler zero ($M(s,u)\to 0$ as $s,u\to 0$), as befits the derivative couplings of Goldstone bosons.

\subsection{High energy: Partial wave unitarity}
On the other hand, we will be agnostic about the physics above the cutoff. Even though we know that $M(s,u)$ becomes meromorphic in the $N\to \infty$ limit, we do not know the position of the poles or their coupling constants. Therefore, at high energies we will expand $M(s,u)$ in partial waves as in \eqref{eq:ImMbasic},
\begin{equation}\label{eq:Mhigh}
    {\rm Im} \,  M_{\text{high}}(s,u)  = s^{\frac{4-D}{2}} \sum_{J} n^{(D)}_J \rho_J(s)\, \legP_J\left( {1+\frac{2u}{s}} \right) \, ,
\end{equation}
with a general spectral density $\rho_J(s)$ that can have support anywhere on $s\geq M^2$. In other words, we are effectively allowing for a cut on the real $s$ axis rather than a collection of poles -- this is not unlike the
conformal bootstrap, where one does not a priori input discreteness of the operator spectrum.
The only assumption we will make about the theory at high energies is that it remains unitary, and in particular that the spectral density satisfies the lower bound $\rho_J(s)\geq 0$.

It is convenient to define the high-energy averages
\be \label{eq:HEavg}
 \avg{(\cdots)} \equiv \frac{1}{\pi} \sum_{J }  n^{(D)}_J
 \int_{M^2}^\infty \frac{dm^2}{m^2} m^{4-D}  \rho_J(m^2) \;(\cdots)\, ,
\ee
which have positive measure by the assumption of unitarity at high energies. Plugging \eqref{eq:Mhigh} into \eqref{eq:disprels2} and using $\legP_J(-x)=(-1)^J\legP_J(x)$ together with crossing symmetry then gives the sum rules
\begin{subequations}\label{eq:disprels3}
\begin{equation}
     \text{SU:}\quad\frac{1}{2 \pi i } \oint_0 ds'\,\frac{M_{\text{low}}(s', u)}{s'^{k+1}} =
    \avg{   \frac{ \legP_J(  1+\frac{2u}{m^2})}{m^{2k}}} \,,  \qquad k=1,2, \dots
\end{equation}
\begin{align}\label{eq:disprels3ST}
     \text{ST:}\quad\frac{1}{2 \pi i } \oint_0 ds'\,\frac{M_{\text{low}}(s', -s'-u)}{s'^{k+1}} &= 
    \avg{ \frac{ (-1)^J\legP_J(  1+\frac{2u}{m^2})}{m^{2k}}} \qquad k=1,2, \dots\,\\
  + &\, \avg{(-1)^k\frac{ m^2(-1)^{J} \, \legP_J(1+\frac{2u}{m^2})}{(m^{2} + u )^{k+1}}}\,. \nonumber
\end{align}
\end{subequations}
Let us stress again that although we have in mind large $N$ pion scattering, these sum rules are more general, and so will be our results. Since we are not using that the amplitude is actually meromorphic, these sum rules apply to any $s\leftrightarrow u$ crossing-symmetric unitary amplitude $M(s,u)$ satisfying the Regge behavior \eqref{eq:Regge} and such that at fixed $u<0$ it is analytic away from the \textit{positive} real $s$ axis. This last condition is perhaps the strongest, expressing  a clear feature (the Zweig rule) of large $N$ kinematics. 

\subsection{SU sum rules}
We have managed to relate the low-energy pole to an average on the high-energy states. Now we just have to plug \eqref{eq:Mlow} into the left hand side of \eqref{eq:disprels3}, expand the right hand side around $u\sim 0$ (the  forward limit) and match the coefficients to obtain separate sum rules for the couplings $g_{n,\ell}$. The first few SU sum rules read:
\begin{subequations}\label{firstfew}
\begin{alignat}{2} 
    k=1\, : & \quad  g_{1, 0} + 2 g_{2, 1} u + g_{3, 1} u^2 + \cdots &&=  
        \avg{ \frac{ \legP_J(1)}{m^2} + 2\frac{ \legP_J^{(1)}(1)}{m^4} u + 2\frac{ \legP_J^{(2)}(1)}{m^6} u^2 + \cdots}\\
    k=2\, : & \quad   g_{2, 0} + g_{3, 1} u + 2 g_{4, 2} u^2 + \dots  &&= 
        \avg{ \frac{ \legP_J(1)}{m^4} + 2\frac{ \legP_J^{(1)}(1)}{m^6} u + 2\frac{ \legP_J^{(2)}(1)}{m^8} u^2 + \cdots}\\
    k=3\, : & \quad   g_{3, 0} + g_{4, 1} u + g_{5, 2} u^2 +  \dots &&= 
        \avg{ \frac{ \legP_J(1)}{m^6} + 2\frac{ \legP_J^{(1)}(1)}{m^8} u + 2\frac{ \legP_J^{(2)}(1)}{m^{10}} u^2 + \cdots} \,.
\end{alignat}
\end{subequations}
Note that changing the number of subtractions $k$ allows us to pick different powers of $s$ in the expansion of $M_{\text{low}}(s,u)$. The general expression for the $k$-subtracted SU sum rule is
\begin{equation}
    \sum_{\ell =0}^{k}   g_{k+\ell, \ell}\,  u^\ell + \sum_{n =2k }^\infty  g_{n, k}\, u^{n-k}  = 
    \sum_{i=0}^\infty \frac{2^i}{i!}\avg{ \frac{ \legP_J^{(i)}(1)}{m^{2(k+i)}}}\,u^i\, , \quad k=1,2, \dots\,,
\end{equation}
where we have used the linearity of the high-energy average \eqref{eq:HEavg}.

The couplings $g_{k,0}$ and $g_{2k,k}$ appear only in the $k$-subtracted sum rule. By matching the coefficients of the Taylor expansion in $u$, we find
\begin{subequations}\label{eq:sumrules}
\begin{equation}
    g_{k, 0} = \avg{\frac{1}{m^{2k}}} \, , \qquad g_{2k, k} = \frac{2^{k-1}}{k!} \avg{   \frac{\legP_J^{(k)  }  (1) }{m^{4k }}   } \,, \quad k=1,2, \dots\,.
\end{equation}
Meanwhile, all the other couplings $g_{n,\ell}$ (with $\ell\neq 0$ and $n\neq 2\ell$) appear precisely in two sum rules (namely the ones with $k=\ell$ and $k=n-\ell$) giving rise to two distinct expressions in terms of high-energy averages,
\begin{equation}
    g_{n , \ell} = \frac{2^{n-\ell}}{(n-\ell)!} \avg{  \frac{\legP_J^{(n - \ell)  }  (1) }{m^{2n }}   }   = \frac{2^{\ell}}{\ell!} \avg{  \frac{\legP_J^{(\ell)  }  (1) }{m^{2n }}   }  \, ,\qquad
    \begin{aligned}
        n &= 3, 4, \dots \, \\
        \ell&=1,\dots,\left[\tfrac{n-1}{2}\right] \,.
    \end{aligned}
\end{equation}
\end{subequations}
Imposing equality of the two expressions gives rise to an infinite set of null constraints
\be \label{Xnull}
{\cal X}_{n , \ell} (m^2, J) \equiv  \frac{2^{n-\ell}}{(n-\ell)!}  \frac{\legP_J^{(n - \ell)  }  (1) }{  m^{2n }}     -  \frac{2^{\ell}}{\ell!}   \frac{\legP_J^{(\ell)  }  (1) }{m^{2n }} \,, \qquad
    \begin{aligned}
        n &= 3, 4, \dots \, \\
        \ell&=1,\dots,\left[\tfrac{n-1}{2}\right] \,,
    \end{aligned}
\ee
whose high-energy average must vanish,
\be
 \avg{ {\cal X}_{n , \ell} (m^2, J) }= 0\, .
\ee
This imposes highly nontrivial constraints on the heavy data; not anything goes for a unitary UV-complete theory.

\subsection{ST sum rules}
Following the same procedure for the ST sum rules, we reach the general expression for the $k$-subtracted sum rule,
\begin{align}
    \sum_{n=k}^\infty  &\sum_{\ell=0}^{[n/2]}g_{n,\ell} \left((-1)^{\ell}\frac{\left(k-(n-\ell)+1\right)_{n-k}}{(n-k)!}+
	(-1)^{n-\ell}\frac{\left(k-\ell+1\right)_{n-k}}{(n-k)!}\right)  u^{n-k}  \qquad k=1,2, \dots\nonumber\\
    &=\sum_{i=0}^\infty \left(\frac{2^i}{i!}\avg{ \frac{ (-1)^J\legP_J^{(i)}(1)}{m^{2(k+i)}}}\,u^i
    +\sum_{j=0}^\infty  (-1)^{k+j}\frac{(j+1)_k}{k!}\frac{2^i}{i!}\avg{\frac{ (-1)^J \legP_J^{(i)}(1)}{m^{2(k+i+j)}}} u^{i+j}\right)\,,
\end{align}
where $(x)_n\equiv \Gamma(x+n)/\Gamma(n)$ is the Pochhammer symbol. Equating the coefficients of $u^{n-k}$ for fixed $n\geq k$,
\begin{align}
	\sum_{\ell=0}^{[n/2]}&g_{n,\ell} \left((-1)^{\ell}\frac{\left(k-(n-\ell)+1\right)_{n-k}}{(n-k)!}+
	(-1)^{n-\ell}\frac{\left(k-\ell+1\right)_{n-k}}{(n-k)!}\right)   \quad\qquad k=1,2, \dots, n\nonumber\\
	&=\frac{2^{n-k}}{(n-k)!}\avg{ \frac{ (-1)^J\legP_J^{(n-k)}(1)}{m^{2n}}}
    +\sum_{i=0}^{n-k}  (-1)^{n-i}\frac{(n-k-i+1)_{k}}{k!}\frac{2^i}{i!}\avg{\frac{ (-1)^J \legP_J^{(i)}(1)}{m^{2n}}}\,.
	\label{STsumrule}
\end{align}
As \eqref{eq:sumrules} already gives us a sum rule for every coupling $g_{n,\ell}$, there is no need to solve for them here. Instead, we can use those sum rules to get rid of the couplings and derive a new infinite set of null constraints. It would look like at each level $n$ one gets $n$ new null constraints, but not all of them are independent. One can check that the first equations with $k=1,\ldots,\left[\tfrac{n}{2}\right]$ suffice. Plugging \eqref{eq:sumrules} into the left hand side of these equations yields the second set of null constraints:
\begin{equation}
	\bigg<\mathcal{Y}_{n,k}(m^2,J)\bigg>=0\;,
\end{equation}
where
\begin{align}
	\mathcal{Y}_{n,k}(m^2,J)=
	&\sum_{\ell=0}^{k}(-1)^{n-\ell} \frac{\left(k-\ell+1\right)_{n-k}}{(n-k)!}\frac{2^\ell}{\ell!} \frac{\legP_J^{(\ell)  }  (1) }{m^{2n }} -\frac{2^{n-k}}{(n-k)!}\frac{(-1)^J\mathcal{P}_J^{(n-k)}(1)}{m^{2n}}\nonumber\\
  -& \sum_{i=0}^{n-k}(-1)^{n-i}
  \frac{(n-k-i+1)_{k}}{k!}\frac{2^i}{i!}
  \frac{ (-1)^J \legP_J^{(i)}(1)}{m^{2n}}\,,\qquad
    \begin{aligned}
        n &= 2, 3, \dots \, \\
        k&=1,\dots,\left[\tfrac{n}{2}\right] \,.
    \end{aligned}
\end{align}
Recalling from \eqref{Xnull} that the null constraints $\mathcal{X}_{n,\ell}$ exist for $\ell=1,\ldots,\left[\tfrac{n-1}{2}\right]$, we conclude that at a given level $n\geq 2$ there are
\begin{equation}
	\bigg[\frac{n}{2}\bigg]+\left[\frac{n-1}{2}\right]=n-1 \text{ null constraints.}
\end{equation}
The first few null constraints are (in arbitrary normalization)
\begin{align}\label{eq:someNC}
	m^4 \mathcal{Y}_{2,1}&= \left(1-(-1)^J\right)(2-D)+\mathcal{J}^2\;, \nonumber\\
	m^6 \mathcal{Y}_{3,1}&= 3\left(1-(-1)^J\right)(2-D)+2\left(1-2(-1)^J\right)\mathcal{J}^2\;,\nonumber \\
	m^6 \mathcal{X}_{3,1}&=2(1-D)\mathcal{J}^2+\mathcal{J}^4\;,\nonumber\\
	m^8 \mathcal{Y}_{4,1}&= 2\left(1-(-1)^J\right)(2-D)D
	+\left(4(-1)^J+D\left(1-5(-1)^J\right)\right)\mathcal{J}^2+2(-1)^J\mathcal{J}^4\;,\nonumber \\
	m^8 \mathcal{X}_{4,1}&=\left(8-18D+D^2\right)\mathcal{J}^2+(8-6D)\mathcal{J}^4+2\mathcal{J}^6\;,\nonumber\\
	m^8 \mathcal{Y}_{4,2}&= -3\left(1-(-1)^J\right)D
	+\left(1-2(-1)^J\right)\mathcal{J}^2\;, 
\end{align}
where $\JJ^2\equiv J\left(J+D-3\right)$ is the quadratic Casimir of the little group $SO\left(D-1\right)$. Note that all of them vanish for $J=0$, implying that spinless heavy states are decoupled from the states with $J\geq 1$. This was expected, as we only used dispersion relations with at least one subtraction due to the Regge behavior \eqref{eq:Regge}.

\subsection{A second look at the high-energy behavior}
\label{subsec:secondhighenergy}
Now we can ask if it would make sense to make boundedness assumptions stronger than \eqref{eq:Regge}. For example, if we assumed that the amplitude died asymptotically in the forward limit, i.e.
\begin{equation}\label{eq:strongRegge}
    \lim_{|s| \to \infty} M(s, u) = 0\,,\qquad u\lesssim 0\,,
\end{equation}
we could write down unsubtracted SU dispersion relations. This means that \eqref{eq:disprelSU} would also hold for $k=0$. Following the same steps as above, this would yield ---among others--- the sum rule
\begin{equation}\label{eq:adler}
    g_{0, 0} = \avg{1}\,,
\end{equation}
where $g_{0,0}$ is the would-be constant term in \eqref{eq:Mlow}. Since pion scattering amplitudes must satisfy the Adler zero $M(0,0)= 0$, we must have $g_{0,0}=0$. But then the sum rule \eqref{eq:adler} cannot be satisfied by any non-trivial theory due to the positivity of the high-energy average $\left<\cdots\right>$. This contradiction proves that \eqref{eq:strongRegge} is too strong an assumption; one should stick to the milder Regge behavior of \eqref{eq:Regge}.

\section{Positivity bounds}
\label{sec:bounds}
It has long been known that dispersion relations link low and high energies in such a way that allows us to bound EFT couplings as a consequence of unitarity at high energies~\cite{Martin1969,Pham:1985cr,Ananthanarayan:1994hf,Pennington:1994kc,Comellas:1995hq,Dita:1998mh}. But only recently (building on much previous work, e.g.~\cite{Adams:2006sv,Manohar:2008tc, Mateu:2008gv, Nicolis:2009qm, Baumann:2015nta, Bellazzini:2015cra, Bellazzini:2016xrt, Cheung:2016yqr, Bonifacio:2016wcb,  Cheung:2016wjt, deRham:2017avq, Bellazzini:2017fep, deRham:2017zjm, deRham:2017imi, Hinterbichler:2017qyt, Bonifacio:2017nnt, Bellazzini:2017bkb, Bonifacio:2018vzv, deRham:2018qqo, Zhang:2018shp, Bellazzini:2018paj, Bellazzini:2019xts, Melville:2019wyy, deRham:2019ctd, Alberte:2019xfh, Alberte:2019zhd, Bi:2019phv, Remmen:2019cyz, Ye:2019oxx, Herrero-Valea:2019hde,Bellazzini:2020cot}) the full set of inequalities that follow from $2\to 2$ scattering have been derived~\cite{Caron-Huot:2020cmc,Tolley:2020gtv}, exploiting the role of null constraints.\footnote{Null constraints are also implicitly present in the beautiful geometric approach of the EFT-hedron \cite{Arkani-Hamed:2020blm,Chiang:2021ziz}, but to the best of our knowledge, it is still difficult to implement their  algorithm to higher orders \cite{Huang:2020nqy}. Also, we have not found a formulation of our problem in terms of the EFT-hedron that accounts for the ST sum rules.} Here we will use the method introduced in \cite{Caron-Huot:2020cmc} to obtain optimal two-sided bounds on normalized ratios of the EFT coefficients $g_{n,\ell}$ using semidefinite programming. We start with a brief review of their approach using a slight reformulation that draws an analogy with the familiar conformal bootstrap. This will prove useful for the modifications that we will introduce in section \ref{sec:rho}. Then we present our results and we compare them to meson phenomenology and to previous work.

\subsection{Dual problem}\label{sec:dualprob}
In the previous section we have expressed the dispersion relations \eqref{eq:disprels} in terms of sum rules and null constraints of the form
\begin{align}
    g_{n,\ell}=&\,\avg{g_{n,\ell}\left(m^2,J\right)}\,, \nonumber\\
    0=&\,\avg{\XX_{n,\ell}\left(m^2,J\right)}\,,\quad 0=\avg{\YY_{n,\ell}\left(m^2,J\right)}\,,
\end{align}
where $\left<\cdots\right>$ denotes the high-energy average introduced in \eqref{eq:HEavg}, which integrates ---with positive measure--- over all masses $m$ above the cutoff scale $M$ and sums over all spins $J=0,1,2\ldots$. The goal is to use these ingredients to obtain two-sided bounds for any given coupling $g_{n,\ell}$ normalized by $g_{1,0}$ and the cutoff as
\begin{equation}
    \tilde g_{n,\ell}=g_{n,\ell}\frac{M^{2(n-1)}}{g_{1,0}}\,.
\end{equation}
To do so, we define the vectors
\begin{equation}
	\vec{v}_\mathbb{1}=\begin{pmatrix}
	1\\ 0 \\0\\ \vdots\\ 0
	\end{pmatrix}\,,\qquad
	\vec{v}_\mathcal{O}=\begin{pmatrix}
	0\\ 1 \\0\\ \vdots\\ 0
	\end{pmatrix}\,,\qquad
	\vec{v}_{\text{HE}}\left(m^2,J\right)=\begin{pmatrix}
	-g_{1,0}\left(m^2,J\right)M^2\\ -g_{n,\ell}\left(m^2,J\right)M^{2n} \\ \mathcal{Y}_{2,1}\left(m^2,J\right)M^4\\\mathcal{X}_{3,1}\left(m^2,J\right)M^6\\ \vdots
	\end{pmatrix}\,,
\end{equation}
where the last vector includes as many null constraints as we like.

These vectors clearly satisfy the equation
\begin{equation} \label{eq:bootstrpeq}
	g_{1,0}\,M^2 \,\vec v_\mathbb{1}+g_{n,\ell}\,M^{2n}\, \vec v_\mathcal{O}+ \avg{\vec v_\text{HE}\left(m^2,J\right)}=0\,.
\end{equation}
This is to be understood as a bootstrap equation analogous to that of the conformal bootstrap where $\vec v_\mathbb{1}$ plays the role of the identity operator, $\vec v_\mathcal{O}$ corresponds to an operator we are singling out and $\vec v_\text{HE}\left(m^2,J\right)$ represents a continuum of heavy operators. With these identifications, we see that the low-energy coefficients are analogous to the OPE coefficients squared and the positivity of $\left<\cdots\right>$ corresponds to the statement of unitarity for the heavy operators.
Obtaining bounds for the low-energy couplings is thus as easy as bounding OPE coefficients in the conformal bootstrap \cite{Caracciolo:2009bx}:

We must look for a ``functional'' $\vec \alpha$ such that
\begin{enumerate}
	\item is normalized as $\vec \alpha \cdot \vec v_\mathcal{O}=\left\{\begin{matrix}
	+1\quad \text{for upper bound}\\
	-1\quad \text{for lower bound,}
	\end{matrix}\right.$
	\item is positive on all heavy states, $\vec \alpha \cdot \vec v_\text{HE}\left(m^2,J\right)\geq 0\quad \forall J,\,m\geq M$,
	\item and maximizes $\vec \alpha \cdot \vec v_\mathbb{1}$.
\end{enumerate}
Once we find such a functional, the positivity of the high-energy average guarantees that
\begin{equation}
    \avg{\vec \alpha \cdot \vec v_\text{HE}\left(m^2,J\right)}\geq 0\,.
\end{equation}
So, by contracting \eqref{eq:bootstrpeq} with $\vec \alpha$ and using the linearity of $\left<\cdots\right>$ we can get the bounds
\begin{equation}
	\vec \alpha_{(-)} \cdot \vec v_\mathbb{1}\leq \tilde g_{n,\ell} \leq -\left(\vec \alpha_{(+)} \cdot \vec v_\mathbb{1}\right)\,,
\end{equation}
where the subscript denotes the choice of normalization for the functional. The maximization condition on the functional makes the bounds as strong as possible.

This can be formulated as a semidefinite problem and solved for example with \texttt{SDPB}~\cite{sdpb}. When implementing it numerically, though, one must deal with infinities. The three infinities in the problem are the masses of the high-energy vectors, their spins and the number of available null constraints. As a semidefinite problem solver, \texttt{SDPB} imposes positivity on polynomials for $x\geq 0$, so replacing $m^2\to M^2\left(1+x\right)$ and removing the common denominator in $v_\text{HE}\left(m^2,J\right)$ readily allows us to cover all the region $m^2\in \left[M^2,\infty\right)$. As far as the spins go, we truncate the infinity by considering $J=0,1,2,\ldots,J_{\text{max}}$ with $J_{\text{max}}$ large enough that further increases of it do not change the results. By numerical experimentation we have seen that convergence in spins depends significantly more on how large $J_{\text{max}}$ is rather than on the density of the grid of spins. One can safely skip many intermediate spins so long as both even and odd spins are included.

The last infinity comes from the number of null constraints. Ideally, one would like to use all the null constraints derived in the previous section, but in practice one must truncate that infinity to construct finite-length vectors. We will consider vectors including all the null constraints up to ``Mandelstam order'' $n$, i.e.\ all the null constraints that follow from the terms in $M_\text{low}(s,u)$ up to order $O(p^{2n})$,
$$\YY_{2,1},\;\YY_{3,1},\;\XX_{3,1},\;\YY_{4,1},\;\XX_{4,1},\;\YY_{4,2},\;\cdots\;\XX_{n,\left[\frac{n-1}{2}\right]},\;\YY_{n,\left[\frac{n}{2}\right]}\,.$$
Up to a given order $n$, the total number of null constraints used is then
\begin{equation}
    \sum_{i=2}^{n}\left(i-1\right)=\frac{1}{2}n\left(n-1\right)\,.
\end{equation}
This makes $n$ analogous to the parameter $\Lambda$ that controls the number of derivatives of conformal blocks used in constructing the vectors for the conformal bootstrap. Just like in that case, increasing $n$ can only improve the bounds.

The above procedure can be used to bound any low-energy coupling, but in the following sections we will mostly be concerned with the first ones. The sum rules we will need are
\begin{equation}\label{eq:fewsumrules}
    g_{1,0}=\avg{\frac{1}{m^2}}\,,\qquad
    g_{2,0}=\avg{\frac{1}{m^4}}\,,\qquad
    g_{2,1}=\frac{1}{(D-2)}\avg{\frac{\JJ^2}{m^4}}\,,
\end{equation}
where $\JJ^2\equiv J\left(J+D-3\right)$. Also, while we will keep our notation general, we set $D=4$ henceforth to make contact with the real world. For ease of notation, we define
\begin{equation}\label{eq:gtilde}
    \tilde g_2 \equiv g_{2,0}\frac{M^2}{g_{1,0}}\,,\qquad \tilde g_2' \equiv 2g_{2,1}\frac{M^2}{g_{1,0}}\,.
\end{equation}
Note that it is safe to normalize by $g_{1,0}$ because the positivity of the high-energy average in \eqref{eq:fewsumrules} ensures that $g_{1,0}>0$ for any nontrivial theory. Using all the null constraints up to Mandelstam order $n=17$ (136 null constraints) we obtain the two-sided bounds
\be \label{bounds}
	0\leq \tilde g_2 \leq 1\,,\qquad 0\leq \tilde g_2' \leq 3.25889135348...\,.
\ee
The lower bounds for these couplings are already obvious from the positivity of their sum rules, and so is the upper bound $\tilde g_2 \leq 1$ if we recall that $m^2\geq M^2$ in the high-energy averages. The upper bound on $\tilde g_2'$, on the other hand, is non-trivial.

\subsection{Exclusion plot}\label{sec:exclPlot}
\begin{figure}[ht]
\centering
\includegraphics[scale=1.1]{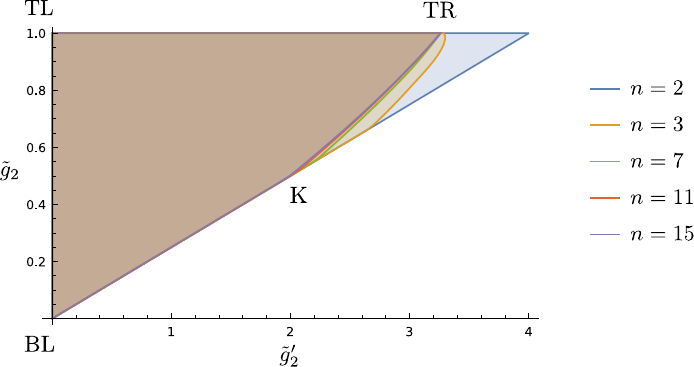}
\caption{Exclusion plots in the space of couplings $(\tilde g'_2,  \tilde g_2 )$ for $D=4$ including all the null constraints up to Mandelstam order $n=2,3,7,11,15$. The colored area is the allowed region in each case. By $n=15$ the plot has visually converged and it shows three corners (labelled TL, TR, BL) and a kink (K).}
\label{fig:exclPlot}
\end{figure}

Following again \cite{Caron-Huot:2020cmc}, we can modify this program to place bounds on $\tilde g_2'$ for every value of $\tilde g_2$ in its allowed range and thus carve out the space of EFTs that are compatible with unitary UV completions. Consider now the vectors
\begin{equation}\label{eq:exclvectors}
	\vec{v}_\mathbb{1}=\begin{pmatrix}
	1\\ 0 \\0\\ \vdots\\ 0
	\end{pmatrix}\,,\quad
	\vec{v}_2=\begin{pmatrix}
	0\\ 1 \\0\\ \vdots\\ 0
	\end{pmatrix}\,,\quad
	\vec{v}_{2'}=\begin{pmatrix}
	0\\ 0 \\1\\ \vdots\\ 0
	\end{pmatrix}\,,\quad
	\vec{v}_{\text{HE}}\left(m^2,J\right)=\begin{pmatrix}
	-M^2/m^2\\ -M^4/m^4 \\ -\frac{2}{D-2} \JJ^2M^4/m^4\\ \mathcal{Y}_{2,1}\left(m^2,J\right)M^4\\ \vdots
	\end{pmatrix}\,,
\end{equation}
satisfying\footnote{We have absorbed a factor $\frac{1}{g_{1,0} M^2}$ in the definition of the high-energy average to simplify the notation. Since $g_{1,0}> 0$, this does not spoil the positivity of $\left<\cdots\right>$.}
\begin{equation} \label{eq:bootstrpeq2}
	\vec v_\mathbb{1}+\tilde g_2\, \vec v_2+\tilde g_2'\, \vec v_{2'}+ \avg{\vec v_\text{HE}\left(m^2,J\right)}=0\,.
\end{equation}
For any fixed $\tilde g_2^{(0)}$, we can look for a functional $\vec \alpha$ that
\begin{enumerate}
	\item is normalized as $\vec \alpha \cdot \vec v_{2'}=\left\{\begin{matrix}
	+1\quad \text{for upper bound}\\
	-1\quad \text{for lower bound,}
	\end{matrix}\right.$
	\item is positive on all heavy states, $\vec \alpha \cdot \vec v_\text{HE}\left(m^2,J\right)\geq 0\quad \forall J,\,m\geq M$,
	\item and maximizes $\vec \alpha \cdot \left(\vec v_\mathbb{1}+\tilde g_2^{(0)} \vec v_2\right)$,
\end{enumerate}
to get the bounds
\begin{equation}
	\vec \alpha_{(-)} \cdot \left(\vec v_\mathbb{1}+\tilde g_2^{(0)} \vec v_2\right)\leq \tilde g_2' \leq -\vec \alpha_{(+)} \cdot \left(\vec v_\mathbb{1}+\tilde g_2^{(0)} \vec v_2\right)\,.
\end{equation}
Repeating these steps for several values of $\tilde g_2$ spanning its allowed range then yields the exclusion plot of allowed theories in the space of couplings $\left(\tilde g_2',\tilde g_2\right)$, shown in figure \ref{fig:exclPlot}.

The figure shows the exclusion plots obtained at different Mandelstam orders, each of them lying inside the previous one, as expected. Although the convergence in $n$ is somewhat slow, around $n=15$ the change becomes imperceptible. The final plot has a trapezoidal shape with three corners and a kink! At the top-right corner (TR), the value of $\tilde g_2'$ matches its upper bound \eqref{bounds}. This and the other two corners (top-left TL and bottom-left BL) can be understood in close analogy to \cite{Caron-Huot:2020cmc}, as we explain in section \ref{sec:rulingin}. What is more intriguing is the kink (K). The optimistic hope is that it corresponds to large $N$ QCD in the chiral limit, but of course this does not have to be the case.

\subsection{The kink}
The lower part of the right-hand bound of figure \ref{fig:exclPlot}, extending from the origin (BL) to the kink (K), corresponds to the line $\tilde g_2'=4\,\tilde g_2$ for any $n$, and the kink moves down this line as we increase $n$. This bound appears already at order $n=2$, where we only use the null constraint $\YY_{2,1}$. Indeed, in this case the optimization problem tells us to look for a functional of the form $\vec \alpha=\left(\alpha_1, \,\alpha_2,\, 1,\,\alpha_4\right)$
satisfying
\begin{equation}\label{eq:SLfunctional}
    \vec\alpha\cdot\vec v_\text{HE}\left(m^2,J\right)
    =-\alpha_1\frac{M^2}{m^2}-\alpha_2\frac{M^4}{m^4}-\JJ^2\frac{M^4}{m^4}+\alpha_4\Big(2\left(1-(-1)^J\right)-\mathcal{J}^2\Big)\frac{ M^4}{m^4}\geq 0\,,
\end{equation}
for all $J$ and $m\geq M$, that maximizes $\alpha_1+\alpha_2\,\tilde g_2^{(0)}$. The solution to this problem is given by
\begin{equation}
    \vec \alpha=\big(0,\, -4,\,1,\,-1\big)
\end{equation}
for any value of $\tilde g_2^{(0)}$, resulting in the bound $\tilde g_2'\leq 4\tilde g_2$. When further null constraints are included, there is a competition between solutions. For small enough $\tilde g_2^{(0)}$ this solution is still optimal, but as $\tilde g_2^{(0)}$ crosses a particular point (the kink), the minimization is suddenly achieved by a functional with nonzero $\alpha_1$ and lower $\alpha_2$. This is essentially a first order phase transition, in contrast to the case of \cite{Caron-Huot:2020cmc} (see also \cite{Chiang:2021ziz}), where the transition looks second order with the discontinuity in the second derivative.

\begin{figure}[ht]
\centering
\begin{subfigure}[b]{0.70\textwidth}
   \includegraphics[width=\linewidth]{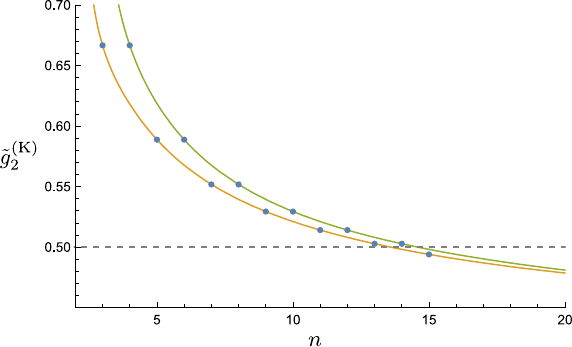}
\end{subfigure}
\qquad
\begin{subfigure}[b]{0.20\textwidth}
   \begin{tabular}{ | c | c | }
   \hline
   $n$ & $\tilde g_2^{\text{(K)}}$\\
   \hline
   3,\;4 & 0.6667\\
   5,\;6 & 0.5888\\
   7,\;8 & 0.5518\\
   9,\;10 & 0.5294\\
   11,\,12 & 0.5141\\
   13,\,14 & 0.5028\\
   15 & 0.4940\\
   \hline
   \end{tabular}
    \vspace{0.6in}
\end{subfigure}
\caption{\label{fig:kinkvalue}Position of the kink along the $\tilde g_2'=4\,\tilde g_2$ line as a function of the Mandelstam order $n$. The orange and green curves are, respectively, fits for the points of odd and even $n$ made with the function $f(n)=\sum_{i=0}^5\frac{a_i}{n^i}$ using the first six values in each case. The fits predict very well the last number in the table for $n=15$ ($n=16$ in the even $n$ case) and they suggest the asymptotic value $a_0\sim 0.42$.}
\end{figure}

Knowing that the kink sits on top of the line $\tilde g_2'=4\,\tilde g_2$, we can determine its precise position by looking for the upper bound on $\tilde g_2$ when a theory is forced to live on this line. This can be achieved by including as an additional null constraint the combination
\begin{equation}
    \tilde g_2'-4\,\tilde g_2\propto\avg{g_{2,1}\left(m^2,J\right)M^4-2g_{2,0}\left(m^2,J\right)M^4}=0\,.
\end{equation}
Figure \ref{fig:kinkvalue} shows the position of the kink computed in this way for increasing values of $n$. Its convergence is somewhat slow and, surprisingly, in steps of two. This indicates that the null constraints that contribute the most in shaping the plot are $\XX_{n,\frac{n-1}{2}}$ (for odd $n$). From the first points in the plot it might seem that the kink is converging towards the exact value $0.5$, which would call for a simple analytic explanation, but at order $n=15$ we obtain
\begin{equation}\label{eq:kinkpos}
    \tilde g_2^{(\text{K})}\simeq 0.4940\,,\qquad \tilde g_2'^{(\text{K})}= 4 \tilde g_2^{(\text{K})}\simeq 1.9760\,, \qquad(n=15)\,.
\end{equation}
This already rules out the value $0.5$ and, in fact, the fit in figure \ref{fig:kinkvalue} suggests that it converges to a value significantly lower; around $\tilde g_2^{(\text{K})}\sim 0.42$. 

\begin{figure}[ht]
    \centering
    \subfloat{{\includegraphics[scale=0.50]{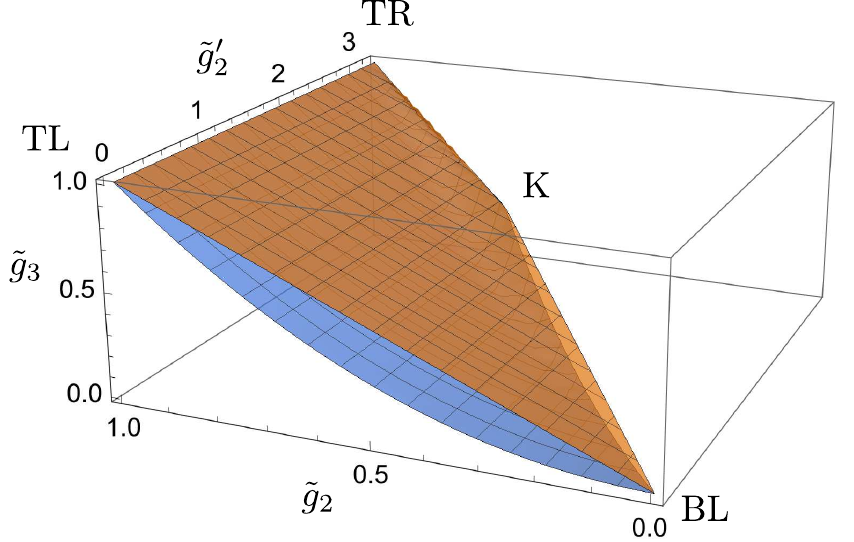}}}
    \subfloat{{\includegraphics[scale=0.51]{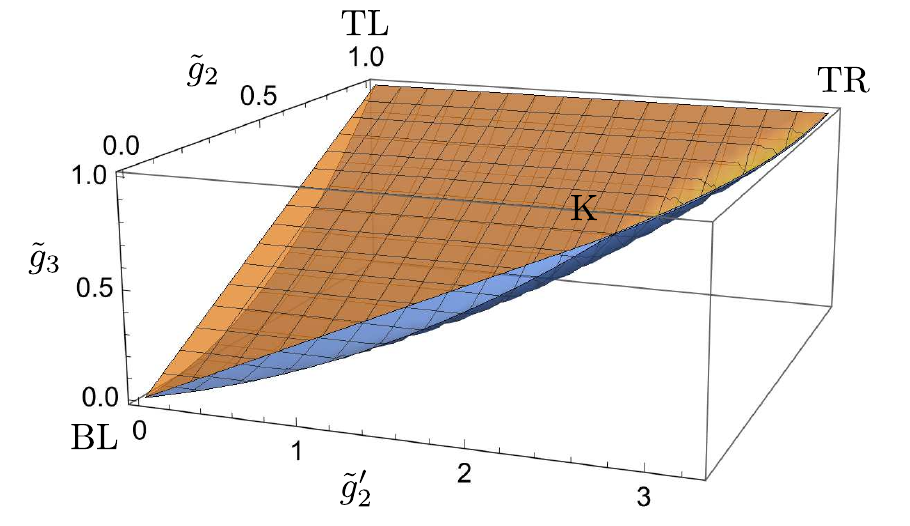}}}
    \caption{Two different views of the three-dimensional exclusion plot in the space $(\tilde g'_2,  \tilde g_2 , \tilde g_3)$. The blue sheet describes the lower bound on $\tilde g_3$ for every value of $\tilde g_2$ and $\tilde g_2'$, the orange sheet is for the upper bound. This exclusion plot was made using all null constraints up to $n=7$.}
    \label{fig:3Dviews}
\end{figure}

Finally, to try to get a better picture of the kink, we have also solved the dual problem in a three-dimensional space using the coupling
\be
	\tilde g_3\equiv g_{3,0}\frac{M^4}{g_{1,0}}\,.
\ee
The results are shown in figure \ref{fig:3Dviews}, where we have included two views of the 3D plot. The blue sheet represents the lower bound on $\tilde g_3$, while the orange sheet corresponds to its upper bound. From the picture on the left, we see that the allowed space at the plane $\tilde g_2'=0$ is given by the convex hull of a parabola (that we will shortly identify) connecting the corners BL and TL along the blue sheet. The other view shows that the theories along the bound $\tilde g_2'=4\,\tilde g_2$ also lie in the convex hull of a parabola, this one running from BL up to the kink K, at which point the blue and orange sheets fuse together.

\subsection{QCD, where art thou?}\label{sec:whereartthou}
We have obtained sharp bounds for the low energy couplings of EFTs for pion scattering. The question now is \textit{how do they compare to known EFTs?} and more importantly, \textit{where is large $N$ QCD?}

\subsubsection{The chiral Lagrangian}
As mentioned above, the EFT for pion scattering is the standard chiral Lagrangian (see e.g.\ \cite{Gasser:1983yg,Gasser:1984gg}). The first terms of the $SU(N_f)$ chiral Lagrangian for vanishing quark mass read
\begin{align}
    \label{eq:chiralL}
	\mathcal{L}_\text{Ch}=&-\frac{f_\pi^2}{4}\tr{\partial_\mu U^\dagger\partial^\mu U}
	+L_1' \left[\tr{\partial_\mu U^\dagger\partial^\mu U}\right]^2
	+L_2' \tr{\partial_\mu U^\dagger\partial_\nu U}\tr{\partial^\mu U^\dagger\partial^\nu U}\nonumber \\
	&+L_3' \tr{\partial_\mu U^\dagger\partial^\mu U\partial_\nu U^\dagger\partial^\nu U}
	+L_4' \tr{\partial_\mu U^\dagger\partial_\nu U\partial^\mu U^\dagger\partial^\nu U}
	+\cdots\,,
\end{align}
where $U(x)=\exp \left[i\frac{2}{f_\pi} T_a\pi^a(x)\right]\in SU(N_f)$ with $T_a$ the $SU(N_f)$ generators in the defining representation. The coefficients $L_i'$ are the low energy couplings that we alluded to before. They are in correspondence with the $g_{n,\ell}$ couplings in \eqref{eq:Mlow} and we will give their precise relation momentarily. Equation \eqref{eq:chiralL} includes all the independent chiral-invariant terms with four derivatives that we need for general $N_f$, but there are two special cases. For $N_f=3$, there is an identity\footnote{For $U\in SU(3)$, \begin{align}\label{eq:SU3identity}
\tr{\partial_\mu U^\dagger\partial_\nu U\partial^\mu U^\dagger\partial^\nu U}\,&=\frac{1}{2}\left[\tr{\partial_\mu U^\dagger\partial^\mu U}\right]^2 \\
&+\tr{\partial_\mu U^\dagger\partial_\nu U}\tr{\partial^\mu U^\dagger\partial^\nu U}-2\tr{\partial_\mu U^\dagger\partial^\mu U\partial_\nu U^\dagger\partial^\nu U}\,.\nonumber
\end{align}} that relates the last term to the other three four-derivative terms, so the $L_4'$ term is dropped and the couplings are conventionally \cite{Gasser:1984gg} renamed as
\begin{equation}
    L_1'\to L_1\equiv L_1'+\frac{L_4'}{2}\,,\qquad L_2'\to L_2\equiv L_2'+L_4'\,,\qquad L_3'\to L_3\equiv L_3'-2L_4'\,.
\end{equation}
For $N_f=2$, a further identity\footnote{For $U\in SU(2)$, apart from \eqref{eq:SU3identity},
\begin{equation}
    \tr{\partial_\mu U^\dagger\partial^\mu U\partial_\nu U^\dagger\partial^\nu U}
    =\frac{1}{2}\left[\tr{\partial_\mu U^\dagger\partial^\mu U}\right]^2 \,.
\end{equation}} relates the remaining terms, so we only keep the first two four-derivative terms with the conventional couplings \cite{Gasser:1983yg}
\begin{equation}
    L_1'\to \frac{\ell_1}{4}\equiv L_1'+\frac{L_3'}{2}-\frac{L_4'}{2}\,,\qquad L_2'\to \frac{\ell_2}{4}\equiv L_2'+L_4'\,.
\end{equation}

Although the traces appearing in \eqref{eq:chiralL} concern the flavor space $SU(N_f)$, each of them corresponds to a quark loop, which in turn is associated to a color trace. In the limit  $N\to\infty$, quark loops are suppressed by powers of $1/N$. Thus, double-trace operators in the chiral Lagrangian are subleading at large $N$. This implies that in the limit $N=\infty$ and for any $N_f$, $L_1'=L_2'=0$; only single-trace operators survive. Moreover, at $N=\infty$ we are allowed to treat chiral perturbation theory at tree level because loop diagrams would introduce additional quark loops. By expanding \eqref{eq:chiralL} in the pion fields and keeping only 4-pion terms it is then straightforward to see that the $\pi\pi\to \pi\pi$ scattering amplitude at large $N$ has the form of \eqref{eq:T} with
\begin{equation}
    A(s|t,u) = \frac{s}{f_\pi^2} + \frac{4}{f_\pi^4}\left(L_3'-L_4'\right)s^2 + \frac{4}{f_\pi^4}L_4'\left(t^2+u^2\right)+\cdots\,.
\end{equation}
Or, in terms of the $SU(2)$ couplings,
\begin{equation}\label{eq:ASU(2)}
    A(s|t,u) = \frac{s}{f_\pi^2} + \frac{2\ell_1}{f_\pi^4}s^2 + \frac{\ell_2}{f_\pi^4}\left(t^2+u^2\right)+\cdots\,.
\end{equation}

Using then \eqref{eq:AtoM} we can compute the corresponding basic amplitude $M(s,u)$, which gives the relation to the $g_{n,\ell}$ by comparing with \eqref{eq:Mlow},
\begin{equation}\label{eq:Ltog}
    g_{1,0}=\frac{1}{2f_\pi^2}\,,\qquad g_{2,0}=\frac{2L_3'+4L_4'}{f_\pi^4}\,,\qquad g_{2,1}=\frac{4L_4'}{f_\pi^4}\,.
\end{equation}
Or, in terms of the $SU(2)$ couplings,
\begin{equation}\label{eq:gtol}
    g_{1,0}=\frac{1}{2f_\pi^2}\,,\qquad g_{2,0}=\frac{2\ell_1+3\ell_2}{2f_\pi^4}\,,\qquad g_{2,1}=\frac{\ell_2}{f_\pi^4}\,.
\end{equation}
For completeness, we also write down the corresponding normalized coefficients:
\begin{equation}\label{eq:gtildetol}
    \tilde g_2 \equiv g_{2,0}\frac{M^2}{g_{1,0}}=\left(2\ell_1+3\ell_2\right)\frac{M^2}{f_\pi^2}\,,\qquad \tilde g_2' \equiv 2g_{2,1}\frac{M^2}{g_{1,0}}=4\ell_2 \frac{M^2}{f_\pi^2}\,.
\end{equation}

\subsubsection{The Skyrme line}
The low-energy couplings of the chiral Lagrangian \eqref{eq:chiralL} are a priori unknown. However, there is a particularly interesting model that fixes some of them; the Skyrme model (see e.g.\ \cite{Zahed:1986qz} for a review  and references). This model has been extensively used to describe baryons as solitons of the chiral Lagrangian. The Lagrangian of the Skyrme model is\footnote{We have not included the topological Wess-Zumino term \cite{Witten:1983tw} because in $D=4$ it is proportional to at least five pion fields and therefore it has no effect on the $2\to 2$ pion scattering amplitude.}
\begin{equation}
    \label{eq:skyrmeL}
	\mathcal{L}_\text{Skyr}=-\frac{f_\pi^2}{4}\tr{\partial_\mu U^\dagger\partial^\mu U}
	+\frac{1}{32e^2}\tr{\left[U^\dagger\partial_\mu U,U^\dagger\partial_\nu U\right]\left[U^\dagger\partial^\mu U,U^\dagger\partial^\nu U\right]}\,,
\end{equation}
where $U(x)\in SU(N_f)$ is the same matrix defined above and $e$ is an unfixed parameter. The second term in \eqref{eq:skyrmeL} is called the ``Skyrme term'' and it is the only chiral-invariant term with four derivatives that is second-order in time derivatives, an important feature for the stability of the solitons.

Since it involves only single-trace terms, the Skyrme model can be studied at large $N$ as is. Comparing with \eqref{eq:chiralL}, we see that the Skyrme model corresponds to the choice of low energy couplings
\begin{equation}
    L_3'=-\frac{1}{16e^2}\,,\qquad  L_4'=\frac{1}{16e^2}\,.
\end{equation}
By \eqref{eq:Ltog} and \eqref{eq:gtilde}, this corresponds to the line 
\begin{equation}
    \tilde g_2'=4\tilde g_2=\frac{1}{e^2}\frac{M^2}{f_\pi^2}\,.
\end{equation}
Remarkably, this is precisely the line on the lower-right bound of figure \ref{fig:exclPlot} connecting the origin (BL) to the kink (K). This means that the Skyrme model saturates the unitarity bounds for pion EFTs. We will henceforth refer to this line as the ``Skyrme line''.

Given that the Skyrme line is only allowed up to the kink, we can use \eqref{eq:kinkpos} to bound $ef_\pi$. Using $m_\rho\simeq 775\,$MeV as our cutoff scale,\footnote{Recall that the cutoff $M$ can be pushed only up to the mass of the first meson in the spectrum; the rho meson in large $N$ QCD. See appendix \ref{app:internalstates} for a discussion on the exchanged mesons in pion scattering at large $N$.} we obtain
\begin{equation}\label{eq:SkyrBound}
    ef_\pi\geq \sqrt{\frac{M^2}{4\tilde g_2^{(\text{K})}}}\simeq 551\,\text{MeV}.
\end{equation}
One can fix the values of $e$ and $f_\pi$ that we would need for the Skyrme model to be an accurate model for baryons by comparing to experimental data. In \cite{Adkins:1983ya}, by fitting the masses of the nucleon and the $\Delta$ baryon, they found $e\simeq 5.45$ and $f_\pi \simeq 64.5\,$MeV (rather far from the accepted value $f_\pi\simeq 92\,$MeV). These results give $ef_\pi\simeq 352\,$MeV, which does not satisfy the unitarity bound \eqref{eq:SkyrBound}. This discrepancy is of course not  worrisome; apart from the fact that our bounds are valid only at large $N$ and in the chiral limit, 
one would not expect the Skyrme model (which for no good reason keeps only terms with at most four derivatives) to give a completely accurate description of the baryons.

\subsubsection{Comparing with experiment}
Let us now see how our results compare to experimental data for real-world QCD. Since loops are not negligible at finite $N$ chiral perturbation theory, the (renormalized) real-world low-energy couplings $\ell_i^r(\mu)$ run with energy \cite{Gasser:1983yg}. We must therefore choose an energy scale when comparing them to our bounds. The relevant energy scale of our problem is the cutoff $M$, which represents the first exchanged meson in the spectrum; the rho meson. Thus, we will compare to experimental results for $\ell_i^r(m_\rho)$. We have found somewhat large discrepancies for the values of these couplings throughout the literature, see table \ref{tab:experiment}.
\begin{table}
\centering
\caption{\label{tab:experiment}Experimental values for the renormalized couplings $\ell_i^r(m_\rho)$ obtained from the quoted references together with the corresponding values for the normalized couplings $\tilde g_2$, $\tilde g_2'$.}
\begin{tabular}{ c | c | c | c || c } 
 $\ell_1^r(m_\rho) \times 10^3$ & $\ell_2^r(m_\rho)\times 10^3$ & $\tilde g_2$ & $\tilde g_2'$ & ref\\
 \hline
 $-6.0\pm 3.9$ & $5.5\pm 2.7$ & $\phantom{-}0.31\pm 0.81$ & $1.55\pm 0.78$ & \cite{Gasser:1983yg}\\
 $-5.4\pm 1.1$ & $5.7\pm 1.1$ & $\phantom{-}0.44\pm 0.27$ & $1.61\pm 0.30$ & \cite{Bijnens:1994ie}\\
 $-4.0\pm 2.1$ & $1.6\pm 1.0$ & $-0.23\pm 0.36$ & $0.45\pm 0.28$ & \cite{Girlanda:1997ed}\\
 $-3.2\pm 2.5$ & $3.1\pm 2.1$ & $\phantom{-}0.22\pm 0.58$ & $0.89\pm 0.60$ & \cite{Amoros:2000mc}\\
\end{tabular}
\end{table}

\begin{figure}[ht]
\centering
\includegraphics[scale=1.1]{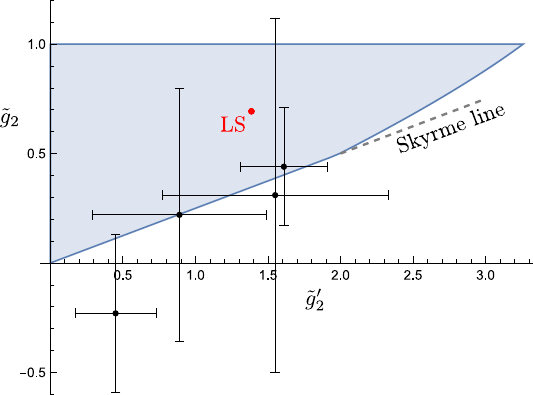}
\caption{Experimental points for real-world QCD (table \ref{tab:experiment}) with their uncertainties compared to the exclusion plot at Mandelstam order $n=15$. They all lie relatively close to the Skyrme line connecting the origin to the kink. The red dot marks the position of the Lovelace-Shapiro amplitude.}
\label{fig:exclPlot_experiment}
\end{figure}

The experimental points for real-world QCD from table \ref{tab:experiment} are plotted in figure \ref{fig:exclPlot_experiment} against our bounds. Despite the discrepancies between the points themselves, they all show some overlap with our allowed region to within uncertainty. This is somewhat remarkable, considering that our bounds really apply to theories in the large $N$ and chiral limits. In fact, the whole allowed region is comparable in size to the area spanned by the different points and their experimental uncertainties. This highlights the power of the bootstrap: just from the basic assumptions of section \ref{sec:setup} and the sole requirement that the theory be unitary, one can restrict the EFT couplings down to values comparable to experimental precision. A first optimistic take is that this
bodes well for pinning down large $N$ QCD only from theoretical considerations. 
Adding  further physical assumptions may turn the allowed region 
into a small numerical island. 

So, where is QCD? Surprisingly, the experimental points in figure \ref{fig:exclPlot_experiment} gather around the Skyrme line. But they spread all along it, and it would certainly be unwarranted to conclude that QCD sits at the kink. They do suggest, though, that QCD (at least in the chiral and large $N$ limits) lies somewhere in our plot close to the lower-right bound, making this region the most interesting part of the plot.

\subsubsection{The Lovelace-Shapiro amplitude}
The Lovelace-Shapiro amplitude \eqref{eq:LS} was an explicit example of a theory satisfying all of our assumptions, so it better be allowed by our bounds. We can compute its position in the space of couplings $(\tilde g_2',\tilde g_2)$ as a consistency check. From \eqref{eq:LS} we extract
\be
\tilde g_{2\, {\rm LS} } = -\frac{ \gamma + \psi(\frac{1}{2})}{2} \simeq 0.693147 \, , \quad  \tilde g'_{2\,  {\rm LS}} = 2  \tilde g_{2\, {\rm LS} } 
\simeq 1.38629 \, ,
\ee
where $\gamma$ is the Euler-Mascheroni constant and $\psi(z)$ the digamma function. This point indeed sits in the interior of the allowed region in figure \ref{fig:exclPlot_experiment}, as it should.

\subsection{Comparing with previous work}\label{sec:previouswork}
The program of constraining pion scattering amplitudes from the simple assumptions of unitarity, crossing symmetry and boundedness has a long history dating back to the '60s \cite{Martin1969}. Naturally, not all of our results are new. For example, the positivity of the low-energy coefficients $\ell_2\geq 0$ and $\ell_1+\ell_2\geq 0$ has been known since time immemorial \cite{Pham:1985cr, Adams:2006sv}. In our language, these translate into $\tilde g_2'\geq 0$ and $4\tilde g_2-\tilde g_2'\geq 0$ (i.e. the ``Skyrme line''). Similarly, some lower bounds on the combination of couplings\footnote{Since the (finite $N$) low-energy couplings of the chiral Lagrangian run with energy due to loops, it is customary to give their values at the mass of the pion using the beta function from one-loop chiral perturbation theory \cite{Leutwyler:1997yr}:
\begin{equation}\label{eq:lbars}
    \bar \ell_1 = \frac{32\pi^2}{1/3}\ell_1^r(\mu)-\log \frac{m_\pi^2}{\mu^2} \,,\qquad
    \bar \ell_2 = \frac{32\pi^2}{2/3}\ell_2^r(\mu)-\log \frac{m_\pi^2}{\mu^2}\,.
\end{equation}
Clearly, this is not well defined in the chiral limit, one should take $m_\pi\simeq 139.57\,$MeV for these formulae.} $\bar \ell_1 + 2\bar \ell_2$ were derived long ago \cite{Ananthanarayan:1994hf,Pennington:1994kc}. Recently, with the improvement of computational power, there has been a resurgence of this program using a variety of methods that have led to a menagerie of new bounds. In this subsection we compare our results with a few recent papers.

In a very nice recent work \cite{Alvarez:2021kpq} (which builds on \cite{Manohar:2008tc,Mateu:2008gv}), they  derived bounds for the chiral Lagrangian using techniques similar to ours. Although the core idea of using dispersion relations to bound the low-energy coefficients is shared by both approaches, there are significant differences that we now outline. On the one hand,
they consider massive pions and include loops in chiral perturbation theory (that correspond to $1/N$ corrections), which makes their results
much closer to nature, while
our target is the Platonic ideal of large $N$ QCD in the chiral limit, of which the real world is but an imperfect shadow.

On the other hand, their method does not exploit all the constraints that the dispersion relations provide. The reason is that they truncate the low-energy amplitude either to $O(p^4)$ or to $O(p^6)$ (respectively,  
NLO or NNLO in chiral perturbation theory). 
Keeping all orders in \eqref{eq:Mlow} allowed us to derive an infinity of null constraints that we took into account completely (up to numerical convergence, which seems quite fast). Null constraints~\cite{Caron-Huot:2020cmc, Tolley:2020gtv}  are the key improvement that allowed us to use all the information hidden in the dispersion relations. In addition, since they work at finite $N$, their Regge behavior is worse than \eqref{eq:Regge} and they need at least two subtractions, reducing the number of constraints  they have access to.

To see explicitly how the improvements from the full set of null relations come about, consider truncating the low-energy amplitude \eqref{eq:Mlow} to order $O(p^4)$ and using only twice-subtracted dispersion relations. Expanding \eqref{eq:disprels3} for $k=2$ around the forward limit yields the sum rules
\begin{equation}
    g_{2,0}=\avg{\frac{1}{m^4}}\,,\qquad g_{2,0}-g_{2,1}=\avg{\frac{(-1)^J}{m^4}}\,.
\end{equation}
Summing them and using  positivity of the high-energy average then gives the bound
\begin{equation}\label{eq:2subtbound}
    \left(4\tilde g_2 -\tilde g_2'\right)\frac{g_{1,0}}{2M^2}=2g_{2,0}-g_{2,1}=\avg{\frac{1+(-1)^J}{m^4}}\geq 0\,.
\end{equation}
Subtracting them yields $g_{2,1}=\avg{\frac{1-(-1)^J}{m^4}}\geq 0$. In terms of the low-energy couplings \eqref{eq:gtol} these bounds correspond to $\ell_1+\ell_2\geq 0$ and $\ell_2\geq 0$, respectively. This is the result at large $N$. To compare to finite $N$ results, we would have to include low-energy loops from the start, but we can see the qualitative effect of the loops by ``running up'' the couplings from $m_\rho$ to $m_\pi$ using \eqref{eq:lbars}. This yields the bounds
\begin{equation}
    \bar\ell_1 + 2\bar \ell_2\geq 3\log \frac{m_\rho^2}{m_\pi^2}\simeq 10.3\,\text{MeV}\,,\qquad
    \bar \ell_2\geq \log\frac{m_\rho^2}{m_\pi^2}\simeq 3.43\,\text{MeV}\,,
\end{equation}
which are comparable to figure 6 of \cite{Alvarez:2021kpq}. Including terms of order $O(p^6)$ and using more subtractions brings in the first null constraint, which we can use to place two-sided bounds on the couplings as in figure 7 of \cite{Alvarez:2021kpq}. But for the complete bounds that reveal the kink we need to keep $M_\text{low}(s,u)$ to all orders and use all the dispersion relations with one or more subtractions.

Another approach that has proven successful in the recent years is the numerical S-matrix bootstrap developed in \cite{Paulos:2017fhb}, which was first applied to massive \cite{Guerrieri:2018uew} and then to massless \cite{Guerrieri:2020bto} pions, and has been pursued further in \cite{Bose:2020cod,Bose:2020shm}. The optimization method we have used to obtain our positivity bounds solves a \textit{dual problem}, where one systematically rules out more and more points as more constraints are introduced. This method, in contrast, solves a \textit{primal problem}, in which points are progressively ruled in. Both methods should be regarded as complementary and they should be pushed until their solutions match, at which point one can be sure that the bounds are optimal. Sadly, our results are not directly comparable to any of these references because we do not tackle exactly the same problem.

\begin{figure}[ht]
\centering
\includegraphics[scale=1.1]{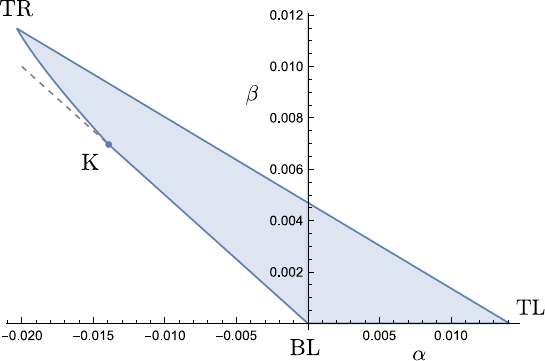}
\caption{Final exclusion plot from figure \ref{fig:exclPlot} in the parametrization \eqref{eq:gtoab}. This plot is to be compared with figure 1 of \cite{Guerrieri:2020bto}. We used the values $M=m_\rho\simeq 775\,$MeV as our cutoff scale and $f_\pi\simeq 92\,$MeV.}
\label{fig:alphabeta}
\end{figure}

The closest one is \cite{Guerrieri:2020bto}, where they also consider massless pions, but they work at one-loop level in chiral perturbation (i.e.\ next-to-leading order in $1/N$) and they do not assume a mass gap $M$ before the first resonance.\footnote{This assumption was crucial for us to define a cutoff separating high from low energies, and it is well justified since we know that the first exchanged meson is the rho.} Yet, it is instructive to rewrite our results in their notation. For the low-energy amplitude, they use the one-loop result
\be \label{AGuerrieri}
A (s | t, u)  = \frac{s}{f_\pi^2} + \frac{1}{f_\pi^4} \left[ \alpha s^2 +  \beta (t^2 + u^2) -\frac{s^2}{32\pi^2}\log\frac{-s}{f_\pi^2}-\frac{t-u}{96\pi^2}\left(t\log\frac{-t}{f_\pi^2}-u\log\frac{-u}{f_\pi^2}\right)\right]+\cdots
\ee
Neglecting loops, we can relate $\alpha$ and $\beta$ to our parameters as
\be
g_1 = \frac{1}{2 f_\pi^2} \, , \quad g_{2,0} = \frac{\alpha + 3 \beta }{2 f_\pi^4} \, , \quad g_{2,1} = \frac{\beta}{f_\pi^4} \, ,
\ee
which gives
\be\label{eq:gtoab}
\alpha=\left(\tilde g_2 -\frac{3}{4}\tilde g_2'\right)\frac{f_\pi^2}{M^2}\,,\qquad \beta=\frac{\tilde g_2'}{4}\frac{f_\pi^2}{M^2}\,.
\ee
Figure \ref{fig:alphabeta} shows our results in this new parametrization so that they can be (morally) compared with figure 1 of \cite{Guerrieri:2020bto}.

Both plots agree on the lower bounds $\beta \geq 0$, $\left(\alpha + 2\beta\right)\geq 0$, which are reached asymptotically in their case. These are precisely the bounds that we derived around \eqref{eq:2subtbound}. In \cite{Guerrieri:2020bto} they give a dispersive argument for these bounds analogous to ours. A clear difference between both figures, though, is that ours also has \textit{upper bounds} for $\alpha$ and $\beta$. This is due to the assumption of the mass gap $M$ corresponding to the mass of the rho. Notably, our plots do not overlap. The reason for this is that we have neglected loops. As discussed above, the effect of the loops from the tree-level point of view is to make the couplings run, so including them would shift our plot and make them overlap. It would be interesting to include low-energy loops (i.e.\  $1/N$ corrections) in our approach from the start and look for better agreement with the results of \cite{Guerrieri:2020bto}.

Finally, an interesting method for constraining scattering amplitudes  was developed in \cite{Haldar:2021rri,Raman:2021pkf} using geometric function theory. It was then applied to pions in \cite{Zahed:2021fkp}. 
This method yields {\it uncorrelated} bounds for each Wilson coefficient, i.e. a set of allowed intervals.
Thus, while they are easier to compute, these constraints do not capture  the rich geometry of the space of healthy EFTs (e.g.\ the kink). We could not compare our results with \cite{Zahed:2021fkp} because it seems essential for them to consider \textit{massive}  pions and also because they do not find bounds for the couplings $\tilde g_2,$ $\tilde g_2'$. This is probably due to the fact that they only use dispersion relations for fully crossing-symmetric combinations of the pion amplitude, which reduces the number of constraints and prevents  from reaching the lowest-lying coefficients.

\section{\boldmath Including the \texorpdfstring{rho}{rho}}\label{sec:rho}
\begin{figure}[htb]
\centering
\includegraphics[scale=0.4]{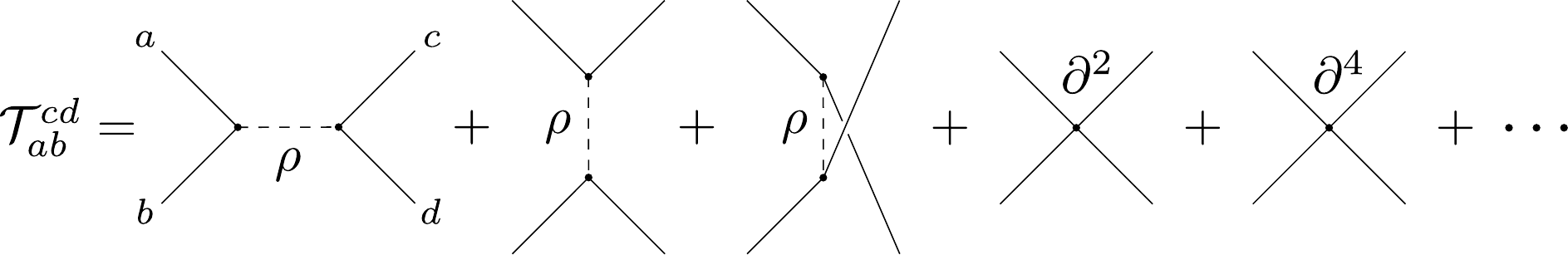}
\caption{Feynman diagrams contributing to the $2\to 2$ scattering of pions for an EFT including the rho. At large $N$ only tree-level diagrams survive. There is the explicit exchange of the rho and an infinite series of 4-point vertices coming from integrating out the heavier mesons in the spectrum.}
\label{fig:diagrams}
\end{figure}

We can make another step towards QCD by modifying our low-energy EFT to account for the exchange of the rho meson, the first state that contributes  to pion scattering at large $N$ (see the discussion in appendix \ref{app:internalstates}). In nature, the $\rho^a_\mu$ meson is an $SU(2)$ isospin triplet of spin $J=1$ and mass $m_\rho$. More generally, we will consider a massive spin-one particle in the adjoint representation of $SU(N_f)$, which we will continue to call $\rho^a_\mu$ ($a=1,\ldots,N_f^2-1$). Such a particle can only interact with two pions via an interaction term of the form
\begin{equation}\label{eq:Lint}
    \mathcal{L}_\text{int}=-\grho\, f_{abc}\pi^a\partial^\mu\pi^b\rho^c_\mu\,,
\end{equation}
as discussed in appendix \ref{app:Nfselectionrules}, and in the $N\to \infty$ limit this is all we need since only tree-level diagrams survive. The low energy $\pi\pi\to\pi\pi$ scattering amplitude for an EFT including such an interaction term (together with a kinetic and mass terms for the rho) is computed with the diagrams of figure \ref{fig:diagrams}, and it has the form of \eqref{eq:T} with
\begin{equation}
    A_{\text{low}}^{(\rho)}(s|t,u)=\grho^2\left(\frac{m_\rho^2+2s}{m_\rho^2-u}+\frac{m_\rho^2+2s}{m_\rho^2-t}\right)
    +\text{analytic}\;.
\end{equation}

Using \eqref{eq:AtoM}, this yields the following new low-energy amplitude,
\begin{equation}\label{eq:MlowRho}
	M_{\text{low}}^{(\rho)}(s,u)=
	\frac{1}{2}\grho^2 \left(\frac{m_\rho^2+2u}{m_\rho^2-s}+\frac{m_\rho^2+2s}{m_\rho^2-u}\right)
	+\sum_{n=0}^{\infty}\sum_{\ell=0}^{[n/2]}\hat g_{n,\ell}\left(s^{n-\ell}u^\ell+u^{n-\ell}s^\ell\right)\,.
\end{equation}
This amplitude is crossing-symmetric in $s\leftrightarrow u$ and it has a pole (for fixed $u$) at $s=m_\rho^2$ with residue $-\frac{1}{2}\grho^2m_\rho^2\,\mathcal{P}_1\left(1+\frac{2u}{m_\rho^2}\right)$, which is consistent with unitarity as long as $\grho^2\geq 0$. Note that we have written hats on top of the $g_{n,\ell}$ couplings to emphasize that these coefficients are different from those in \eqref{eq:Mlow}. Integrating out the rho would bring back the original coefficients $\hat g_{n,\ell}\to g_{n,\ell}$. At the level of the amplitude, integrating out the rho is as simple as Taylor-expanding the poles around $s, u\sim 0$, which we can do safely for energies below the mass of the rho. The precise relation between the couplings is\footnote{The Adler zero is accounted for by the coupling $\hat g_{0,0}=-\frac{1}{2}\grho^2$.}
\begin{align} \label{eq:gghat}
    &g_{1,0}=\hat g_{1,0}+\frac{3}{2}\frac{\grho^2}{m_\rho^{2}}\,, 
 	&&g_{n,1}=\hat g_{n,1}+\frac{\grho^2}{m_\rho^{2n}} \quad n=2,3,\dots\,,\nonumber \\
	&g_{n,0}=\hat g_{n,0}+\frac{1}{2}\frac{\grho^2}{m_\rho^{2n}} \quad n=2,3,\dots\,,\quad
	&&g_{n,\ell}=\hat g_{n,\ell} \phantom{+\frac{\grho^2}{m_\rho^{2n}}} \;\,\,\left\{
    \begin{aligned}
        n &= 4,5, \dots \, \\
        \ell&=2,\dots,\left[\tfrac{n}{2}\right] \,.
    \end{aligned}\right.
\end{align}

In the previous section, we used the original EFT amplitude \eqref{eq:Mlow} for energies up to the cutoff scale $M$. When considering QCD, this cutoff can be pushed only up to $m_\rho$. Beyond this scale, one must use the new amplitude $M_{\text{low}}^{(\rho)}$ including the rho pole. Despite having the explicit rho, this amplitude is still an EFT with all the higher mesons integrated out, so it can in turn only be used up to a new cutoff scale $M'> m_\rho$ (the mass of the next meson in the spectrum).

The amplitude \eqref{eq:MlowRho} describes the low-energy physics of any theory with a rho as the first meson exchanged by the pions and, among them, is large $N$ QCD. So we can play the same game as in section \ref{sec:disprels} with it to carve out the space of such theories allowed by unitarity. This should bring us closer to QCD. Moreover, with this low-energy amplitude we gain access to the more experimentally-available quantities $\grho^2$, $m_\rho$. We can try to put bounds on these quantities as well.

\subsection{``New'' sum rules and null constraints}
\begin{figure}
\centering
\begin{subfigure}[b]{0.45\textwidth}
   \includegraphics[scale=0.6]{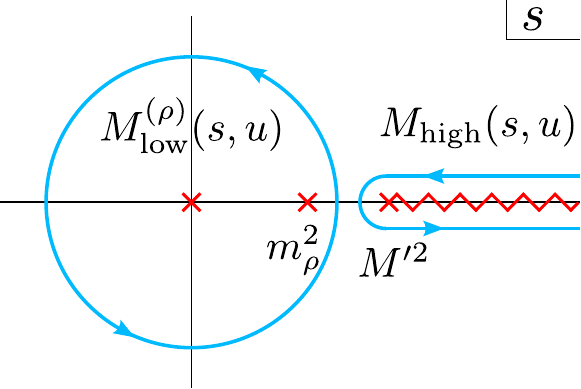}
   \caption{rho at low energies}
\end{subfigure}
\quad
\begin{subfigure}[b]{0.45\textwidth}
   \includegraphics[scale=0.6]{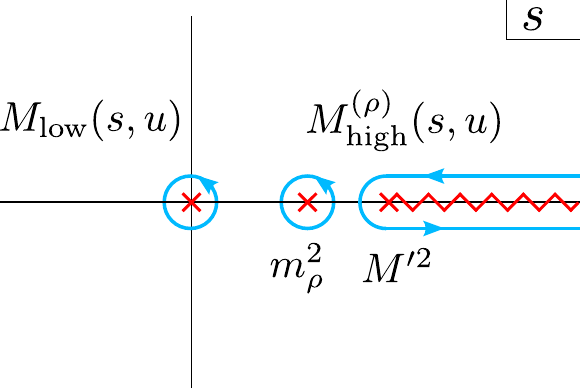}
   \caption{rho at high energies}
\end{subfigure}
\caption{There are two equivalent ways of dealing with the rho: (a) including it at low energies or (b) keeping it at high energies. We will go with the latter. The analogous contours for the ST sum rules also have poles on the negative $\text{Re}\,s$ axis.}
\label{fig:rhocontours}
\end{figure}

The naive way of treating this new amplitude is to separate low and high energies by the new cutoff $M'$ so that the contour at low energies steps on two poles; zero and $m_\rho^2$ (see figure \ref{fig:rhocontours}(a)). In this approach we would simply have to use \eqref{eq:MlowRho} in the left hand side of the dispersion relations \eqref{eq:disprels3} and consider high-energy averages integrating over masses $m\geq M'$. Expanding in $u$ would then yield new null constraints and sum rules for the couplings $\hat g_{n,\ell}$, $\grho$ that we could use in a dual problem like the one in section \ref{sec:dualprob}. However, it is much more efficient and illuminating to keep the rho pole in the high-energy side of the dispersion relations. That is, to use the fully analytic amplitude \eqref{eq:Mlow} (with $g_{n,\ell}$ couplings) around the origin and consider as high-energy spectrum an isolated rho pole together with a continuum starting above $M'$ (see figure \ref{fig:rhocontours}(b)).

This is achieved by shifting the high-energy spectral density by
\begin{equation}
    n_J^{(D)}\rho_J(s)\longrightarrow \frac{\pi}{2} \,\grho^2\delta(s-m_\rho^2)\,\delta_{J,1}\,m_\rho^{D-2}+n^{(D)}_J\rho_J'(s)\,,
\end{equation}
where $\rho_J'(s)$ now has support above the new cutoff $M'$. This makes a new term pop out from the heavy averages accounting for the isolated rho,
\begin{equation}
\label{eq:newavg}
	\avg{F\left(m^2,J\right)}\longrightarrow  \frac{1}{2}\grho^2\, F\left(m_\rho^2,J=1\right) +\avg{F\left(m^2,J\right)}' \,,
\end{equation}
where the prime indicates that the averaging is done for masses $m\geq M'$. With this replacement it is straightforward to derive the new sum rules and null constraints from the original ones. A few examples are
\begin{equation}
    g_{1, 0} = \frac{1}{2}\frac{\grho^2}{m_\rho^{2}}+\avg{\frac{1}{m^{2}}}' \, , \quad
    g_{2, 0} = \frac{1}{2}\frac{\grho^2}{m_\rho^{4}}+\avg{\frac{1}{m^{4}}}' \, , \quad
    g_{2, 1} = \frac{1}{2}\frac{\grho^2}{m_\rho^{4}}+\frac{1}{(D-2)}\avg{\frac{\JJ^2}{m^4}}'\,.
\end{equation}
These results could be rearranged into sum rules for the new couplings $\hat g_{n,\ell}$ using \eqref{eq:gghat}, but as we now discuss, they will be more useful as they are.

\subsection{New exclusion plot}\label{sec:newExcPlot}
The easiest way to compare the allowed region for theories including a rho with the original plot (figure \ref{fig:exclPlot}) is to look for bounds on the same couplings $g_{n,\ell}$ rather than the new couplings $\hat g_{n,\ell}$. We will construct the same plot as before but assuming the high-energy spectrum of figure \ref{fig:rhocontours}(b) for different values of $M'$. Thus, we will be asking \textit{how does the allowed region for pion EFTs change as we restrict to theories with a rho as their first meson in the spectrum?} And, \textit{how does that change as we increase the gap after the rho?}

Consider the same vectors as in \eqref{eq:exclvectors} together with
\begin{equation}
    \vec v_\rho \equiv \vec v_\text{HE}\left(m_\rho^2,J=1\right)\,,
\end{equation}
and normalize them by $m_\rho$ (i.e.\ replace $M\to m_\rho$). These vectors now satisfy the bootstrap equation
\begin{equation} \label{eq:bootstrpeqRho}
	\vec v_\mathbb{1}+\tilde g_2\, \vec v_2+\tilde g_2'\, \vec v_{2'}+\tilde g_\rho^2\vec v_\rho+ \avg{\vec v_\text{HE}\left(m^2,J\right)}'=0\,,
\end{equation}
where the normalized couplings are defined as
\begin{equation}\label{eq:normcouplingsRho}
    \tilde g_2 \equiv g_{2,0}\frac{m_\rho^2}{g_{1,0}}\,,\qquad 
    \tilde g_2' \equiv 2g_{2,1}\frac{m_\rho^2}{g_{1,0}}\,,\qquad 
    \tilde g_\rho^2 \equiv \frac{1}{2}\frac{\grho^2}{g_{1,0}\,m_\rho^2}\,.
\end{equation}
Compared to \eqref{eq:bootstrpeq2}, in this bootstrap equation the heavy states consist of the isolated $\vec v_\rho$ and the continuum $\vec v_\text{HE}$. So we can construct the new exclusion plot by following the same program from section \ref{sec:exclPlot} with the positivity condition (step 2) updated to
\begin{equation}
    \vec \alpha\cdot\vec v_\rho\geq 0\,,\qquad
    \vec \alpha \cdot \vec v_\text{HE}\left(m^2,J\right)\geq 0\quad \forall J,\,m\geq M'\,,
\end{equation}
for fixed values of $M'$ and $m_\rho$.

\begin{figure}[htb]
\centering
\includegraphics[scale=1.1]{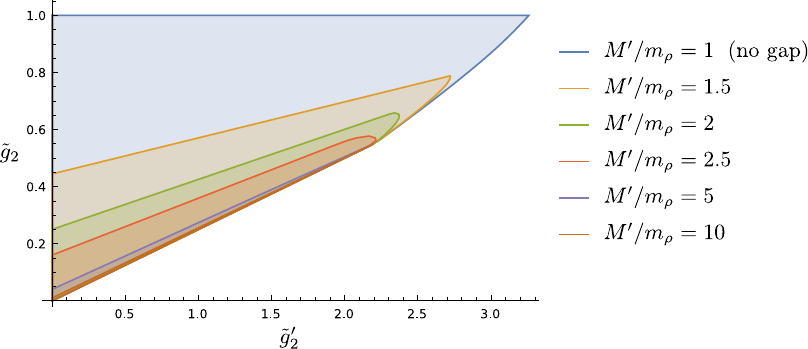}
\caption{Exclusion plot in the space of couplings $(\tilde g_2',\tilde g_2)$ when the high-energy spectrum contains an isolated rho below a cutoff $M'$, for different values of $M'/m_\rho$. All the plots were computed at Mandelstam order $n=7$. At $M'/m_\rho=1$ the allowed region matches figure \ref{fig:exclPlot}, and when $M'/m_\rho\to\infty$ it shrinks towards the Skyrme line connecting the origin to the kink.}
\label{fig:RHOexclPlot}
\end{figure}

Out of the two mass scales of this problem, only the ratio $M'/m_\rho$ is meaningful, but the bounds change by an overall rescaling depending on which mass we use to normalize the couplings in \eqref{eq:normcouplingsRho}. We have chosen to normalize everything on $m_\rho$ so that raising $M'$ corresponds to enlarging the gap after a fixed rho rather than moving down the rho under a fixed cutoff. Figure \ref{fig:RHOexclPlot} shows how the exclusion plot changes as a function of this gap.

For $M'/m_\rho=1$, there is no gap after the rho and hence the plot reduces to the original one (figure \ref{fig:exclPlot}). This is the most permissive assumption we can make, since it allows for virtually any UV completion, and so it yields the biggest allowed region. As we raise $M'$, we restrict ourselves to theories with a larger and larger gap and the region shrinks. Already for relatively small gaps, the allowed region shrinks significantly towards the lower-right part of the plot, indicating that this is where the theories with a rho as their first internal state congregate. Astonishingly, as we push $M'$ very high, the allowed region shrinks onto the Skyrme line connecting the origin to the kink, which seems to survive all the way to $M'/m_\rho\to\infty$.

This is an interesting result. It implies that one can construct ``healthy theories'' (i.e.\ crossing-symmetric, unitary and spin-one Regge bounded) out of a single rho and infinitely heavy mesons. Moreover, these theories must lie on the Skyrme line. So, at the kink there must live \textit{at least} a funny UV completion of a spin-one particle,\footnote{We pursue this idea further in section \ref{sec:belowKink}.} but this is not necessarily the whole story. Our bounds show the allowed region for theories that \textit{meet our assumptions}; they have nothing to say about theories that do not satisfy them. So it could well be that the kink is also populated by other theories with a much different spectrum.

For QCD, the cutoff $M'$ can only be pushed as high as the next meson in the spectrum, and one might naively conclude that we will rule out QCD when $M'/m_\rho$ is taken high enough. But we cannot make such a statement. Again, our bounds ensure that any theory compatible with the assumed spectrum will live inside the allowed region, but they do not force all the theories that fall within the bounds to have the assumed spectrum. For example, a healthy theory with a fixed gap of --say-- $M'/m_\rho=1.5$ will lie inside the light orange area of figure \ref{fig:RHOexclPlot}, but it can either fall inside or outside the green region. Therefore, while we have something of a candidate for a theory living at the kink, the possibility that large $N$ QCD  sits on top of it remains open.\footnote{In fact, the significance of the rho meson was emphasized long ago by observing that the experimental low-energy couplings are almost saturated by the rho pole alone \cite{Ecker:1988te,Donoghue:1988ed,Pennington:1994kc}. So perhaps it is not that far fetched that large $N$ QCD might sit on top of this funny theory that involves just a rho meson.}

\subsection{The rho coupling}\label{sec:rhocoupling}
Let us continue our quest for QCD by studying what values of the rho coupling $\grho$ are allowed by unitarity. To do so, one might be tempted to rearrange the new null constraints resulting from \eqref{eq:newavg} into a function $\grho^2(m^2,J)$ whose high-energy average gives $\grho^2$ and then use it in a dual problem like in section \ref{sec:dualprob}. However, with our reformulation of the dual problem in terms of bootstrap equations, we do not have to work so hard to bound $\grho^2$. Indeed, the vectors
\begin{equation}\label{eq:rhovectors}
	\vec{v}_\mathbb{1}=\begin{pmatrix}
	1\\ 0 \\ \vdots\\ 0
	\end{pmatrix}\,,\qquad
	\vec{v}_{\text{HE}}\left(m^2,J\right)=\begin{pmatrix}
	-g_{1,0}\left(m^2,J\right)m_\rho^2\\  \mathcal{Y}_{2,1}\left(m^2,J\right)m_\rho^4\\\mathcal{X}_{3,1}\left(m^2,J\right)m_\rho^6\\ \vdots
	\end{pmatrix}\,,\quad
	\vec v_\rho \equiv \vec v_\text{HE}\left(m_\rho^2,J=1\right)\,,
\end{equation}
defined with the same sum rules and null constraints from section \ref{sec:disprels}, satisfy the bootstrap equation
\begin{equation} \label{eq:bootstrpeqRho_coupling}
	\vec v_\mathbb{1}+\tilde g_\rho^2\vec v_\rho+ \avg{\vec v_\text{HE}\left(m^2,J\right)}'=0\,.
\end{equation}
This has the form of \eqref{eq:bootstrpeq} with $\vec v_\rho$ the operator whose ``OPE coefficient'' we want to bound. We can thus obtain bounds for $\tilde g_\rho^2$ directly by applying the same program from section \ref{sec:dualprob} to these vectors. The key point is that by normalizing the functional $\vec \alpha$ on $\vec v_\rho$ we avoid having to derive a sum rule for $\grho^2$.

\begin{figure}[htb]
\centering
\includegraphics[scale=1.1]{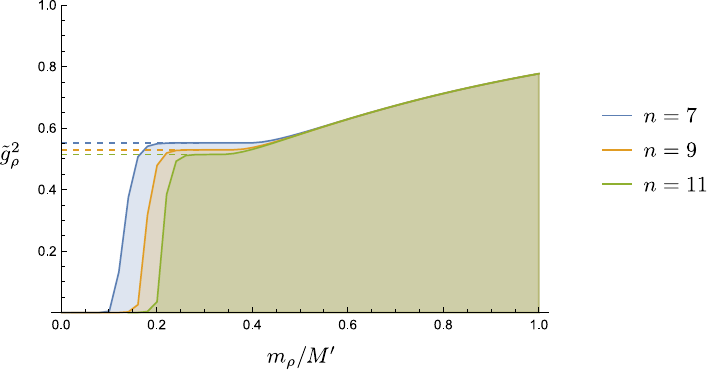}
\caption{Upper bounds for the normalized coupling of the rho, $\tilde g_\rho^2\equiv \frac{1}{2}\frac{\grho^2}{g_{1,0}\,m_\rho^2}$, as a function of the gap between the rho and the higher mesons in the spectrum (starting above the cutoff scale $M'$). The different plots were made using all the null constraints up to Mandelstam order $n=7,9,11$. The sharp edges to the left of the plots are a numerical artifact (see figure \ref{fig:spin_convergence}); the dashed lines delineate the true bounds.}
\label{fig:RHOcoupling}
\end{figure}

Either way, we can systematically apply this program using \texttt{SDPB} \cite{sdpb} to place bounds on $\tilde g_\rho^2$ for different fixed values of the cutoff $M'$. This yields the upper bounds shown in figure \ref{fig:RHOcoupling}. The highest allowed value for $\tilde g_\rho^2$ happens for $M'=m_\rho$ (when there is no gap), where we have
\begin{equation}\label{eq:grhobound}
    \tilde g_\rho^2 \leq 0.776674...\,.
\end{equation}
The lower bounds one gets from this program are always negative, but we know that $\grho^2\geq 0$ for the rho pole to be unitary, so they are meaningless.

\begin{figure}[htb]
\centering
\includegraphics[scale=1.1]{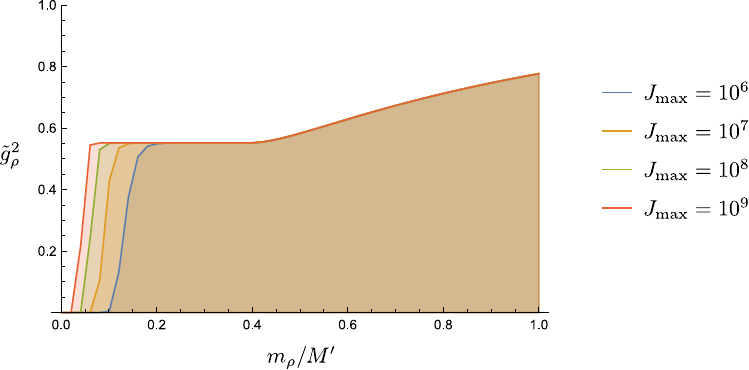}
\caption{Convergence in spins for the plot in figure \ref{fig:RHOcoupling} at Mandelstam order $n=7$. The right part of the plot converges quickly in spins. In contrast, the steep descent at the left moves further to the left as we increase $J_\text{max}$, indicating that the plateau extends all the way to the $y$-axis when convergence is achieved.}
\label{fig:spin_convergence}
\end{figure}

A surprising feature of figure \ref{fig:RHOcoupling} is the sudden jumps at the left part of the plots. This is purely a numerical artifact from the truncation in spins that we discussed in section \ref{sec:dualprob}. As shown in figure \ref{fig:spin_convergence}, increasing the spin cutoff $J_\text{max}$ pushes the bound further to the left, so the true bounds have a plateau that extends all the way to $m_\rho/M'=0$. Thus, we find once again that there must exist some healthy theory with just a rho and all the higher mesons pushed to infinity, like we saw in figure \ref{fig:RHOexclPlot}.

Returning to figure \ref{fig:RHOcoupling}, we see that the bounds separate into two regions of distinct behavior; a smooth curve to the right of $m_\rho/M'\sim 0.4$ and a plateau to the left. While the curve has already converged in $n$, the plateau has not.\footnote{Exploring the plateau for higher values of $n$ is numerically very expensive as more and more spins must be included.} Surprisingly, we have found that for each value of $n$ the value of $\tilde g_\rho^2$ at the plateau matches exactly with the value of $\tilde g_2$ at the kink for the corresponding plot in figure \ref{fig:exclPlot}! Which raises the quesion: \textit{what is the origin of this connection?} and \textit{are any of these features related to large $N$ QCD?} We will shed some light into this mysterious relation in section \ref{sec:rulingin}, but for the moment this observation suggests that the convergence of the plateau is also given by figure \ref{fig:kinkvalue}. The plateau would then converge towards   $\tilde g_\rho^2\sim 0.42$.

\subsection{Hidden local symmetry and vector meson dominance}\label{sec:HLS}
The standard chiral Lagrangian \eqref{eq:chiralL} is an EFT only for pions. In general, to include the rho we would have to include all the interaction terms allowed by the symmetries of the problem together with the corresponding unknown coefficients. There is a notorious model, whose fundamental significance is still very much unclear,
where the rho meson is treated as the dynamical gauge boson of a ``hidden local symmetry'' (HLS) of the chiral Lagrangian \cite{Bando:1984ej,Bando:1987br} (see \cite{Komargodski:2010mc} for a nice discussion). This ansatz reduces the number of unfixed coefficients, giving relations between different physical quantities that agree surprisingly well with phenomenological observations. Let us see what our bounds can say about this model.

In a nutshell, if the matrix $U(x)\in SU(N_f)$ of the chiral Lagrangian is rewritten as a product of two special unitary matrices $U(x)=\xi_L(x)\xi_R(x)^\dagger$, apart from the global chiral symmetry $SU(N_f)_L\otimes SU(N_f)_R$ acting as
\begin{equation}
    \xi_L(x)\to g_L \xi_L(x)\,,\qquad
    \xi_R(x)\to g_R \xi_R(x)\,,
\end{equation}
the system enjoys an additional $\left[SU(N_f)\right]_\text{local}$ symmetry
\begin{equation}
    \xi_L(x)\to  \xi_L(x)h(x)\,,\qquad
    \xi_R(x)\to  \xi_R(x)h(x)\,.
\end{equation}
Of course, this ``hidden symmetry'' just indicates that the new parametrization is redundant, but now we can incorporate the corresponding dynamical gauge field $\rho_\mu^a$ and interpret it as the rho meson. The idea is that its mass should be  generated (somehow) via the Higgs mechanism.

The Lagrangian is then obtained by writing down all the $L\leftrightarrow R$ symmetric terms compatible with both the global and local symmetries. At two-derivative order, it reads
\begin{equation}\label{eq:LHLS}
    \mathcal{L}_\text{HLS}=-\frac{1}{2 g^2}\tr{F_{\mu\nu}F^{\mu\nu}}-\frac{f_\pi^2}{4}\tr{\left(\rho_\mu^L-\rho_\mu^R\right)^2}-a\frac{f_\pi^2}{4}\tr{\left(\rho_\mu^L+\rho_\mu^R\right)^2}+\cdots\,,
\end{equation}
where
\begin{equation}
    \rho_\mu^L=\rho_\mu^aT_a - i\xi_L^\dagger \partial_\mu \xi_L\,,\qquad
    \rho_\mu^R=\rho_\mu^aT_a - i\xi_R^\dagger \partial_\mu \xi_R\,.
\end{equation}
There are two free parameters; the ``gauge coupling'' $g$ and an arbitrary coefficient $a$. The normalization of the second term is fixed in terms of $f_\pi$ by requiring that this Lagrangian reproduces the leading term of the chiral Lagrangian \eqref{eq:chiralL} upon integrating out the rho. 

By picking the unitary gauge $\xi_L(x)=\xi_R^\dagger(x)\equiv \exp \left[\frac{i}{f_\pi}T_a\pi^a(x)\right]$ and expanding in pion fields it is easy to see that $\mathcal{L}_\text{HLS}$ includes a single $\pi\pi\rho$-interaction term, which has the form of \eqref{eq:Lint} with coupling
\begin{equation}
    \grho=\frac{1}{2}ga\,.
\end{equation}
There is also a mass term for the gauge boson $\rho_\mu^a$, due to the higgsing of $\left[SU(N_f)\right]_\text{local}$, with mass
\begin{equation}
    m_\rho^2=ag^2f_\pi^2\,.
\end{equation}
Apart from these two quantities, the assumption that the rho meson arises as the gauge boson for a HLS relates other couplings to the basic paramaters $g$ and $a$. Namely, the three- and four-rho couplings are proportional to $g$ and $g^2$ respectively, while the $\rho\rho\pi\pi$ coupling is $\sim a g^2$. However, in the $2\to 2$ scattering of pions at large $N$ only $\grho$ and $m_\rho$ show up, cf.\ \eqref{eq:MlowRho}. Since there is a total of two independent parameters, $m_\rho$ and $\grho$ are independent and we can use \eqref{eq:LHLS} to describe \textit{any} amplitude allowed by our bounds. So our results cannot test HLS per se. To do so, one would need to consider new processes, such as $\pi\pi\to\rho\rho$ or $\rho\rho\to\rho\rho$ to gain access to the remaining couplings and check if the relations imposed by HLS are allowed by unitarity.  Considering the full system of mixed $2\to 2$ amplitudes involving pions and rhos is an important direction for future work.

\begin{figure}[htb]
\centering
\includegraphics[scale=1.1]{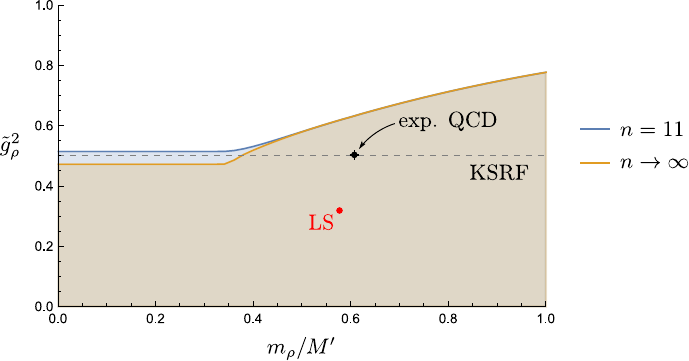}
\caption{Location of different amplitudes inside the bounds for $\tilde g_\rho^2$. The blue curve is the bound computed at Mandelstam order $n=11$ while the orange one is the projected convergence of the bound obtained from fitting the lower values of $n$. The gray dashed line corresponds to the KSRF relation \cite{Kawarabayashi:1966kd,Riazuddin:1966sw}, while the red and black dots mark, respectively, the location of the Lovelace-Shapiro amplitude and real-world QCD. For QCD, the experimental error bars are comparable to the size of the dot.}
\label{fig:RHOcoupling_exp}
\end{figure}

There is nevertheless one thing we can assess: the parameter tuning  that corresponds to ``rho dominance''. This is the choice  $a=2$ in the HLS Lagrangian, as  motivated by several phenomenological observations \cite{Bando:1984ej}. First, for this value of $a$, the HLS Lagrangian predicts $\grho=g$, which explains the universality of the rho coupling \cite{Sakurai}. Second, the mass and the rho coupling satisfy the celebrated KSRF relation \cite{Kawarabayashi:1966kd,Riazuddin:1966sw},
\begin{equation}\label{eq:KSRF}
    m_\rho^2=2\grho^2f_\pi^2\,,
\end{equation}
observed empirically. And third, (after extending \eqref{eq:LHLS} to account for interactions with photons) one sees that for $a=2$ the electromagnetic form factor of the pion (in this ultrasimplified model, of course) only receives contributions from the rho; this is known as rho dominance (or vector meson dominance)~\cite{Sakurai}.
With this choice, the normalized coupling of the rho takes the value\footnote{Note that although this is the result for the choice $a=2$ in the HLS model, it readily follows from the KSRF relation \eqref{eq:KSRF}. So we are probing just one of the consequences of the model for $a=2$. To probe independently universality and rho dominance, we would need to consider scattering processes with external rhos and external photons respectively.}
\begin{equation}\label{eq:rhodominance}
    \tilde g_\rho^2\equiv \frac{1}{2}\frac{\grho^2}{g_{1,0}m_\rho^2}=\frac{a}{4}=0.5\,,
\end{equation}
where we have used $g_{1,0}=\frac{1}{2f_\pi^2}$, as derived in \eqref{eq:Ltog} for the chiral Lagrangian. This result is plotted against our bounds in figure \ref{fig:RHOcoupling_exp}.

In this figure we have included, apart from the bound at order $n=11$, the extrapolation to $n\to\infty$ obtained from fitting the data in figure \ref{fig:RHOcoupling} to get a rough idea of the final shape of the bound once convergence in $n$ is achieved. From the relation between the plateau and the position of the kink in figure \ref{fig:exclPlot} noted above, though, we actually expect the plateau to move further down to a value around $\tilde g_\rho^2\sim 0.42$ when more values of $n$ are used. Since the plateau inevitably converges to a value below $0.5$, the KSRF line intersects with the bound and thus \eqref{eq:KSRF} is only allowed for values of $m_\rho/M'$ greater than $\sim 0.38$.

We can further study the HLS model by integrating out the rho from \eqref{eq:LHLS} so as to recover the chiral Lagrangian with a particular combination of low energy couplings. As discussed e.g.\ in \cite{Zahed:1986qz}, this yields precisely the Skyrme model \eqref{eq:skyrmeL}. To integrate out a ``heavy'' rho we must plug its equation of motion,
\begin{equation}
    \rho_\mu^aT_a = \frac{i}{2}\left(\xi_L^\dagger \partial_\mu \xi_L+\xi_R^\dagger \partial_\mu \xi_R\right)\,,
\end{equation}
back into \eqref{eq:LHLS}. With this replacement, the last term in \eqref{eq:LHLS} vanishes exactly and the second term (which is actually independent of $\rho_\mu^a$) matches the leading term of the chiral Lagrangian. Meanwhile, the kinetic term for the rho (at leading order in derivatives) reproduces the Skyrme term
\begin{equation}
    -\frac{1}{2g^2}\tr{F_{\mu\nu}F^{\mu\nu}}=\frac{1}{32g^2}\tr{\left[U^\dagger\partial_\mu U,U^\dagger\partial_\nu U\right]\left[U^\dagger\partial^\mu U,U^\dagger\partial^\nu U\right]}\,.
\end{equation}
So, by integrating out the rho and truncating at four-derivative order, we indeed recover the Skyrme model \eqref{eq:skyrmeL} with the free parameter identified with the gauge coupling; $e=g$.

Thus, in the space of two-derivative couplings (figure \ref{fig:exclPlot}), this theory sits on the Skyrme line at a value
\begin{equation}
    \tilde g_2=\frac{1}{4g^2}\frac{M^2}{f_\pi^2}=\frac{a}{4}\,.
\end{equation}
For the special choice of $a=2$, we get $\tilde g_2=0.5$ and so the HLS model ``at the point of rho dominance'' sits close to the kink of figure \ref{fig:exclPlot}. However, according to our numerics (see figure \ref{fig:kinkvalue}), $\tilde g_2=0.5$ is not quite the kink; which we expect to converge to a value below 0.5 (to approximately $0.42$ according to our $n \to \infty$ extrapolation). In fact, the bound at $n=15$ \eqref{eq:kinkpos} already rules out this point. This means that \eqref{eq:LHLS} with the choice $a=2$ is not compatible with unitarity as is; $\mathcal{L}_{\text{HLS}}$ is too simple of a model. Of course, higher-derivative corrections to it would change the values of $\tilde g_2,\tilde g_2'$
and may restore unitarity.

\subsection{Comparing with experiment}
We end this section by locating real-world QCD in the plot of the rho coupling bound. There are several ways of determining $\grho$ \cite{Sakurai:1966zza}. The most direct one is to use experimental results for the decay width of $\rho\to\pi\pi$ and compare to
\begin{equation}
    \Gamma_{\rho\to\pi\pi}=\frac{\grho^2}{48\pi}m_\rho\left(1-\frac{4m_\pi^2}{m_\rho^2}\right)^{\frac{3}{2}}\,.
\end{equation}
From the current measured values $\Gamma_{\rho\to\pi\pi}=149.1\pm 0.8\,$MeV, $m_\pi\simeq 139.57\,$MeV \cite{PDG}, we get $\grho^2=35.7\pm 0.2$, which in turn yields
\begin{equation}
    \tilde g_\rho^2=\grho^2\frac{f_\pi^2}{m_\rho^2}=0.503\pm 0.003\,.
\end{equation}
This agrees very well with the KSRF relation \eqref{eq:rhodominance}.

As for the value of the cutoff $M'$, it can only be pushed up to the next meson in the spectrum after the rho. As discussed in appendix \ref{app:internalstates}, for large $N$ QCD this is the $f_2 (1270)$, which has mass $M'=1275.5\pm0.8\,$MeV \cite{PDG}.\footnote{It so happens that the first non-exotic meson after the rho is the  $f_2$, which has spin two. We note in passing that a spin-zero state would have decoupled 
from the optimization problem leading to figure~\ref{fig:RHOcoupling}. Indeed, since all the null constraints vanish for $J=0$ (recall \eqref{eq:someNC}), the vectors $\vec v_\text{HE}(m^2,J=0)$ are proportional to $\vec v_\mathbb{1}$ and demanding positivity of the functional on them becomes trivial. So spin-zero particles do not contribute in constraining the rho coupling. The ultimate reason behind this fact is that we only used dispersion relations with at least one subtraction, due to the spin-one Regge behavior \eqref{eq:Regge}.
Note however 
that the bounds on $\tilde g_2$, $\tilde g_2'$ (figure \ref{fig:RHOexclPlot}) {\it are} sensitive to scalars. Indeed, in contrast with the null constraints, not all the sum rules $g_{n,\ell}(m^2,J)$ vanish for $J=0$.}

With these coordinates in hand, we can place real-world QCD 
in the allowed region for the rho coupling, see figure \ref{fig:RHOcoupling_exp}. For reference, we have also added the point that corresponds to the Lovelace-Shapiro amplitude, which has $m_\rho/M'=1/\sqrt{3}$, $\tilde g_\rho^2=1/\pi$. Comfortingly, both points are allowed by our bounds, but QCD appears to be rather far from the boundary and so we cannot make a precise connection with any feature of the plot. Note, however, that although the experimental error bars for real-world QCD are very small, one should always keep in mind that we are comparing it to the large $N$ theory in the chiral limit, so there is room for larger discrepancies.

\section{Understanding the geometry of the bounds}\label{sec:rulingin}
It is an outstanding question to understand the geometry of positivity bounds from a more analytic point of view. While we have seen some intriguing connections to old hadron phenomenology that might indicate that QCD sits in the vicinity of the kink, we have also found evidence that a simpler theory lives at that point. In  \cite{Caron-Huot:2020cmc} it was shown that some positivity bounds for EFTs can be saturated by very simple amplitudes satisfying crossing, unitarity and Regge boundedness. So it could be that the kink is just a spurious solution that prevents us from reaching more interesting parts inside our plot. In this section we organize the evidence for this hypothesis and engineer an example that rules in a significant part of the exclusion plot. We would like to stress, though, that even if such a simple amplitude saturating the bounds is eventually found, it could still be that \textit{different} theories (like large $N$ QCD) sit on top of it. The way to go would then be to make further assumptions that are true for large $N$ QCD but not for the spurious solution and see whether the kink survives or not.

\subsection{Analytically ruling in}
As discussed in section \ref{sec:previouswork}, there is a complementary method to the \textit{dual problem} that we have used; the \textit{primal problem}. The difference being that the former progressively rules out more points of the space of EFTs while the latter rules them in. In a primal problem, one looks for examples of EFTs that satisfy our assumptions in order to say that a given point is allowed. When the bounds of both methods agree, we know that they are optimal. This approach has been applied numerically to pion scattering in \cite{Guerrieri:2018uew,Guerrieri:2020bto,Bose:2020cod,Bose:2020shm}, but in \cite{Caron-Huot:2020cmc} they introduced an alternative ---and very insightful--- way of answering this question, which consists in ruling in amplitudes \textit{analytically}. The basic idea is that positivity bounds might be saturated by simple functions $M(s,u)$ satisfying crossing symmetry, unitarity and the assumed Regge behavior. If we manage to find them, not only do we learn that the bounds are optimal, but we also gain a conceptual understanding of the geometry of the bounds.

Following \cite{Caron-Huot:2020cmc} closely, we tackle first the top bound of figure \ref{fig:exclPlot}; $\tilde g_2\leq 1$. From the sum rules 
\begin{equation}\label{eq:g20sumRule}
    g_{1,0}=\avg{\frac{1}{m^2}}\,,\qquad
    g_{2,0}=\avg{\frac{1}{m^4}}\,,
\end{equation}
it is clear that $\tilde g_2=\frac{g_{2,0}M^2}{g_{1,0}}$ will only reach $1$ on amplitudes that \textit{only} have states at $m=M$. Any pole at a higher mass would lower $\tilde g_2$. Apart from the constant amplitude, there are two independent rational amplitudes that are unitary, $s \leftrightarrow u$ symmetric, spin-one Regge bounded and have all states at $s = m^2$:
\begin{subequations}\label{eq:RuledIn1}
\begin{align}
	M_{\text{spin-0}}&\;=\;\frac{m^2}{m^2-s}+\frac{m^2}{m^2-u}\,,\\
	M_{su-\text{pole}}&\;=\;\frac{m^4}{\left(m^2-s\right)\left(m^2-u\right)}-\alpha_{J=0}^{(D)}M_\text{spin-0}\,.
	\label{eq:Msu-pole}
\end{align}
\end{subequations}
The former corresponds to the exchange of a single particle of mass $m$ and spin $J=0$, while the spectrum of the latter consists of an infinite tower of spins $J\geq 1$ at mass $m$. In \eqref{eq:Msu-pole}, $\alpha^{(D)}_{J=0}$ is chosen such that it cancels completely the spin-zero contribution.

The contribution from each spin to a given pole is obtained by expanding the residue (at fixed $u$) in Gegenbauer polynomials,
\begin{equation}\label{eq:GegenExp}
    M(s,u) \sim \frac{m^2\sum_{J=0}^{\infty}\gamma_J^{(D)}\legP_J\left(1+\frac{2u}{s}\right)}{m^2-s}
    \,,
\end{equation}
and it can be extracted by using the orthogonality relations of $\legP_J (x)$, \cite{Correia:2020xtr}. Specifically, we have
\begin{equation}\label{eq:gammacoef}
    m^2\gamma_J^{(D)}=-\frac{1}{2}\mathcal{N}_D\, n_J^{(D)}\int_{-1}^1 dx\, (1-x^2)^\frac{D-4}{2}\legP_J(x)\, \text{Res}_{s=m^2}\Big[M\big(s,u(x)\big)\Big]\,,
\end{equation}
where $x=1+\frac{2u(x)}{s}$ and $\mathcal{N}_D=\frac{(16\pi)^\frac{2-D}{2}}{\Gamma\left(\frac{D-2}{2}\right)}$. Applying this for $J=0$ to the first term of \eqref{eq:Msu-pole} shows that we must choose 
\begin{equation}\label{eq:alpha0}
    \alpha_{J=0}^{(D)}=\frac{2}{3}\, {}_2F_1\left(\frac{1}{2},1,\frac{D-1}{2},\frac{1}{9}\right)\,\qquad (\text{for }D\geq 3)\,,
\end{equation}
\begin{equation*}
\alpha_{J=0}^{(3)}  = \frac{1}{\sqrt{2}} \, , \quad \alpha_{J=0}^{(4)}  = \log 2 \, , \quad \alpha_{J=0}^{(5)}  =  12 - 8 \sqrt{2} \, ,\, \dots\,,
\end{equation*}
for the subtraction to cancel the contribution from spin 0. The reason for this subtraction is to saturate the unitarity condition; anything greater than \eqref{eq:alpha0} would break unitarity for the spin-less pole. Similarly to \cite{Caron-Huot:2020cmc}, positivity for the spins $J\geq 1$ can be verified with the Froissart-Gribov formula (reviewed e.g.\ in section 2.4 of \cite{Correia:2020xtr}) by noting that \eqref{eq:gammacoef} becomes proportional to $Q_J^{(D)}(3)$, which is a positive function for $D\geq 3$.

\begin{figure}[ht]
\centering
\includegraphics[scale=1.1]{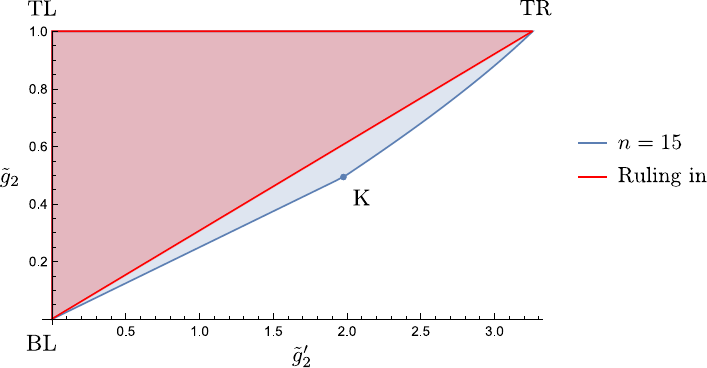}
\caption{Part of the plane $(\tilde g_2',\tilde g_2)$ ruled in by the amplitudes \eqref{eq:RuledIn1} compared to the numerical ruling out at Mandelstam order $n=15$. The three corners (BL, TL and TR) are ruled in, but there remains a sliver containing the kink (K).}
\label{fig:rulingin1}
\end{figure}

By expanding the amplitudes \eqref{eq:RuledIn1} at low energies and comparing to \eqref{eq:Mlow} we find their coordinates in the space of couplings,
\begin{subequations}
\begin{alignat}{3}
	M_{\text{spin-0}}:&\qquad \tilde g_2 =\frac{M^2}{m^2} \, , \qquad && \tilde g'_2  =0 \,,\qquad &&\tilde g_3 =\frac{M^4}{m^4}\,,\\
	M_{su-\text{pole}}:&\qquad \tilde g_2 =\frac{M^2}{m^2} \, , \qquad && \tilde g'_2  =\left(\frac{1}{1-\alpha^{(D)}_{J=0}}\right)\frac{M^2}{m^2} \,, \qquad &&  \tilde g_3 =\frac{M^4}{m^4}\,.
\end{alignat}
\end{subequations}
These amplitudes are crossing symmetric, unitary and have the correct Regge behavior for any mass $m$, but we can only accept them as long as their poles are above the cutoff $M$. Thus, each of them describes a straight line on the plane $(\tilde g_2',\tilde g_2)$ stretching from the origin, to $\tilde g_2=1$. Since positive linear combinations of unitary amplitudes remain unitary, these two amplitudes rule in the whole convex hull of their corresponding lines, shown in red in figure \ref{fig:rulingin1}. The ruled-in region covers an impressively large part of the exclusion plot. In particular, it saturates the left and top bounds and it rules in the three corners of our plot; $\text{BL}=(0,0)$, $\text{TL}=(0,1)$ and $\text{TR}=(\tfrac{1}{1-\log 2},1)\simeq (3.25889135327...,1)$, in perfect agreement with \eqref{bounds}. There remains, however, a sliver of the exclusion plot to be ruled in, much like in \cite{Caron-Huot:2020cmc}.

This agreement is not restricted to the plane $(\tilde g_2',\tilde g_2)$. Moving up in dimensions to the space $(\tilde g_2',\tilde g_2,\tilde g_3)$, we find that the lines of the two amplitudes in \eqref{eq:RuledIn1} become parabolas connecting the origin BL to the corners TL and TR. They fall exactly on the blue sheet of figure \ref{fig:3Dviews}, and the orange sheet is given by their convex hull. Again, there remains a sliver (containing the kink (K)) to be ruled in.

Despite being healthy amplitudes by our criteria, this does not mean that \eqref{eq:RuledIn1} can  be realized in physical large $N$ gauge theories. Rather, they should be regarded as spurious solutions that saturate the bounds and keep us from exploring more interesting regions. One could get rid of them by incorporating further assumptions.

Nevertheless, from this exercise we begin to understand how the healthy amplitudes are arranged in the space of couplings according to their spectra: amplitudes with all their poles at infinity lie at the origin and we find lower and lower masses as we move away from it, until we reach the top bound $\tilde g_2=1$, where all the masses in our amplitudes hit the cutoff $M$. In addition, from the sum rule
\begin{equation}\label{eq:g21sumRule}
    g_{2,1}=\frac{1}{(D-2)}\avg{\frac{\JJ^2}{m^4}}\,,\qquad \Big(\JJ^2\equiv J(J+D-3)\Big)
\end{equation}
it is clear that the axis $\tilde g_2'=0$ is inhabited by amplitudes that only involve scalars. Higher spins move them away from the axis.

The question is what happens in the remaining sliver, and what amplitude(s) lives at the kink. We now address this question using the extra information that figures \ref{fig:RHOexclPlot} and \ref{fig:RHOcoupling} provide. We will discuss the two regions of the right bound separately; first the part below the kink (from BL to K) and then the part above it (from K to TR). In what follows we are going to refer back to figures \ref{fig:RHOexclPlot} and \ref{fig:RHOcoupling} multiple times, so this would be a good point to familiarize oneself with them.

\subsection{Bound below the kink}\label{sec:belowKink}
Figure \ref{fig:RHOexclPlot} was made by assuming a spectrum with a (fixed) isolated rho at the original cutoff $M\equiv m_\rho$ and a continuum after a new (moving) cutoff $M'$. As $M'\to \infty$, the allowed region shrinks all the way to the lower-right bound connecting the origin to the kink; the Skyrme line,
\begin{equation}\label{eq:SkyrLine}
    \tilde g_2'=4\tilde g_2\,.
\end{equation}
As noted at the end of section \ref{sec:newExcPlot}, this implies that on this line there must live (at least) a theory with just a rho in the spectrum and all the higher mesons (if any) pushed to infinity.

The amplitude for the exchange of a single spin-one particle reads
\begin{equation}\label{eq:Mspin1}
    M_{\text{spin-1}}\;=\;\frac{m_\rho^2+2u}{m_\rho^2-s}+\frac{m_\rho^2+2s}{m_\rho^2-u}\,,
\end{equation}
since the spin-one Gegenbauer is $\legP_1\left(1+\frac{2u}{m_\rho^2}\right)=1+\frac{2u}{m_\rho^2}$.
This amplitude is $s\leftrightarrow u$ symmetric and unitary, but it does not satisfy spin-one Regge behavior due to the factor of $s$ in the numerator of the $u$-channel term. It grows too fast with energy. Thus, this amplitude cannot be ruled in as is. Luckily, there is a straightforward fix for this; we just have to add crossing-symmetric poles at a much higher scale $m_\infty$,
\begin{equation}\label{eq:Mspin1UV}
    M_{\text{spin-1}}^{\text{(UV)}}\;=\;\frac{m_\rho^2+2u}{m_\rho^2-s}\left(\frac{m_\infty^2}{m_\infty^2-u}\right)+\frac{m_\rho^2+2s}{m_\rho^2-u}\left(\frac{m_\infty^2}{m_\infty^2-s}\right)\,.
\end{equation}
Now, the extra power of $s$ in the denominator damps the amplitude at high energies, curing the Regge behavior. At low energies, if $m_\infty$ is taken large enough (formally $m_\infty\to\infty$), this amplitude reduces to \eqref{eq:Mspin1}. But the $m_\infty$ pole cannot be removed altogether since we must step on it at some point to save the Regge behavior.

Of course, we can only accept the amplitude $M_{\text{spin-1}}^{\text{(UV)}}$ if it is unitary. To check this we must expand the residue of each pole in Gegenbauers as in \eqref{eq:GegenExp} and verify that all the coefficients are positive. At $m_\rho$, all the $\gamma_J^{(D)}$ are positive and they all tend to zero as $m_\infty\to \infty$ except for $\gamma_1^{(D)}\to 1$, as expected. At the other pole, $m_\infty$, all the coefficients are again positive.\footnote{Like with the $M_{su-\text{pole}}$ amplitude, positivity can be checked here with the Froissart-Gribov formula \cite{Correia:2020xtr,Caron-Huot:2020cmc}. In this case, \eqref{eq:gammacoef} becomes proportional to $Q_J^{(D)}\left(1+\frac{2 m_\infty^2}{m_\rho^2}\right)$ for the coefficients at the rho pole and to $Q_J^{(D)}\left(1+\frac{2 m_\rho^2}{m_\infty^2}\right)$ at the $m_\infty$ pole. The coefficients are positive in both cases because $Q_J^{(D)}(x)$ is a positive function in the range $1<x<\infty$ for $D\geq 3$.} We conclude that \eqref{eq:Mspin1UV} is a healthy amplitude and we should embrace it. In a sense, the $m_\infty$ pole plays the role of a Higgs boson in UV-compleing the spin-one amplitude, but in this case we have an infinite tower of spins. Once again, though, being a ``healthy'' amplitude by our standards does not make it a \textit{physical} theory.

The low energy couplings for this UV-completed spin-one amplitude (in the limit $m_\infty\to~\infty$) are
\begin{equation}\label{eq:gnlSpin1}
    M_{\text{spin-1}}^{\text{(UV)}}:\qquad \tilde g_2 =\frac{1}{3}\frac{M^2}{m_\rho^2} \, , \qquad  \tilde g'_2  =\frac{4}{3}\frac{M^2}{m_\rho^2} \,,\qquad \tilde g_3 =\frac{1}{3}\frac{M^4}{m_\rho^4}\,,
\end{equation}
which satisfy \eqref{eq:SkyrLine}. So this amplitude rules in (part of) the Skyrme line! However, since \eqref{eq:Mspin1UV} is only allowed for masses $m_\rho\geq M$, it only rules in the Skyrme line up to $(\frac{4}{3},\frac{1}{3})$, which is significantly below the kink (even when taking into account the slow convergence from figure \ref{fig:kinkvalue}). Still, it is remarkable that this bound is saturated by such a simple amplitude. What is actually not that surprising is that \eqref{eq:Mspin1UV} matches the Skyrme model, since we saw in section \ref{sec:HLS} that the Skyrme model can be obtained from integrating out a single rho meson.

\begin{figure}[ht]
\centering
\includegraphics[scale=1.1]{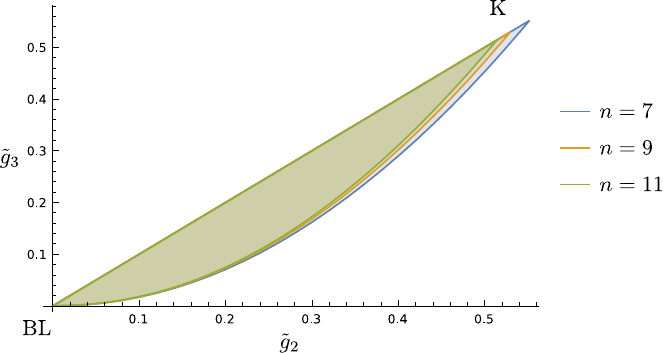}
\caption{Exclusion plot in the space $(\tilde g_3,\tilde g_2)$ along the Skyrme line $\tilde g_2'=4\tilde g_2$ extending from the origin (BL) to the kink (K) of figure \ref{fig:exclPlot}, for Mandelstam orders $n=7,9,11$. The lower bounds are fitted well by the parabolas $\tilde g_3=\frac{1}{\tilde g_2^{\text{(K)}}(n)} \,\tilde g_2^2$ with the values of $\tilde g_2^{\text{(K)}}(n)$ given in figure \ref{fig:kinkvalue}.}
\label{fig:g30plot}
\end{figure}

As for an amplitude that reaches the actual kink, it could be a simple modification of \eqref{eq:Mspin1UV}. Like in the case of \eqref{eq:Msu-pole}, where one has to subtract the spin-zero amplitude to saturate unitarity, one can imagine that subtracting something from \eqref{eq:Mspin1UV} will do the job. From the low energy point of view, we would need the subtraction to lower the value of $g_{1,0}$ but leave all the other $g_{n,\ell}$ unchanged. Indeed, if $g_{1,0}$ decreased by
\begin{equation}\label{eq:g10New}
    g_{1,0}=\frac{3}{m_\rho^2}\quad\longrightarrow\quad \frac{3-A}{m_\rho^2}\,,
\end{equation}
we would get the normalized couplings
\begin{equation}\label{eq:janplitude}
    M_{\text{spin-1}}^{\text{(UV)}}-M_{\text{(?)}}:\quad \tilde g_2 =\frac{1}{3-A}\frac{M^2}{m_\rho^2} \, , \quad  \tilde g'_2  =\frac{4}{3-A}\frac{M^2}{m_\rho^2} \,,\quad \tilde g_3 =\frac{1}{3-A}\frac{M^4}{m_\rho^4}\,.
\end{equation}
Such an amplitude would still sit on the Skyrme line \eqref{eq:SkyrLine}, but it would run further up the line to $(\frac{4}{3-A},\frac{1}{3-A})$, which would match the kink for $A\simeq 0.9757$.\footnote{This result is from the position of the kink at Mandelstam order $n=15$. Taking into account the convergence from figure \ref{fig:kinkvalue}, the final kink should lie around $\tilde g_2^{\text{(K)}}\sim 0.42$, which would give $A\sim 0.62$.} The exclusion plot in the plane $(\tilde g_2,\tilde g_3)$, shown in figure \ref{fig:g30plot}, provides further evidence that \eqref{eq:janplitude} are the correct low-energy coefficients, since the bound in that plot is fitted very well by the parabola $\tilde g_3=(3-A) \tilde g_2^2$.

 But what can we subtract from $M_\text{spin-1}^{(UV)}$? The behavior of figure \ref{fig:RHOexclPlot} as $M'\to\infty$ tells us to keep the rho as the only meson in the finite-energy spectrum, so we can only play with the poles at infinity. That is, we can modify $M_\text{spin-1}^{(UV)}$ by amplitudes whose poles (or cuts, if you wish) are eventually pushed to infinity.\footnote{Along these lines, it is worth mentioning that \eqref{eq:Mspin1UV} is by no means unique. It is easy to find different UV completions for the spin-one particle that are equally valid. For instance, the following two amplitudes are unitary, crossing-symmetric and have spin-one Regge behavior;
\begin{align}
	M_{\text{spin-1}}^{\text{(UV)}'}&\;=\;\left(\frac{m_\rho^2+2u}{m_\rho^2-s}+\frac{m_\rho^2+2s}{m_\rho^2-u}\right)\frac{m_\infty^4}{(m_\infty^2-s)(m_\infty^2-u)}\,,\\
	M_{\text{spin-1}}^{\text{(UV)}''}&\;=\;\frac{m_\rho^2-2m_\infty^2\log\left(1-\frac{u}{m_\infty^2}\right)}{m_\rho^2-s}+\frac{m_\rho^2-2m_\infty^2\log\left(1-\frac{s}{m_\infty^2}\right)}{m_\rho^2-u}\,.
\end{align}
The low-energy coefficients of both of these amplitudes match \eqref{eq:gnlSpin1}, so we would still need to subtract something that lowers $g_{1,0}$ if we want to reach the kink.} From the general form of the sum rules \eqref{eq:sumrules}, it looks like an amplitude with poles only at $m_\infty$ will never contribute to the low-energy couplings when compared with the rho. But if it came with an overall coefficient proportional to $m_\infty^2/m_\rho^2$, it would modify $g_{1,0}$ while leaving all the other couplings unchanged. This is exactly what we would need. Unfortunately, we have not been able to find such a subtraction preserving crossing symmetry and the Regge behavior that does not violate unitarity.

\subsection{Bound above the kink}\label{sec:aboveK}
\begin{figure}
\centering
\begin{subfigure}[b]{0.49\textwidth}
   \includegraphics[width=\linewidth]{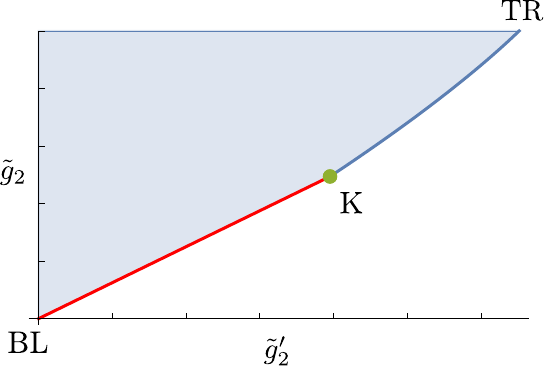}
   \caption{exclusion plot}
\end{subfigure}
\begin{subfigure}[b]{0.49\textwidth}
   \includegraphics[width=\linewidth]{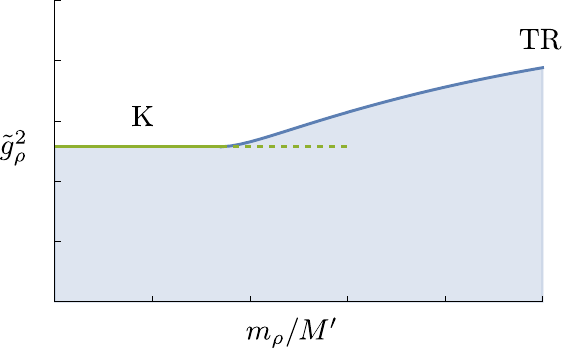}
   \caption{rho coupling plot}
\end{subfigure}
\caption{\label{fig:MatchingPlot}Schematic representation of (a) the exclusion plot in the space of four-derivative couplings and (b) the upper bound on the rho coupling. Matching colors should be identified; i.e.\ the kink (K) on the left corresponds to the plateau on the right and the counterpart of the bound above the kink in (a) is the smooth blue curve in (b), both extending to the top-right corners (TR) of the plots, which are identified.}
\end{figure}

Let us now study what happens along the bound above the kink. For that, we turn to figure \ref{fig:RHOcoupling}. This plot was made with the same spectral assumptions as figure \ref{fig:RHOexclPlot}, so there should be a direct map between them. That is, we should be able to locate the amplitude with the highest possible $\tilde g_\rho^2$ for every $M'/m_\rho$ (figure \ref{fig:RHOcoupling}) on the plane $(\tilde g_2',\tilde g_2)$ (figure \ref{fig:RHOexclPlot}). To help explain the relation between these plots, we have included figure \ref{fig:MatchingPlot}, where we have color-coded the parts of the plot that we will identify. For ease of notation, we will refer to figure \ref{fig:MatchingPlot}(a) as ``exclusion plot'' and to figure \ref{fig:MatchingPlot}(b) as ``rho coupling plot''.

We start with the case $M'/m_\rho=1$, when there is no gap above the rho. The amplitude that saturates the bound on the rho coupling at this point is the very same $M_{su-\text{pole}}$ given in \eqref{eq:Msu-pole}. Indeed, its rho coupling is given by
\begin{align}\label{eq:alpha1}
    \frac{1}{2}\grho^2=\alpha_{J=1}^{(D)}=\,&\frac{D}{4}\, {}_2F_1\left(-\frac{1}{2},1,\frac{D+1}{2},\frac{1}{9}\right)\,\qquad (\text{for }D\geq 3) \nonumber \\
    +\,&\frac{10-9D}{36}\, {}_2F_1\left(\frac{1}{2},1,\frac{D+1}{2},\frac{1}{9}\right)\,,
\end{align}
\begin{equation*}
\alpha_{J=1}^{(3)}  = 3\sqrt{2} - 4\, , \quad \alpha_{J=0}^{(4)}  = 3\left(\log 8 - 2\right) \, , \quad \alpha_{J=0}^{(5)}  =  8\left(17-12\sqrt{2}\right) \, ,\, \dots\,,
\end{equation*}
which, when combined with $g_{1,0}=\frac{1-\alpha_{J=0}^{(D)}}{m^2}$, yields (for $D=4$) the normalized coupling
\begin{equation}
    \tilde g_\rho^2=\frac{3(\log 8 -2)}{1-\log 2}\simeq 0.776674...\,,
\end{equation}
in perfect agreement with \eqref{eq:grhobound}. In the previous section we showed that this simple amplitude rules in the top-right corner (TR) of the exclusion plot, so we learn that the top-right corners of both plots in figure \ref{fig:MatchingPlot} must be identified.

As we increase the gap after the rho by raising $M'/m_\rho$, we start moving down the blue line of the rho coupling plot. For every value of $M'/m_\rho$, the amplitude that saturates the bound on $\tilde g_\rho^2$ sits at the top-right corner (TR) of the corresponding plot in figure \ref{fig:RHOexclPlot}. Since this corner moves down the exclusion plot following closely the original bound, we expect it to match the bound exactly when convergence in $n$ is achieved. Thus, the blue line of figure \ref{fig:MatchingPlot}(a) should be identified with the blue line of figure \ref{fig:MatchingPlot}(b).

Finally, when $M'/m_\rho$ is large enough, we reach a plateau in the rho coupling plot that remains constant all the way to $M'/m_\rho\to\infty$. In the exclusion plot, this happens when the top-right corner (TR) meets the kink (K), where it stops moving. So the kink (K) of figure \ref{fig:MatchingPlot}(a) corresponds to the plateau of figure \ref{fig:MatchingPlot}(b) (both painted green in the figures). This explains the mysterious relation that we pointed out towards the end of section \ref{sec:rhocoupling}: we observed that the value of the plateau in figure \ref{fig:RHOcoupling} for every $n$ matches \textit{exactly} the position of the corresponding kink on the Skyrme line (given in figure \ref{fig:kinkvalue}). As discussed above, the kink is described very well by a UV completion of the rho with a subtraction at infinity that only changes $g_{1,0}$ by \eqref{eq:g10New}. The normalized rho coupling for such an amplitude would be $\tilde g_\rho^2=\frac{1}{3-A}$, which would match exactly the value of $\tilde g_2$ at the kink (see \eqref{eq:janplitude}).

To sum up, the evidence we just presented suggests that the bound above the kink in the exclusion plot corresponds to the upper bound in the right part of the rho coupling plot, while the kink is identified with the plateau. The rho coupling plot thus provides a way to study the bound above the kink of the exclusion plot, which is otherwise hard to study due to its slow convergence in $n$. Our goal for the remainder of this subsection will be to determine \textit{what amplitudes saturate the upper bound in the rho coupling plot away from the endpoints}.

Our claim is that this bound is saturated by amplitudes with the following spectrum:
\begin{enumerate}
    \item a fixed rho at the cutoff $m_\rho=M$,
    \item an infinite tower of higher spins ($J\geq 2$) at the second cutoff $m_\text{hs}=M'$ combining into
    $$\frac{m_\text{hs}^4}{(m_\text{hs}^2-s)(m_\text{hs}^2-u)}-\alpha_{J=0}^{(D)}M_\text{spin-0}-\alpha_{J=1}^{(D)}M_\text{spin-1}\,,$$
    \item and all higher mesons pushed to infinity.
\end{enumerate}
The first observation is that this spectrum has the correct endpoints. At one end, when $m_\text{hs}=m_\rho$, the spectrum matches $M_{su-\text{pole}}$; the amplitude at the top-right corner (TR). At the other end, when $m_\text{hs}\to \infty$, we are left with a single rho and a bunch of particles at infinity; the spectrum that we identified for the plateau. In between, the upper bound of the rho coupling plot should be covered by tuning $m_\rho/m_\text{hs}$.

To prove the claim, we just have to redo the rho coupling plot with this more refined spectral assumption and check that we obtain exactly the same bound. If this is the case, this guarantees that there exists at least an amplitude with this spectrum saturating the bound. This plot can be done following the recipe of section \ref{sec:rhocoupling} but adding the explicit contribution from the $su$-pole, i.e.\ high energy averages should now be shifted by
\begin{equation}
	\avg{F\left(m^2,J\right)}\longrightarrow  \frac{1}{2}\grho^2\, F\left(m_\rho^2,J=1\right) +\avg{F\left(m^2,J\right)}_{su-\text{pole}}+\avg{F\left(m^2,J\right)}'' \,,
\end{equation}
where $\left<\cdots\right>''$ averages over masses above a new cutoff $M''>m_\text{hs}$ and the relevant averages at the $su$-pole can be derived from
\begin{align}
    \avg{\frac{ \legP_J(  1+\frac{2u}{m^2})}{m^{2k}}}_{su-\text{pole}} &= \frac{-1}{2 \pi i } \oint_{m_\text{hs}^2} ds'\,\frac{M(s', u)}{s'^{k+1}} \\
    &=\frac{m_\text{hs}^2}{m_\text{hs}^2-u}\frac{1}{m_\text{hs}^{2k}}-\frac{\alpha_{J=0}^{(D)}}{m_\text{hs}^{2k}}
    -\alpha_{J=1}^{(D)}\frac{\legP_1\left(  1+\frac{2u}{m_\text{hs}^2}\right)}{m_\text{hs}^{2k}}\,.\nonumber
\end{align}
Solving the new dual problem with \texttt{SDPB} \cite{sdpb} for a fixed $m_\rho$ and varying $m_\text{hs}\in[m_\rho,\infty)$ yields an exclusion plot \textit{identical} to figure \ref{fig:RHOcoupling} (identifying $m_\text{hs}=M'$). Moreover, the plot is independent of the cutoff $M''$; it can be pushed as high as desired and the plot remains the same. This proves our claim; the bound of the rho coupling plot is indeed saturated by an amplitude with the spectrum specified above, and so should be the bound above the kink in the exclusion plot, since we identified them.

\subsection{Ruling in (part of) the sliver}
With all this information, we can engineer an allowed analytical solution that rules in a larger area than figure \ref{fig:rulingin1}. Let us discuss it in detail. We start from the $su$-pole amplitude \eqref{eq:Msu-pole} at a mass $m_\text{hs}$ and we remove the spin-one contribution from it. Then, we add the spin-one pole back to the amplitude at a lower mass $m_\rho$. Since a bare spin-one amplitude spoils the spin-one Regge behavior, we must use the UV-completed version \eqref{eq:Mspin1UV}. This is where the masses at infinity become essential. Thus, we consider the amplitude
\begin{equation}\label{eq:Mguess}
    M_\text{guess}=M_{su-\text{pole}}(m_\text{hs})-\alpha_{J=1}^{(D)}M_{\text{spin-1}}^{(UV)}(m_\text{hs}) +  \lambda\, M_{\text{spin-1}}^{(UV)}(m_\rho)\,,
\end{equation}
where we denote in parentheses the location of the poles (the dependence on $m_\infty$ is left implicit). This amplitude depends on a parameter $\lambda$ that in principle could be a function of the masses. From our discussion above we only know $\lambda\to\alpha_{J=1}^{(D)}$ as $m_\text{hs}\to m_\rho$, but it is otherwise unfixed. We will fix it by demanding that the full amplitude be unitary.

$M_\text{guess}$ will be unitary if and only if each of the poles $m_\rho$, $m_\text{hs}$, $m_\infty$ is so (in the sense of \eqref{eq:GegenExp}). Unitarity at $m_\rho$ follows directly from section \ref{sec:belowKink} provided that $\lambda \geq 0$. For the pole at $m_\text{hs}$, all the coefficients $\gamma_{J}^{(D)}$ are positive except for the $J=0$ one, which is negative. This contribution has its origin in $M_{\text{spin-1}}^{(UV)}(m_\text{hs})$ and it dies as $m_\infty\to\infty$, but we cannot accept it because we conceptually want $m_\infty$ to take large but finite values. Thankfully, we can completely remove this contribution by subtracting $M_\text{spin-0}$ with the correct coefficient\footnote{The exact form of this coefficient is
\begin{align}
    \beta_{J=0}^{(D)}\,&=\frac{\frac{1}{2}m_\text{hs}^2}{m_\text{hs}^2+m_\infty^2}
    \left[\frac{D}{D-1} {}_2F_1\left(\frac{-1}{2},1,\frac{D+1}{2},\left(1+\frac{2m_\infty^2}{m_\text{hs}^2}\right)^{-2}\right)\right. \\
    &-\left.\left(1-\frac{1}{D-1}\left(1+\frac{2m_\infty^2}{m_\text{hs}^2}\right)^{-2}\right){}_2F_1\left(\frac{1}{2},1,\frac{D+1}{2},\left(1+\frac{2m_\infty^2}{m_\text{hs}^2}\right)^{-2}\right)\right]\,,
\end{align}
but we will not need it since it dies as $\beta_{J=0}^{(D)}\sim \frac{1}{m_\infty^2}$ when $m_\infty\to\infty$ and thus the ruled-in bound will be independent from it.} since $M_\text{spin-0}$ alone already has a good Regge behavior. So far, we have seen that $M_\text{guess}$ can be made unitary at $m_\rho$ and $m_\text{hs}$. At the last pole, $m_\infty$, the leading term of every coefficient in the limit $m_\infty\to\infty$ is proportional to $\gamma_J^{(D)}\sim (\lambda -\alpha_{J=1}^{(D)})$ with positive coefficients. So, the negative contributions from $M_{\text{spin-1}}^{(UV)}(m_\text{hs})$ can be balanced off by picking $\lambda=\alpha_{J=1}^{(D)}$. This choice is enough to guarantee unitarity as the subleading terms in the limit $m_\infty\to\infty$ are positive.
 
\begin{figure}[ht]
\centering
\includegraphics[scale=1.1]{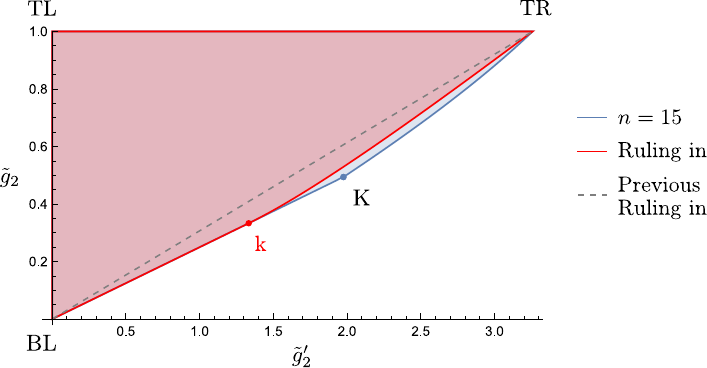}
\caption{Further ruling in achieved with the amplitude \eqref{eq:Mfinal}. It rules in part of the sliver of figure \ref{fig:rulingin1} but there is a bit that still remains.}
\label{fig:rulingin2}
\end{figure}

\begin{figure}[ht]
\centering
\includegraphics[scale=1.1]{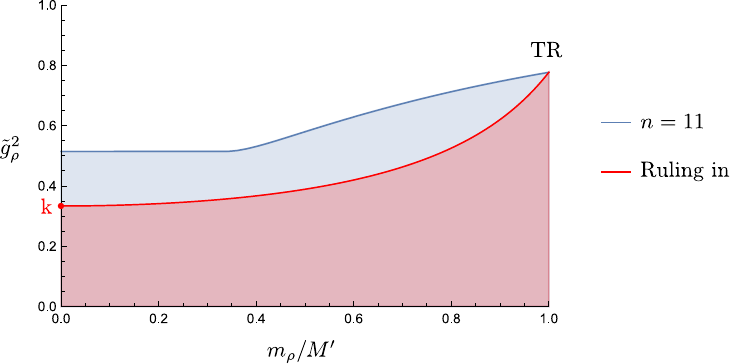}
\caption{Upper bound on $\tilde g_\rho^2$ ruled in by the amplitude \eqref{eq:Mfinal}, compared to the numerical ruling out at Mandelstam order $n=11$. This amplitude rules in the top-right corner, but it is far from the numerical bound everywhere else.}
\label{fig:rulingin3}
\end{figure}

In conclusion, the amplitude
\begin{equation}\label{eq:Mfinal}
    M_\text{final}=M_{su-\text{pole}}(m_\text{hs})-\alpha_{J=1}^{(D)}\left(M_{\text{spin-1}}^{(UV)}(m_\text{hs}) -\beta_{J=0}^{(D)}M_\text{spin-0}(m_\text{hs})\right)+  \alpha_{J=1}^{(D)}\, M_{\text{spin-1}}^{(UV)}(m_\rho)
\end{equation}
is unitary, crossing-symmetric and spin-one Regge bounded. It is therefore acceptable and we can use it to analytically rule in parts of the space of theories. Its low-energy coefficients (in the limit $m_\infty\to\infty$) are
\begin{alignat}{2}
	M_{\text{final}:}\qquad
	g_{1,0} \,&= \frac{1-\alpha_0-3\alpha_1}{m_\text{hs}^2}+\frac{3\alpha_1}{m_\rho^2}\,,\qquad 
	&&\phantom{\frac{1}{2}}\,\,g_{2,0} = \frac{1-\alpha_0-\alpha_1}{m_\text{hs}^4}+\frac{\alpha_1}{m_\rho^4}\,, \\
	2g_{2,1} \,&= \frac{1-4\alpha_1}{m_\text{hs}^4}+\frac{4\alpha_1}{m_\rho^4}\,,
	&&\frac{1}{2}\grho^2 = \alpha_1\,,\nonumber
\end{alignat}
where we have adopted a simplified notation for the $\alpha_{J}^{(D)}$ coefficients. The new area ruled in by this amplitude is plotted in figures \ref{fig:rulingin2} and \ref{fig:rulingin3}.

On the $(\tilde g_2',\tilde g_2)$ plane (figure \ref{fig:rulingin2}), $M_\text{final}$  enlarges the ruled-in region by ruling in part of the sliver of figure \ref{fig:rulingin1}! The new ruling-in bound splits into two parts as we expected. First, if we keep the rho fixed at the cutoff $m_\rho=M$ and we tune $m_\text{hs}$, we cover the upper part of the bound (connecting TR to k). This part starts at $m_\text{hs}=M$, where the amplitude reduces to $M_{su-\text{pole}}$, and it ends at $m_\text{hs}\to\infty$, where it corresponds to $M_\text{spin-one}^{(UV)}$. Second, by keeping $m_\text{hs}=\infty$ and tuning $m_\rho$ we cover the lower segment (connecting k to BL). Unfortunately, there still remains a small part of the sliver to be ruled in, which happens to contain the kink. To push the ruling in up to the kink, one would have to find an amplitude whose low-energy couplings reduce to \eqref{eq:janplitude}.

On the $(m_\rho/M',\tilde g_\rho^2)$ plane (figure \ref{fig:rulingin3}), in contrast, the bound is made of just one part, where $m_\rho=M$ is fixed and $m_\text{hs}(\equiv M')\in[M,\infty)$. At the top-right corner (TR), we correctly recover $M_{su-\text{pole}}$, but at the other end we reach $\tilde g_\rho^2=\frac{1}{3}$, corresponding to $M_\text{spin-1}^{(UV)}$, which is not quite at the numerical value. In fact, apart from the top-right corner, there is no part of the numerical bound that is saturated by $M_\text{final}$, and in particular this simple amplitude does not account for the sharp change in behavior from the curve to the plateau. It would be very interesting to find an amplitude that does so.

\subsection{A note on the spectrum from SDPB}
\begin{figure}[ht]
\centering
\includegraphics[width=\textwidth]{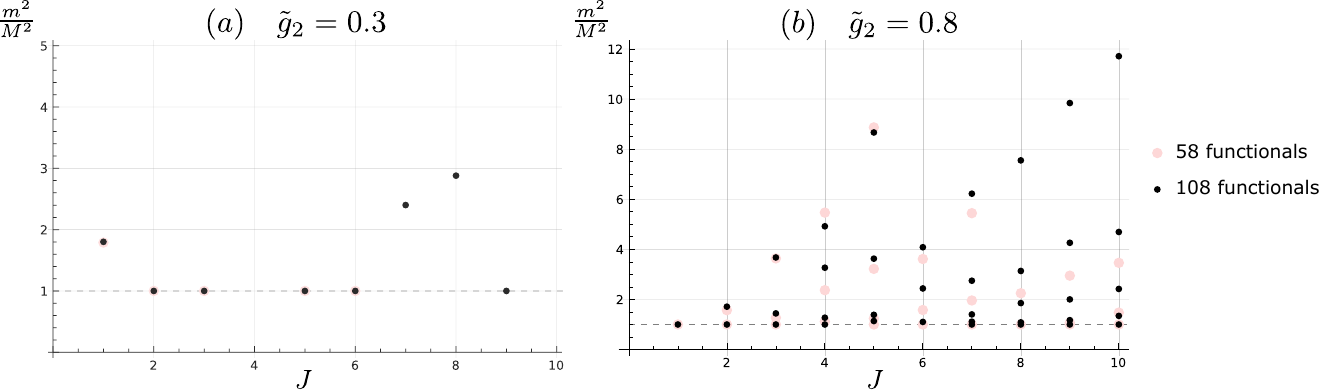}
\caption{Spectrum on the bound for representative points (a) below ($\tilde g_2=0.3$) and (b) above ($\tilde g_2=0.8$) the kink. Black dots were computed using a 108-dimensional space of functionals whereas for the red dots we used 58-dimensional functionals. For very low spins they seem to have converged, but the dots differ for higher spins.}
\label{fig:spectrum}
\end{figure}

When trying to identify the spectrum of the theories that saturate our numerical bounds we encountered a technical point that we think is worth mentioning. In principle, one could directly use the results from \texttt{SDPB} to solve for the spectrum numerically. The optimization problem of section \ref{sec:dualprob} looks for a ``functional'' $\vec{\alpha}$ that satisfies $\vec \alpha\cdot\vec v_\text{HE}\left(m^2,J\right)\geq 0$ for every mass and spin. So, to get the spectrum that saturates the bound one just needs to look for the zeroes of the functional, i.e.\ solve for the $m^2$ and $J$ that make $\vec \alpha\cdot\vec v_\text{HE}\left(m^2,J\right)$ vanish. However, this procedure generates plots like those of figure \ref{fig:spectrum}, which look very different from the spectrum described above.\footnote{At first sight, the spectrum of figure \ref{fig:spectrum}(b) might look closer to what one expects for QCD, with a set of different Regge trajectories, but the dots are on the wrong side of the plot. The Regge trajectories from string theory (and those observed in nature) are of course such that lower masses have fewer spins, not  the other way around.} This is somewhat puzzling because theories that saturate a bound are generally expected to be unique.

One possibility is that some of the zeroes might be numerical artifacts such that when they are removed one recovers the spectrum discussed above, and the truth is that the zeroes of the functional do not tell the whole story. To see whether the corresponding particles are relevant in the theory, one has to look at their coupling constants; particles whose couplings are too small effectively drop out. The coupling constant for each pole can be retrieved from \texttt{SDPB} just like one extracts the OPE coefficients in the conformal bootstrap \cite{Komargodski:2016auf,Simmons-Duffin:2016wlq}. Doing so reveals that the coupling constants of many points of figure \ref{fig:spectrum} are down by several orders of magnitude, indicating that they might indeed be in some sense spurious and should be removed.

This explanation appears to resolve the mismatch in figure \ref{fig:spectrum}(a). The coupling constants in this case are $\sim 10^{-2}$ for the $J=1$ point, $\sim 10^{-5}$ for $J=2$ and $\sim 10^{-9}$ for all other spins. On the other hand, we know that \eqref{eq:Mspin1UV} saturates the bound at this point (see figure \ref{fig:rulingin2}). The spectrum of this amplitude consists of a single rho at a finite mass and everything else pushed to infinity. So, the two spectra match if we disregard all the poles of spin $J>2$ in figure \ref{fig:spectrum}(a), which is justified by how small their couplings are.

Similarly, for the spectrum above the kink (figure \ref{fig:spectrum}(b)) many zeroes have very small couplings, but even when disregarding them, the spectrum does not quite match our guess from section \ref{sec:aboveK}. This could be a matter of convergence; that the spectra will match when sufficient null constraints are included, but it could also be that there really is some degeneracy in the solutions that saturate the bound.\footnote{It does not have to be an \textit{exact} degeneracy---which are usually associated to numerical convergence issues in \texttt{SDPB}, but just an approximate degeneracy whereby different solutions saturate bounds that differ by an imperceptible amount.} It is conceivable that more than one theory might yield the same coefficients since, after all, we are first expanding amplitudes at low energies and then projecting down to a two-dimensional slice. The resolution of this puzzle remains unclear to us.

In any case, it has proven useful for figuring out \textit{a possible spectrum} that saturates the bounds to play with gaps, i.e.\ recomputing the bound(s) with more detailed spectral assumptions. If the bound gets lower, the assumption was wrong. But if the bound remains the same, we can rest assured that there exists \textit{at least} an amplitude with that spectrum saturating the bound. Despite the guesswork this involves, we have seen that it provides insight into the geometry of the bounds. It would be interesting to follow this approach for other bounds in the literature like those in \cite{Caron-Huot:2020cmc} or \cite{Caron-Huot:2021rmr} (in the latter they found a spectrum from \texttt{SDPB} very similar to figure \ref{fig:spectrum}). We expect that similar simplified spectra will be found, which might eventually help us rule in the bounds analytically.

\section{Outlook}
\label{sec:outlook}

In this paper we have 
taken a first step in a systematic program to carve out the space of large $N$ gauge theories. We have focused on the meson sector, assuming that the lowest-lying mesons are massless  scalars in the adjoint representation of $SU(N_f)$, and whose $2\to 2$ scattering amplitude enjoys better than spin-one Regge growth,
as would be the case in large $N$ QCD. 
We have found rigorous two-sided bounds on low-energy EFT parameters. 

The exclusion regions  display intriguing features, including 
a tantalizing kink, but we cannot yet decide whether large $N$ QCD
sits there. 
We have managed to ``explain'' at least a portion of the exclusion boundary in terms of a simple unphysical ansatz for the amplitude (a UV completion of the tree-level rho exchange), but by itself this does not rule out the possibility that large $N$ QCD may sit at the kink. 
It is perhaps not too far fetched to imagine that if large $N$ QCD is in any sense solvable or integrable, such a conspiracy (i.e. that the contribution of mesons above the rho cancels out in the lowest chiral Lagrangian coefficients) might indeed occur. Another logical possibility is that the kink is indeed unphysical,
and that to corner large $N$ QCD one will need to input further physical assumptions. It would be very desirable to have some guidance from  large $N$ lattice data.

Clearly this is just the beginning. 
A natural next step is to consider the full system of $2 \to 2$ amplitudes with 
pions and rhos as external states.
In the conformal bootstrap, mixed correlators are essential to zoom in on physical theories, as in the famous example of  the tiny island corresponding to the 3D Ising model~\cite{Kos:2014bka}. There is no guarantee that things will work out so beautifully in the large $N$ bootstrap, but we are determined to try.
It would also be of interest to see how $1/N$ corrections (which require including low-energy loops in the EFT) affect the geometry of the bounds. 
The generalization to the glueball sector is of obvious relevance. One would have to deal with the small technical complication that the ligthest states are not massless.

The approach to  carving out the space of large $N$  theories  illustrated in this paper
 is very versatile. One can contemplate generalizations to gauge theories in other spacetime dimensions (notably $D=3$), or with different symmetry assumptions.  For example, large $N$ confining supersymmetric gauge theories should  be a very useful playground.

It is tempting to speculate that there may be some interplay 
between the spacetime hadronic bootstrap that we have described in this paper,
and the conformal bootstrap of the (putative) worldsheet string theory
 (see \cite{Huang:2020nqy} for some intriguing ideas in this direction). In a similar vein, it would also be great to make contact with another 
 approach to large $N$ QCD that has been pursued in recent years, based on the 2D EFT on the long string flux tube, see~e.g.~\cite{Aharony:2013ipa, Dubovsky:2015zey, Dubovsky:2016cog, EliasMiro:2019kyf,  EliasMiro:2021nul}.

We look forward to explore some of these directions. With some luck, we'll be able to corner large $N$ QCD or some other theory of physical interest.

\acknowledgments
Our work is supported in part by NSF grant PHY-1915093.
We thank Chris Beem, Simon Caron-Huot, Justin Kaidi, Zohar Komargodski,  Dalimil Maz\'a\v c, Martin Ro\v cek, Slava Rychkov and Balt van Rees for useful discussions and suggestions. 

\appendix

\section{Internal states in pion scattering}\label{app:internalstates}
In this appendix we review which mesons can appear as intermediate states in $2\to 2$ pion scattering. Symmetries and conserved quantities impose selection rules that forbid some combinations of quantum numbers for the exchanged mesons. After reviewing the classification of mesons in the quark model, which is exact at large $N$, we derive the selection rules for the scattering of the familiar $SU(2)$ pions. We defer the analysis for general $N_f$ to appendix \ref{app:Nf}. Needless to say, the results of this appendix are completely standard  -- we just apply the  conservation laws of strong interactions (reviewed e.g.\ in \cite{meyer}) to the case at hand.

In the quark model mesons are treated as $q\bar q$ bound states, and this becomes a rigorous statement at large $N$. As such, they come in irreducible representations of the spin $\textbf{S}$, the orbital angular momentum $\textbf{L}$ and the total angular momentum $\textbf{J}=\textbf{L}\oplus\textbf{S}$. They are therefore classified by the quantum numbers $S$, $L$, $J=|L-S|,|L-S+1|,\ldots,|L+S|$. In addition, in the chiral limit of vanishing quark masses there is the flavor symmetry $SU(N_f)$. For $N_f=2$, this defines an angular momentum algebra dubbed isospin $\textbf{I}$. For $N_f=3$, there is an additional Cartan generator and therefore states are also labelled by their hypercharge $Y$. For higher $N_f$, more quantum numbers appear. Since quarks are in the fundamental representation of the flavor symmetry $\Box$, mesons are classified by the irreducible representations $\Box \otimes \bar \Box=\text{adj}\oplus \bullet$. For $SU(2)$, the possibilities are $I=0,1$, i.e.\ the $\textbf{1}$ and the $\textbf{3}$. For $SU(3)$, we can have $\textbf{3}\otimes\bar{\textbf{3}}=\textbf{1}\oplus\textbf{8}$. Finally, mesons are also eigenstates of the parity operator $\textbf{P}$, with eigenvalue
\begin{equation}
    P=(-1)^{L+1}\,.
\end{equation}

Strong interactions preserve the total angular momentum $\textbf{J}$, parity $\textbf{P}$ and the $SU(N_f)$ flavor symmetry. We can use the corresponding conservation laws to constrain the possible quantum numbers for the exchanged states in pion scattering. In general we need the representation theory of $SU(N_f)$ (developed in appendix \ref{app:Nf}) for this, but one can give an ad hoc derivation of the $SU(2)$ selection rules by defining another symmetry preserved by strong interactions; $G$-parity. $\textbf{G}$ consists of a combination of charge conjugation $\textbf{C}$ and a flip of isospin projection $I_z\to-I_z$. Only states with $I_z=0$ are eigenstates of $\textbf{C}$, whereas all $SU(2)$ mesons are eigenvalues of $\textbf{G}$. The eivenvalues of these operators are related to the angular momentum eigenvalues by
\begin{align}
	C&=(-1)^{L+S}\\
	G&=C(-1)^{I}=(-1)^{L+S+I}
\end{align}
We write these eigenvalues as $(I^G)J^{PC}$. Clearly, not every combination of quantum numbers is allowed. Writing out the possibilities for the angular momenta we find that $J^{++}$ exists for any ``spin'' $J$; $J^{+-}$ only for odd $J$; $J^{-+}$ only for even $J$ and $J^{--}$ for any $J\neq 0$.

Conventionally, the meson families receive different names depending on their isospin representation, see Table \ref{tab:spectrum}. For $SU(2)$ mesons, we only have the first row (corresponding to the $\textbf{3}$) and the ``prime'' of the second row (i.e.\ the $\textbf{1}$). For $SU(3)$, the ``prime'' is still the $\textbf{1}$, but the $\textbf{8}$ is made of the first row, the other element in the second row and two copies of the doublet in the third row. This last row contains states with nonvanishing hypercharge (i.e.\ nonzero strangeness) and $I=\frac{1}{2}$ that are not invariant under $\textbf{G}$. We denote these states by $(\frac{1}{2})J^P$, and their naming convention is $K_J$ for $J^P=0^-,1^+,2^-,\ldots$ and $K_J^*$ for $J^P=0^+,1^-,2^+,\ldots$.
\begin{table}
\caption{\label{tab:spectrum}Standard naming convention for the $SU(3)$ (and $SU(2)$) mesons.}
$$\begin{array}{c||c c c c}
 (I^G)J^{PC} & \begin{array}{c} S=1\\L=\text{odd}\\ \forall J  \end{array} & \begin{array}{c} S=0\\L=\text{odd}\\J=\text{odd} \end{array} & \begin{array}{c} S=0\\L=\text{even}\\J=\text{even} \end{array} & \begin{array}{c} S=1\\L=\text{even}\\J\neq 0 \end{array} \\
 \hline
 \hline
 I=1 & \begin{array}{c} a_J\\(1^-)J^{++} \end{array} & \begin{array}{c} b_J\\(1^+)J^{+-} \end{array} & \begin{array}{c} \pi_J\\(1^-)J^{-+} \end{array} &\begin{array}{c} \rho_J\\(1^+)J^{--} \end{array}\\
 \hline
 I=0 & \begin{array}{c} f_J,f'_J\\(0^+)J^{++} \end{array} & \begin{array}{c} h_J,h'_J\\(0^-)J^{+-} \end{array} & \begin{array}{c} \eta_J,\eta'_J\\(0^+)J^{-+} \end{array} &\begin{array}{c} \omega_J,\phi_J\\(0^-)J^{--} \end{array}\\
 \hline
 I=\frac{1}{2} &\begin{array}{c} K_J^*,K_J\\(\frac{1}{2})J^{+} \end{array} & \begin{array}{c} K_J\\(\frac{1}{2})J^{+} \end{array} & \begin{array}{c} K_J\\(\frac{1}{2})J^{-} \end{array} &\begin{array}{c} K_J^*,K_J\\(\frac{1}{2})J^{-} \end{array}
\end{array}$$
\end{table}

\subsection*{\boldmath Selection rules for \texorpdfstring{$SU(2)$}{SU(2)}}
Let us consider $\pi\pi\to\pi\pi$, where the $SU(2)$ pion\footnote{We have gone back to the usual notation where pion refers only to the isospin triplet. Now the $SU(3)$ massless octet contains pions, kaons and the eta.} $\pi$ has quantum numbers $(1^-)0^{-+}$. For an intermediate state $X$ to appear in this process, the subprocess $\pi\pi\to X$ should be allowed. Or, equivalently, $X$ should be able to decay into two pions. From the conservation laws we learn:
\begin{itemize}
	\item[($\textbf{J}$)] The pions have $J=0$, so the initial angular momentum must come entirely from the relative angular momentum $\ell$ and $X$ must have ``spin'' $J=\ell$.
	\item[($\textbf{I}$)] The initial state can have isospin $1\otimes 1=0\oplus 1\oplus 2$, but there are no single-meson states with isospin $I=2$, so $X$ must have isospin $I=0$ or $1$. This rules out the last row from Table \ref{tab:spectrum}.
	\item[(\textbf{P})] Since the initial state has parity $P=(-1)^2(-1)^\ell$, $X$ must have parity $P=(-1)^J$. That is, it can only have positive parity for even spin and negative parity for odd spin. This immediately rules out the two columns at the center of Table \ref{tab:spectrum}. In addition, it restricts the possible $J$ for the remaining columns; $J$ must be even in the first column and odd in the last one.
	\item[($\textbf{G}$)] The $G$-parity of the initial state is fully determined from that of the pions, $G=(-1)(-1)=+1$. We conclude that $X$ must have positive $G$ and therefore the $a_J$ and $\omega_J,\phi_J$ families are ruled out.
\end{itemize}
All in all, we have found that the only mesons that can be exchanged in the $2\to 2$ scattering of $SU(2)$ pions are
\begin{equation}\label{eq:su(2)states}
    \left\{
    \begin{matrix}\rho_J& \text{for }J=\text{odd}\;\\
    f'_J, (f_J)	&\text{for }J=\text{even}.
    \end{matrix}\right.
\end{equation}
Looking at experimental meson tables \cite{PDG} we find the first resonances: $f_0(500)$, $\rho(700)$, $f_0(980)$, $f_2(1270)$, etc. However, the current understanding is that the $f_0(500)$ (previously known as the $\sigma$) and the $f_0(980)$ are not $q \bar q$ states but exotic mesons (more precisely, meson-meson resonances)~\cite{Pelaez:2015qba, PDG}. Thus, at large $N$ the rho 
(which has spin one) is the first meson exchanged in pion scattering,  followed by  the $f_2(1270)$ (which has spin two).

This discussion concerned the scattering of $SU(2)$ pions embedded in $SU(3)$. When we consider the full scattering of $SU(3)$ ``pions'' (i.e.\ including the kaons and the eta), the exchanged families get extended to the full multiplets. In the chiral limit, all the elements in the multiplets should have the same mass, but in nature the $SU(3)$ (and already the $SU(2)$) are explicitly broken by the differences in the masses of the light quarks. We have, for example, that the $a_2(1320)$ experimentally shows up at a different mass than the $f_2(1270)$.

\section{\boldmath General \texorpdfstring{$N_f$}{Nf}}\label{app:Nf}
\begin{figure}[ht]
\vspace{0.1in}
\centering
\includegraphics[scale=0.4]{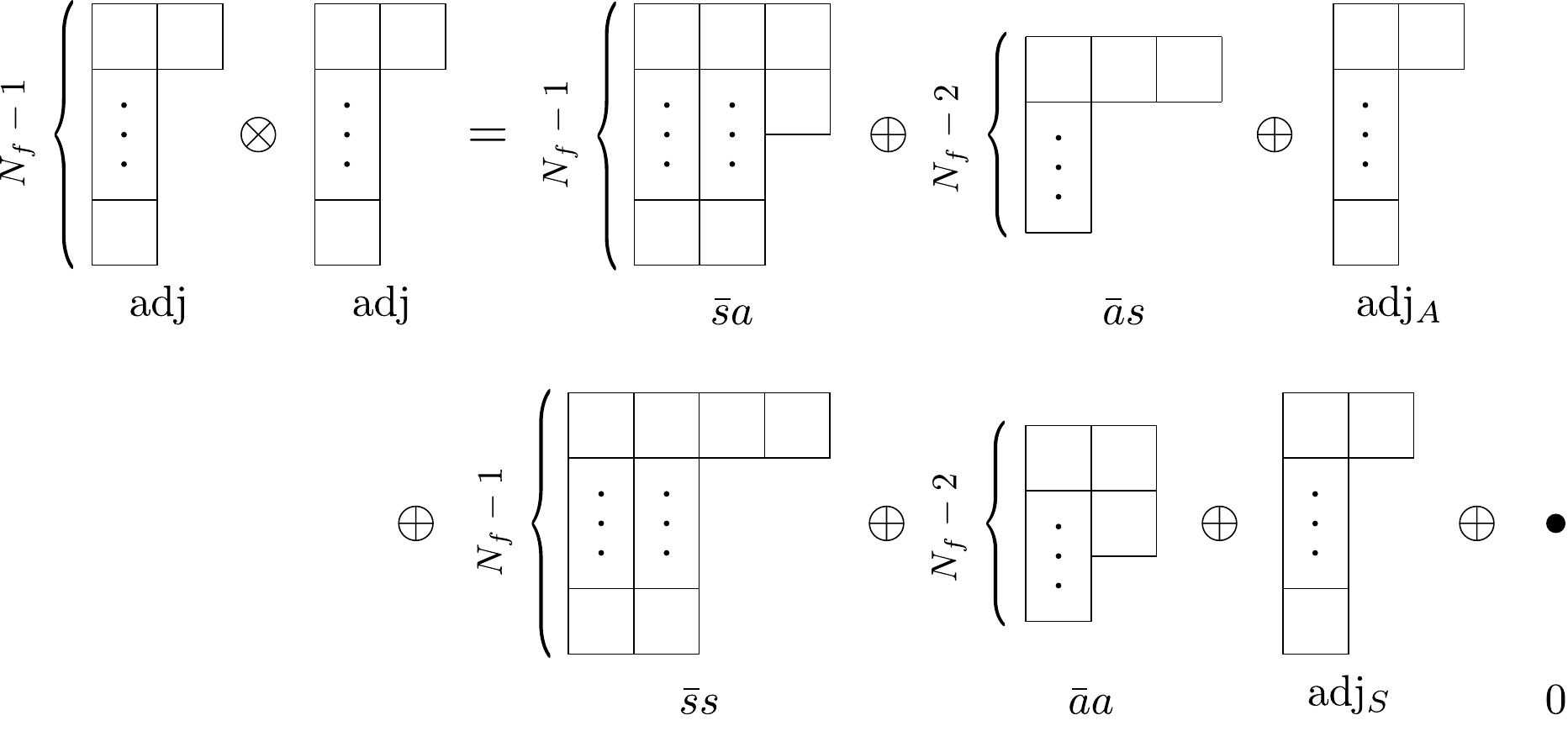}
\caption{Young tableaux for the $SU(N_f)$ irreps in the tensor product of two adjoint irreps together with their names. The irreps in the first row come from antisymmetrizing the two adjoint indices while the ones in the second row come from their symmetrization. For $N_f=3$, the $\bar a a$ does not appear and the decomposition is $\textbf{8}\otimes\textbf{8}=\left(\textbf{10}\oplus \textbf{10}\oplus \textbf{8}\right)_A\oplus\left(\textbf{27}\oplus\textbf{8}\oplus\textbf{1}\right)_S.$ For $N_f=2$, the missing irreps are $\bar s a$, $\bar a s$, $\bar a a$ and $\text{adj}_S$, leaving $\textbf{3}\otimes\textbf{3}=\textbf{3}_A\oplus\left(\textbf{5}\oplus\textbf{1}\right)_S$; the isospin decomposition of section \ref{sec:zweig}.}
\label{fig:Ytableau}
\end{figure}

For a general number of flavors, the analog of \eqref{eq:Idecomp} reads
\begin{equation}
    {\cal T}   =   \sum_{\RR}   \MM^{\RR} (s | t, u) \PP^{\RR}_s\,,
\end{equation}
where the sum runs over the irreps $\RR\in \text{adj}\otimes\text{adj}$, described in figure \ref{fig:Ytableau}. The $s$-channel projectors $\PP^{\RR}_s$ can be found for example in \cite{BandaGuzman:2020wrz} and, in terms of the parametrizations of section \ref{sec:parametriz}, the new amplitudes $\MM^{\RR} (s | t, u)$ are given by\footnote{In the special cases of $N_f=2,3$, some of the irreps of figure \ref{fig:Ytableau} do not show up. The corresponding projectors $\PP^{\RR}_s$ are identically zero, so the expressions of \eqref{eq:channels} remain valid for these cases. One should just ignore the amplitudes for the channels that drop out from the decomposition.}
\begin{subequations}\label{eq:channels}
\begin{align}
    \label{eq:0channel}
    \MM^{0} (s | t, u)=&\,2\frac{N_f^2-1}{N_f}A(s | t, u) + \frac{N_f^2-2}{N_f}\Big(A(t | s, u) + A(u | s, t)\Big)\nonumber\\
    =&\,2\frac{N_f^2-1}{N_f}\Big(M(s,t)+M(s,u)\Big)-\frac{2}{N_f}M(t,u)\,,\\
    \MM^{\text{adj}_S}(s|t,u)=&\frac{N_f^2-4}{N_f}A(s|t,u)+\frac{N_f^2-8}{2N_f}\Big(A(t|s,u)+A(u|s,t)\Big)\nonumber\\
    =&\,\frac{N_f^2-4}{N_f}\Big(M(s,t)+M(s,u)\Big)-\frac{4}{N_f}M(t,u)\,,\\
    \MM^{\text{adj}_A}(s|t,u)=&\,\frac{N_f}{2}\Big(A(u|s,t)-A(t|s,u)\Big)=N_f\Big(M(s,u)-M(s,t)\Big)\,,\\
    \label{eq:symChannel}
    \MM^{\bar s s}(s|t,u)=&\,-\MM^{\bar a a}(s|t,u)=A(t|s,u)+A(u|s,t)=2M(t,u)\,,\\
    \MM^{\bar s a}(s|t,u)=&\, \MM^{\bar a s}(s|t,u)=0\,.
    \end{align}
\end{subequations}
Note that (at large $N$) these amplitudes are not all independent. Also, the reason why the amplitudes in the $\bar s a$, $\bar a s$ channels vanish exactly is again the single-trace structure of the large $N$ amplitude. The $t\leftrightarrow u$ symmetry of $\MM^{\RR}(s|t,u)$ is related to the symmetrization of the adjoint indices, namely, it is symmetric for $\RR=0,\text{adj}_S,\bar a a,\bar s s$ and antisymmetric for $\RR=\text{adj}_A$. Under $s\leftrightarrow t$, the amplitudes in the different channels mix.

The argument now goes as in the main text. At large $N$, the amplitude $M(s,u)$ is a meromorphic function with poles corresponding to the exchange of physical mesons in the spectrum. Since mesons are exactly $q\bar q$ states in this limit, the internal states can only carry irreps appearing in $\Box \otimes \bar \Box=\text{adj}\oplus \bullet$. Thus, of the channels of figure \ref{fig:Ytableau}, the physical intermediate states can only carry the irreps $\RR=0,\text{adj}_A,\text{adj}_S$. The amplitudes $\MM^{\bar s s}(s|t,u)$ and $\MM^{\bar a a}(s|t,u)$ cannot have poles for $s>0$ (at fixed $u<0$). Since both of these amplitudes are proportional to $M(t,u)$, we conclude that the result from section \ref{sec:zweig} generalizes to any $N_f$: the basic amplitude \textit{$M(s,u)$ (for fixed $u<0$) has poles only on the positive real s axis}.

\subsection*{Unitarity}
Following section \ref{sec:unitarity} we can now expand the $\RR$-channel amplitudes in partial waves as
\be
{\rm Im} \, 
  \MM^{\RR} (s | t, u)  =
s^{\frac{4-D}{2}} \sum_{J} n^{(D)}_J  \rho^{(\RR)}_J(s)\,  \legP_J\left(1+\frac{2u}{s}\right)\,,
\ee
and demand that each of them be unitary for the full amplitude ${\cal T}_{ab}^{cd}$ to be so. By comparing the symmetry properties of $\MM^{\RR} (s | t, u)$ and the Gegenbauers under $t\leftrightarrow u$ crossing we see that odd spins contribute only to $\RR=\text{adj}_A$ while even spins contribute both to $\RR=0,\text{adj}_S$. Thus, if we expand the basic amplitude in partial waves as in \eqref{eq:ImMbasic} with
\begin{subequations}
\begin{alignat}{2}
 \rho_J(s) &  =   \frac{N_f/4}{N_f^2-1}\rho^{(0)}_J(s) = \frac{N_f/2}{N_f^2-4}\rho^{(\text{adj}_S)}_J(s)  \qquad && s > 0 \,, \;J \;{\rm even}\, ,\\
\rho_J(s) &  =   \frac{1}{2N_f}\rho^{(\text{adj}_A)}_J(s)  \qquad && s > 0 \,, \;J \;{\rm odd}\, ,
\end{alignat}
\end{subequations}
unitarity for the full amplitude implies positivity of $M(s,u)$, just like in \eqref{eq:unitarity} but now for any $N_f$.

\subsection*{Selection rules}\label{app:Nfselectionrules}
To summarize, from the above discussion we know that at large $N$ and for any $N_f$, the even-spin intermediate states in $\pi\pi\to\pi\pi$ scattering can carry any of the irreps $\RR=0,\text{adj}_S$ while odd-spin states must carry $\RR=\text{adj}_A$. From the explicit form of the projectors
\begin{equation}
    {\PP^{(0)}_s}_{ab}^{cd}=\frac{1}{N_f^2-1}\delta_{ab}\delta^{cd}\,,\quad
    {\PP^{(\text{adj}_S)}_s}_{ab}^{cd}=\frac{N_f}{N_f^2-4}d_{abe}d^{cde}\,,\quad
    {\PP^{(\text{adj}_A)}_s}_{ab}^{cd}=\frac{1}{N_f}f_{abe}f^{cde}\,,
\end{equation}
we learn that the $\RR=0$ family couples to the pions with a three-point vertex $\sim \delta_{ab}$, the $\RR=\text{adj}_A$ couples with $\sim f_{abc}$ and the $\RR=\text{adj}_S$ with $\sim d_{abc}$. Using
\begin{equation}
    f_{abe}f_{cde}=\frac{2}{N_f}\Big(\delta_{ac}\delta_{bd}-\delta_{ad}\delta_{bc}\Big)+d_{ace}d_{bde}-d_{bce}d_{ade}\,,
\end{equation}
we see that the exchange of an $\text{adj}_A$ alone is enough to generate the group theory factors of \eqref{eq:T}, but neither the $0$ or the $\text{adj}_S$ are good on their own. What we need to reproduce the factors from \eqref{eq:T} in the symmetric side is the sum of both. This means that, at large $N$, the $0$ and the $\text{adj}_S$ are degenerate as intermediate states and we always get one when we have the other! Borrowing the parity analysis from appendix \ref{app:internalstates}, which still holds, we get that the intermediate states have parity $P=(-1)^J$. All in all, for general $N_f$ the intermediate states that show up in $\pi\pi\to\pi\pi$ are
\begin{equation}	
    \left\{
    \begin{matrix}\RR=\text{adj}_A& \text{ for }J=\text{odd}, & P=-\;\\
    \RR=0\oplus \text{adj}_S	&\text{ for }J=\text{even}, & P=+.
    \end{matrix}\right.
\end{equation}

This readily reproduces the results of appendix \ref{app:internalstates}, where we derived the $SU(2)$ selection rules. Indeed, for odd $J$ (and negative parity) the internal states form an isospin triplet ($\RR=\text{adj}_A$), while for even $J$ (and positive parity) we have the singlet ($\RR=0$). These correspond, respectively to the families $\rho_J$ and $f_J'$. One advantage of this derivation is that it does not rely on $G$-parity. Its role has been taken by the observation that the symmetry of the adjoint indices is related to the parity of $J$. Another obvious advantage of this derivation is its generality. We now see that, indeed, for the scattering of $SU(3)$ pions, the exchanged mesons extend \eqref{eq:su(2)states} to the whole multiplets: for even $J$ we have all the mesons in the first column of table \ref{tab:spectrum} and for odd $J$ all the mesons in the last column except for the singlet.

\bibliographystyle{ytphys}
\bibliography{references_pions_paper}

\end{document}